\documentclass[a4paper,11pt]{article}
\usepackage{jheppub} 

\usepackage[dvipsnames]{xcolor}
\usepackage[normalem]{ulem}
\usepackage{amsmath,amsfonts,amssymb,mathrsfs}
\usepackage{mathtools}
\usepackage{bbm}
\usepackage[inline]{enumitem}
\usepackage{multirow}
\usepackage{xspace}
\usepackage{mathrsfs}
\DeclareMathAlphabet{\mathscr}{U}{rsfs}{m}{n}
\usepackage{nicefrac}
\usepackage{comment}
\usepackage{cancel}

\newcommand\papertitle{Entanglement in Presence of Topological Interfaces and Dualities}

\usepackage[colorlinks=true]{hyperref}
\hypersetup{
    bookmarks=true,         
    unicode=false,          
    pdftoolbar=true,        
    pdfmenubar=true,        
    pdffitwindow=false,     
    pdfstartview={FitH},    
    pdftitle={\papertitle}, 
    pdfauthor={Christian Northe, Riccardo Poletti, Paolo Rossi},
    pdfnewwindow=true,      
    colorlinks=true,        
    linkcolor=teal,         
    citecolor=Maroon,          
    filecolor=magenta,      
    urlcolor=teal          
}

\newcommand{\change}[2]{#2}
\newcommand{\id}{\mathbbm{1}}
\def\a{\alpha}
\def\b{\beta}

\newcommand{\p}	{\partial}
\newcommand{\bp}{\bar{\partial}}

\newcommand{\R}	{\mathbb{R}}
\newcommand{\C}	{\mathbb{C}}
\newcommand{\Z}	{\mathbb{Z}}
\newcommand{\N}	{\mathbb{N}}

\newcommand{\cA}{\mathcal{A}}
\newcommand{\cB}{\mathcal{B}}
\newcommand{\cC}{\mathcal{C}}

\newcommand{\cF}{\mathcal{F}}
\newcommand{\cG}{\mathcal{G}}
\newcommand{\cH}{\mathcal{H}}
\newcommand{\cI}{\mathcal{I}}

\newcommand{\cP}{\mathcal{P}}

\newcommand{\cS}{\mathcal{S}}

\DeclareMathOperator{\tr}{tr}

\DeclareMathOperator{\Hom}{Hom}
\DeclareMathOperator{\End}{End}
\DeclareMathOperator{\rep}{Rep}

\newcommand{\bra}[1]	{\langle{#1}\vert}
\newcommand{\ket}[1]	{\vert{#1}\rangle}

\newcommand{\ketbra}[2]	{\ket{#1}\bra{#2}}
\newcommand{\corr}[1]   {\left\langle{#1}\right\rangle}

\newcommand{\bz}    {\bar{z}}

\newcommand{\bphi}  {\bar{\phi}}

\newcommand{\ba}    {\bar{a}}
\newcommand{\bJ}    {\bar{J}}
\newcommand{\bL}    {\bar{L}}
\newcommand{\bh}    {\bar{h}}

\newcommand{\bq}    {\bar{q}}
\newcommand{\btau}  {\bar{\tau}}

\newcommand{\bT}    {\bar{T}}
\newcommand{\bi}    {\bar{\iota}}

\newcommand{\tw}    {\tilde{w}}
\newcommand{\tq}    {\tilde{q}}
\newcommand{\ttau}  {\tilde{\tau}}

\newcommand{\cc}        {\mathsf{c}} 
\newcommand{\gf}        {\mathsf{g}} 
\newcommand{\dir}       {\mathsf{D}}
\newcommand{\neu}       {\mathsf{N}}
\newcommand{\gfd}[1]    {\gf_{\dir}^{(#1)}}
\newcommand{\gfn}[1]    {\gf_{\neu}^{(#1)}}

\newcommand{\bbra}[1]	{\langle\!\langle{#1}\lVert}
\newcommand{\bket}[1]	{\lVert{#1}\rangle\!\rangle}
\newcommand{\ibra}[1]	{\langle\!\langle{#1}\lvert}
\newcommand{\iket}[1]	{\lvert{#1}\rangle\!\rangle}

\newcommand{\proj}[1]   {|\hspace{-0.1em}|\,{#1}\,|\hspace{-0.1em}|}

\newcommand{\pr}[2]{%
  \left\|
  \begin{smallmatrix}
    #2 \\
    #1
  \end{smallmatrix}
  \right\|
}

\newcommand{\fus}       {\mathsf{N}}
\newcommand{\fuse}      {\star}
\newcommand{\fusIB}     {K}
\newcommand{\qdim}      {\mathsf{d}}

\newcommand{\modS}      {\cS}

\renewcommand{\th}      {\text{th}}

\newcommand{\iu}{\mathsf{i}}

\newcommand{\confFam}   {\cI}

\newcommand{\niab}      {\mathsf{n}^i_{\alpha\beta}}

\newcommand{\D}         {D} 
\newcommand{\Do}        {\hat{\D}} 
\newcommand{\bD}        {\bar{\D}} 
\newcommand{\DB}        {\check{\D}} 
\newcommand{\bDB}       {\check{\bar{\D}}} 
\newcommand{\dbo}[3]    {\DB^{#1\,#2}_{\;\;#3}}
\newcommand{\bdbo}[3]    {\bDB^{#1\,#2}_{\;\;#3}}
\newcommand{\db}[4]     {\DB^{#1\,#2}_{#3\,#4}}
\newcommand{\go}[1]      {\hat{g}_{#1}}

\newcommand{\I}         {I}
\newcommand{\Io}        {\hat{\I}}
\newcommand{\bI}        {\bar{\I}}
\newcommand{\bIo}       {\hat{\bI}}
\newcommand{\IB}        {\check{\I}}
\newcommand{\bIB}       {\check{\bar{\I}}}
\newcommand{\ib}[4]     {\IB^{#1\,#2}_{#3\,#4}}
\newcommand{\bib}[4]    {\check{\bar{\I}}^{#1\,#2}_{#3\,#4}}
\newcommand{\ibo}[3]    {\IB^{#1\,#2}_{\;\;#3}}
\newcommand{\bibo}[3]   {\bIB^{#1\,#2}_{\;\;#3}}
\newcommand{\Hjunc}[3]  {\cH_{#1|#2}^{#3}}
\newcommand{\njunc}[3]  {\mathsf{n}_{#1|#2}^{#3}}
\newcommand{\TopInt}[1]    {\cF_{#1}}

\newcommand{\dua}       {\sigma}
\newcommand{\stabR}[1]     {S_{#1}^r}
\newcommand{\stabL}[1]     {S_{#1}^l}
\newcommand{\stab}[1]     {S_{#1}}
\newcommand{\T}          {\mathsf{T}}
\newcommand{\To}        {\hat{\T}}
\newcommand{\bTo}        {\hat{\bar{\T}}}

\newcommand{\fusing}[3] {\mathsf{F}^{(#1)#2}_{#3}}
\newcommand{\fusingG}[3]{\mathsf{G}^{(#1)#2}_{#3}}

\newcommand{\fusMa}[5]  {%
  \mathsf{M}_{#1}^{#2}
  \scalebox{0.5}{$
    \begin{bmatrix}
      #3 \\ #4 \\ #5
    \end{bmatrix}
  $}
}
\newcommand{\bfusMa}[5]  {%
  \mathsf{\bar{M}}_{#1}^{#2}
  \scalebox{0.5}{$
    \begin{bmatrix}
      #3 \\ #4 \\ #5
    \end{bmatrix}
  $}
}

\newcommand{\ab}        { {\alpha\beta} }

\newcommand{\wi}        {\mathsf{W}} 
\newcommand{\cyl}       {\mathrm{cyl}}

\newcommand{\X}         {X}

\newcommand{\prob}      {\mathsf{p}}

\newcommand\quotes[1]  {``{#1}"}
\renewcommand\emph[1]  {\textbf{#1}}

\newcommand\secref[1]	{section~\ref{#1}\xspace}

\newcommand\figref[1]	{figure~\ref{#1}\xspace}
\newcommand\appref[1] {appendix~\ref{#1}\xspace}

\title{\papertitle}

\author[a]{Christian Northe,}
\author[a]{Riccardo Poletti}
\author[a]{and Paolo Rossi}

\affiliation[a]{CEICO, Institute of Physics of the Czech Academy of Sciences,
Na Slovance 2, 182 00 Prague 8, Czech Republic}

\emailAdd{northe@fzu.cz, polettiricc@gmail.com, rossip@fzu.cz}

\abstract{Entanglement through interfaces has attracted considerable attention in $2d$ conformal field theory (CFT). However, it is known that field-theoretic predictions based on the existing framework are in general incompatible with numerical results \cite{Roy_2022,Rogerson:2022yim,Roy:2023wer,Sinha:2023hum}. A new framework for entanglement through topological defects was recently proposed in \cite{Northe:2025zmv}. It provides a general description of entanglement through topological defects and successfully reproduces the numerical results for the Ising model in all tested cases and regimes. The key insight is that the relevant quantum correlations are encoded in twisted states, allowing for the construction of the full reduced density matrix (RDM). In this work we pursue two objectives. First, we provide new examples by studying defects in the free boson CFT. Second, we extend the framework to topological interfaces connecting two, possibly distinct, CFTs. Of particular interest are interfaces relating dual theories. We show that the RDM for a duality interface is the projection of the vacuum reduced density matrix onto a single symmetry sector, closely paralleling the framework of symmetry resolution. Unlike symmetry resolution, however, the projection is imposed by the physical interface itself, demonstrating that duality interfaces reflect quantum correlations back into the entangling interval. We establish this mechanism for diagonal and non-diagonal rational CFTs as well as the free boson CFT. Relative entropy allows us to quantify the distinguishability of the duality interface RDM from the vacuum RDM.}

\begin{document}
\maketitle
\flushbottom

\section{Introduction}
\label{sec:intro}
Generalized symmetry has revolutionized our understanding of conserved quantities over the past decade \cite{gaiotto2015generalized}. By now, the study of generalized symmetry has \change{lead}{led} to progress in many branches of theoretical research across high-energy theory, quantum information theory, condensed matter theory and mathematics, see the reviews \cite{schafer2024ictp, Shao:2023gho, bhardwaj2024lectures, costa2024simons} and references therein. Rather than viewing symmetries as groups acting on fields, focus falls on topological operators, their defect networks and their fusion. In contrast to group elements, these operations need not have an inverse, coining the term non-invertible symmetry. In 1+1 dimensions in particular, topological line operators, associated with topological interfaces, have been studied intensely in the framework of rational CFT \cite{Petkova:2000ip, Fuchs:2002cm, fuchs2004tftII, fuchs2004tftIII, Fuchs:2004xi, Fuchs:2007tx, Frohlich:2004ef, Frohlich_2007, Runkel:2007wd}. 

Dualities relate two theories, which can potentially be very distinct. Perhaps the most prominent example is the AdS/CFT correspondence which relates a CFT in $d$ dimensions to a gravitational theory in $d+1$ dimensions \cite{maldacena}. Evidently, these two theories are very distinct in nature. Kramers-Wannier duality, on the other hand, relates a high-temperature Ising model to a low-temperature Ising model. Similarly $\T$-duality relates bosonic theories in 1+1 dimensions compactified at radius $R$ to one at radius $1/2R$. In string theory this implies that small compact dimensions are essentially indistinguishable from large compact dimensions. Besides deepening our understanding of the space of consistent theories \cite{Aganagic:2001uw, Karch:2016sxi, Karch:2016aux, Karch:2018mer, Karch:2019lnn}, dualities provide a powerful tool for exploring otherwise inaccessible regimes of physics by relating them to more familiar and tractable ones.

One important insight, presented first in \cite{Frohlich:2004ef} and formalized fully in \cite{Frohlich_2007}, was that some dualities could be treated on the same footing as symmetries, i.e. by extended topological operators -- a highly non-trivial claim. After all, symmetries are transformations within a single theory, while dualities are a priori maps between distinct theories. In addition, they inherit all favorable features of topological interfaces, leading to superb analytic control.

It is important to understand the effect of topological operators, or more generally interfaces, even non-topological ones, on quantum correlations. This has several reasons. Besides the modern perspective of non-invertible symmetry, (non-topological) interfaces have historically played an important role in modeling impurities \cite{Affleck:1995ge}, which appear naturally and ubiquitously in the laboratory. Moreover, renewed interest in the study of measurements in quantum critical states has also led to the implementation of interfaces \cite{PhysRevB.108.165120, Sun:2023hlu, Hoshino:2024kxk}. 

Entanglement in the presence of interfaces has been intensely analyzed in $1+1$-dimensional CFT. Two classes of setups are typically distinguished. The first concerns impurities, modeled by conformal interfaces, inside the entangling interval. Because the entangling interval is typically, but not exclusively, placed symmetrically around the impurity, this is sometimes called \quotes{symmetric entropy}. Some examples are found in \cite{Sorensen:2009zqb,Afxonidis:2024gne,Chiodaroli:2010ur,Erdmenger:2020hug, affleck2009quantumimpurityproblemscondensed,Liu:2024oxg,Northe:2025qcv}. The second class of problems focuses on entangling intervals terminating on interfaces and quantifies quantum correlations through the interface. A vast body of literature has been established on this so-called \quotes{interface entropy}, beginning with a comprehensive theory \cite{Sakai_2008, Brehm:2015lja, Gutperle:2017enx, Brehm_2016, Gutperle_2016, gutperle2024note}, which enabled many applications, including holography \cite{Karch:2023evr, Karch:2024udk, Tang:2023chv, Karch:2022vot}, thermal systems, junctions and domain walls \cite{Calabrese:2011ru,Capizzi:2022uni,Capizzi:2022xdt,Capizzi:2023mly,capizzi2023domain}, critical lattice systems \cite{Eisler:2012xry, Peschel_2005, Capizzi:2023vsz} and out-of-equilibrium problems \cite{Capizzi:2022igy, Wen_2018}.

Unfortunately, central predictions made by the theory of interface entropies \cite{Brehm_2016, Gutperle_2016} have been contradicted by numerical ab-initio studies of critical quantum states including the Ising model \cite{Roy_2022, Rogerson:2022yim}, Luttinger liquid \cite{Roy:2023wer} and three-state Potts model \cite{Sinha:2023hum}. Further conceptual incompatibilities of the assumptions made in the foundational paper \cite{Sakai_2008} with a modern entanglement spectrum perspective have recently been addressed in \cite{Northe:2025zmv}. In order to amend these shortcomings, the latter paper also presents a new formalism for the evaluation of entanglement through (non-invertible) topological defects. In all cases and regimes that it was tested, it reproduced the Ising model's numerics in \cite{Roy_2022} successfully and offers field-theoretic explanations for the form of the numeric results. Crucially, the theory developed in \cite{Northe:2025zmv} is centered around constructing RDMs in the presence of defects from which much more information of the subsystem can be extracted than just the entanglement entropy.

In order to quantify the entanglement through a topological defect in a CFT, the authors of \cite{Northe:2025zmv} propose to study the entanglement in twisted states. This forces one to consider states which do not lie in the bulk spectrum of the CFT. In consequence, entanglement through such defects is quantified by an excited state. While the leading contribution to the R\'enyi entropy is still given by the celebrated Cardy-Calabrese term \cite{Calabrese_2009}, the next-to-leading order contains the signature of the defect and twisted excitation. A central aspect of this analysis is that the defect network can influence the value of the R\'enyi entropies significantly. For example, the entropies differ a priori when the defect runs through the entangling interval, i.e. a symmetric entropy, or intersects the entangling edge, i.e. an interface entropy, even though the twisted primary under scrutiny is the same. 

Our aim in the present paper is to extend the theory of \cite{Northe:2025zmv} in two ways. First, we provide further examples for entanglement in the presence of defects, namely grouplike defects in free boson theories. We find that the entanglement reduces to that of the vacuum state. Grouplike defects only reassign fields, suggesting that quantum correlations can at best be shuffled with respect to those in the vacuum state, not however lost nor gained. This expectation has previously been borne out in the Ising model \cite{Roy_2022, Northe:2025zmv}.

The second, and main purpose of this paper, is to 
extend the theory of \cite{Northe:2025zmv} to topological interfaces, i.e. the case that the two theories -- we refer to them as phases -- glued by the line operator may be distinct. In this general case, twist fields are disqualified as pure quantum  states describing the physical setup. Instead, we argue here that entanglement must be quantified by interface-changing fields. Furthermore, this extends the pool of possible interface networks dramatically, and we present the RDMs which lead to inequivalent replica geometries. For simplicity, we consider RDMs for a single entangling interval. 

To illustrate the formalism, we turn to the natural extension of interface entropies on periodic systems. Because the two theories adjacent to the interface are distinct, two interfaces must pierce the constant time slice to give rise to a meaningful state, see \figref{fig:DefectInterfaceState} below. In order to study quantum correlations through the interface, we have these interfaces intersect the entangling edges of a subsystem. Hence, the entangling interval is in one phase, theory $Q$ say,  and the outside in the other, theory $P$. We consider the configuration of lowest possible energy, leading us directly to the study of the identity field harbored on the interface, as we explain below. 

In the light of the importance of dualities in theoretical physics, a particularly interesting setup arises when the two phases $P$ and $Q$ are dual to each other, and can be related by a topological duality interface. We therefore extend the general theory for the action of duality interfaces in bulk CFTs \cite{Frohlich:2004ef, Frohlich_2007} to the presence of boundaries. This is necessary from an information theoretic point of view because entanglement spectra are related to boundary state spaces \cite{Ohmori_2015, Cardy_2016}. 

A core result of our paper is to show that the resulting interface RDMs for duality interfaces act as projectors on the entanglement spectrum. Each duality interface comes equipped with two groups, which individually form a subsymmetry of the two adjacent phases. The RDMs project onto charge sectors of precisely these symmetries contained in the entanglement spectrum. Therefore, the setup mirrors that of symmetry resolution of entanglement \cite{Goldstein:2017bua, Zhao_2021, Weisenberger_2021, Castro-Alvaredo:2024azg, northe2023entanglement, Ares:2026vjt, Northe:2025qcv, Yan_2025, Bonsignori_2020, Murciano_2022, Murciano_2020, Heymann_2025, Choi:2024wfm, Choi:2024tri, Kusuki:2023bsp, Bhattacharyya:2025tmg, Das:2024qdx, Saura-Bastida:2024yye, capizzi2022symmetry, capizzi2022symmetry2, capizzi2023symmetry} for said subgroups. It is important to note however, that here the formalism arises naturally due to the interfaces dressing the physical setup. In contrast, in symmetry resolution the projectors are imposed by hand. 

The physical consequences are that the entanglement spectrum in presence of the duality interfaces is diluted to a fixed-charge subsector of the vacuum's entanglement spectrum. The uncertainty in the presence of duality interfaces, i.e. the entanglement entropy, is hence smaller than in a vacuum state, as we explain below in detail. Given these clear relations between the RDMs in presence of the duality interfaces and that of a vacuum state,  we are naturally led to quantify their difference by employing relative entropy. We find that the vacuum state's RDM can approximate the duality interface RDM at the cost of a finite error, while the latter is too narrow to approximate the former. It simply lacks the other charge sectors that it projects away.

Before outlining the structure of our paper, we briefly review the factorization procedure giving rise to the entanglement spectrum and all information measures employed in this paper. 
\subsection*{Bulk RDM and information measures}

One prescription to factorize QFT Hilbert spaces into subspaces associated with spatial domains is through a homomorphism \cite{Ohmori_2015},
\begin{equation}\label{iota_def}
 \iota_{\underline{\a}}: \cH \to \cH_{\underline{\a}}^A\otimes\cH_{-\underline{\a}}^B,
 \qquad
 \iota_{\underline{\a}}:\ket{\phi}\mapsto\iota_{\underline{\a}}\ket{\phi}\,,
 \qquad
 \ket{\phi}\in\cH
\end{equation}
It uses the path-integral to propagate \change{}{the} incoming hypersurface associated to ${\cal H}$ to an outgoing hypersurface which is the disjoint union of the geometries associated to ${\cal H}^{A},{\cal H}^{B}$. In this paper we concentrate on the case that $A$ is a single entangling interval. The homomorphism $\iota_{\underline{\a}}$ depends on a choice of boundary conditions $\underline{\a}=\{\a_1,\a_2,\cdots\}$ imposed at the entangling surfaces between $A$ and $B$, and the notation $-\underline{\a}$ indicates the orientation reversal, when contemplating system $B$. The $\a_i$ label the various boundary conditions required at the entangling edges of disconnected intervals, which need be compatible with an interface network suitable to the physical system \cite{Northe:2025zmv}. This is elaborated on in detail in \secref{secDisorderStates} and \secref{secDualityIntRDM}.

The conceptual power of the map \eqref{iota_def} lies in explicitly providing the Hilbert space $\cH^A$ describing a generic subsystem. For the vacuum state of a CFT reduced to a single interval, $\cH^A$ even coincides with the entanglement spectrum \cite{Cardy_2016}, granting immediate access to all information contained on $A$. 

In the absence of interfaces, an RDM associated with a pure primary state $\ket{\phi}\in\cH$ in the bulk describing a single interval $A$ is captured by the strip operator \cite{Ohmori_2015,Northe:2025qcv} 
 \begin{equation}\label{RDM}
 \rho_{\ab}^{\phi}
 =
 \frac{1}{Z^{\phi}_{\ab}(q)}
 \raisebox{-.5\height}{\includegraphics[scale=.15]{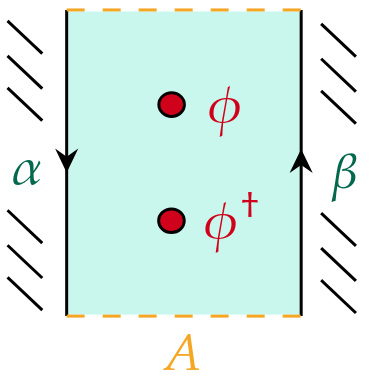}}\,,
 \qquad
 Z^{\phi}_{\ab}(q)
 =
 \tr_{\ab}\left[q^{H_{\ab}} \phi(w(0))\phi^\dagger(w(\infty))
 \right]
\end{equation}
Entanglement (modular) time runs from bottom to top, and is driven by the strip Hamiltonian $H_\ab$; for the bulk identity field $\phi=\id$ it corresponds to entanglement Hamiltonian up to a shift \cite{Cardy_2016}. The Hilbert space \eqref{iota_def} in this case is the strip Hilbert space $\cH_\ab^A$ with conformal boundary conditions $\a,\,\b$ at the edges, and $\rho_\ab^\phi\in\End(\cH_\ab^A)$. The normalization secures $\tr_\ab\left[\rho_{\ab}^{\phi}\right]=1$. A derivation of this result is recapitulated in the appendix \ref{app:EE_bulkRDMs}. The boundary labels $\ab$ imply an RDM so that we shall not explicitly add a label $A$, as is customary. However, when the state $\phi\in\cH$ and and boundaries $\a,\b$ are unspecified, we simply write $\rho_A\in\End(\cH^A)$.

Quantum correlations in pure states, specifically our ignorance on an RDM $\rho_A$, can be quantified by means of the \change{the}{} von Neumann entropy and Rényi entropy
\begin{equation}\label{Entropies}
S(\rho_A)
=
-{\rm Tr}(\rho_A\log\rho_A),
\qquad
S_n(\rho_A)
=
\frac{1}{1-n}\log{\rm Tr}(\rho_A^{n})
\end{equation}
The former is reproduced by the latter via $S(\rho_A)=\lim_{n\to1}S_n(\rho_A)$. If \change{$\rho_A$}{the global state $\rho$} is pure then $S_n(\rho_{A})=S_n(\rho_{B})$, and if it is furthermore separable, then $S_{n}(\rho_{A(B)})=0$.
This means that $S_n$ are good measures of entanglement in pure states. For mixed states in $\cH$, however, $S_n(\rho_{A})\neq S_n(\rho_{B})$
so one employs more suited quantities to describe entanglement, see for instance \cite{LiuVertexStates} and references therein.

In the presence of a symmetry $G_A$ on the subsystem $A$, states in $\cH_A$ decompose into irreducible representations $r_a$ of $G_A$
\begin{equation}\label{SymmetrySplitting}
 \cH^A=\bigoplus_{a\in\rep(G_A)}r_a\otimes \cH^a
\end{equation}
where $\cH^a$ is a multiplicity space. For QFTs, as studied in this work, $\cH^a$ is typically infinite-dimensional. If furthermore $\rho_A$ commutes with all charges $Q_A$ representing $G_A$ on $\cH_A$, the RDM decomposes block-diagonally
\begin{align}
 \rho_A
 =
 \bigoplus_{a\in\rep(G_A)}\prob_a\,\rho_A(a)\,,
 \qquad 
 \rho_A(a)
 =
 \frac{\rho_A\,\Pi_a}{\prob_a}
\end{align}
where $\Pi_a$ projects onto $r_a$, $\prob_a$ is the probability of measuring a state in the representation $r_a$ and $\rho_A(a)$ is the normalized RDM for representation $a$. The information count hidden in $\rho_A(a)$ is quantified -- in analogy to the full RDM $\rho_A$ -- by $S_n(\rho_A(a))$, and is called symmetry-resolved R\'enyi entropy. Its limit $\lim_{n\to1}S_n(\rho_A(a))$ is the symmetry-resolved entanglement entropy.

The last measure employed below, relative entropy, quantifies a distance\footnote{\change{}{Relative entropy is not a metric since it is not symmetric. We use the terminology \quotes{distance} as a shorthand for the mathematically correct term which is \quotes{divergence}.}} between two density matrices $\rho,\sigma\in\End(\cH)$. Here $\cH$ can be either the global Hilbert space or the interval Hilbert space $\cH^A$. In contrast to the R\'enyi entropy, relative entropy is UV-finite and is accessed via a replica trick \cite{lashkari2016modular},
\begin{equation}\label{RelEntropy}
 S(\rho||\sigma)
 =
 \lim_{n\to1}
 \frac{1}{1-n}\log\left(\frac{\tr(\rho\sigma^{n-1})}{\tr[\rho^n]}\right)
 =
  \lim_{n\to1}
 -\frac{\p}{\p n}\frac{\tr(\rho\sigma^{n-1})}{\tr[\rho^n]}
\end{equation}
Relative entropy quantifies how much information is lost when approximating the true state $\rho$ of a system by some other density matrix $\sigma$. Because relative entropy is generally not symmetric in $\rho$ and $\sigma$, the information loss is not the same when exchanging the two RDMs.

\subsection*{Structure of the paper}
The rest of this manuscript is organized as follows. In \secref{secTopInt} we review and settle our notation on topological defects, interfaces and boundaries in 1+1 dimensional rational CFT. As special cases to be employed in the rest of the work, we look at Verlinde lines in general diagonal theories and $U(1)$ preserving interfaces in the free compact boson theory.

In \secref{secDisorderStates} we recapitulate the definition of RDMs in the presence of topological defects from \cite{Northe:2025zmv}. This approach reformulates entanglement in the presence of defects as entanglement in twisted states. Extending on the example of the Ising model given in \cite{Northe:2025zmv}, we compute R\'enyi entropy through grouplike defects in the free boson theory. We show that entanglement entropy does not feel the presence of grouplike defects in the free boson theory, indicating that quantum correlations are only reshuffled, not lost across grouplike defects. 
This is also the case in the Ising model \cite{Northe:2025zmv}. 

In \secref{secDualityIntRDM}, we generalize our construction to define RDMs in presence of many topological defects or interfaces. For definiteness, we look at interfaces separating two phases and mainly look at entanglement in the lowest energy field, namely the identity field on the topological interface. We then focus on duality interfaces, and show that duality RDMs realize projectors on certain representations of the stabilizer subgroups associated with the interface. We show this separately in the case of diagonal theories, like the critical Ising model, and then in non-diagonal theories, like free boson phases related by $\T$-duality. We also include a comment on zero-temperature limits and how to obtain pure state RDMs from the RDM associated to thermal states in the presence of defects. 

Lastly, in \secref{secInfoThyDuality} we discuss in detail the information theory of the duality interface RDMs constructed in the previous section. We extract their entanglement spectra and show that these are diluted to fixed-charge subsectors of a vacuum state entanglement spectrum. We then derive the associated R\'enyi entropies and show that, to next-to-leading order, these are independent of the specific charge (more generally, representation) describing the subsystem and are smaller than the entropies of the vacuum state. This falls in line with the theory of symmetry resolution of entanglement \cite{Kusuki:2023bsp, northe2023entanglement}. Here, it clearly implies that information is reflected back at the interface into the subsystem. We furthermore utilize relative entropy to quantify how much these RDMs change due to the presence of the duality interfaces.

We conclude with a discussion and outlook in \secref{sec:conclusion}. Some detailed calculations and supplementary information are relegated to the appendices. In \appref{app:EE_bulkRDMs}, we review the construction of RDMs for bulk global states. 
In \appref{app:ExtraFreeBoson} we include more technical details about topological defects in the free boson theory.

\section{Topological Interfaces}
\label{secTopInt}
In this section, we review all preliminaries on topological interfaces required to read this paper.

\begin{figure}
    \centering
    \includegraphics[width=0.5\linewidth]{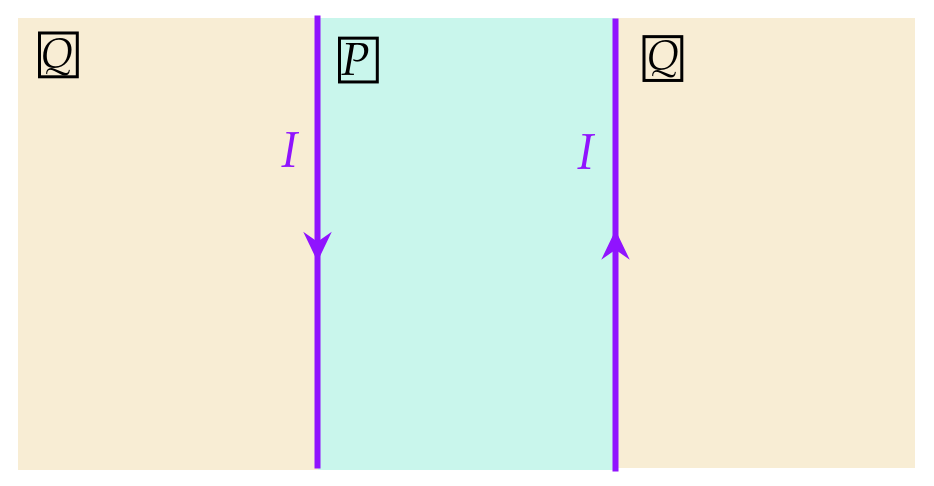}
    \caption{From right to left: Phase $Q$ is mapped into phase $P$ by a $PQ$ interface $\I$ and back by the same orientation-reversed interface. The arrow on the left interface may be inverted at the price of relabeling $\I\to\bI$. $\bI$ is a $QP$ interface.}
    \label{fig:phases}
\end{figure}

Two possibly distinct CFTs, called \quotes{phases} $P$ and $Q$, may be connected by an interface $\I$. When the phase to \change{}{the} left of the interface $\I$ is $P$ and the phase to its right is $Q$, we speak of a $PQ$ interface\footnote{This is in analogy to the terminology \quotes{$PQ$-bimodule} in \cite{Frohlich_2007}.}. When the theories to either side of the interface are the same, $Q=P$, we speak of a defect $\D$. A possible physical configuration is depicted in \figref{fig:phases}.

$PQ$ interfaces give rise to linear operators $\Io: \cH^Q\to\cH^P$,
\begin{equation}
\vcenter{\hbox{\includegraphics[height=3cm]{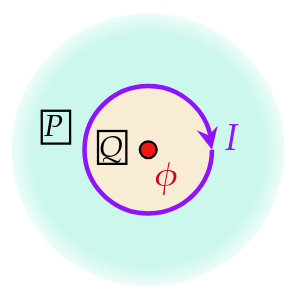}}}
=:
\vcenter{\hbox{\includegraphics[height=1.5cm]{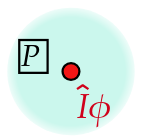}}}
\,,
\qquad
\Io\ket{\phi}^Q
=:
\ket{\Io\phi}^P
\end{equation} 
where the superscript on the ket indicates which Hilbert space the state lies in. Similarly, an interface $\bI$ with opposite orientation gives rise to a linear operator $\bIo:\cH^P\to\cH^Q$,
\begin{equation}
\vcenter{\hbox{\includegraphics[height=3cm]{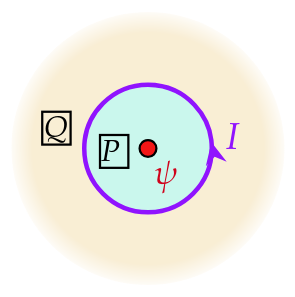}}}
=:
\vcenter{\hbox{\includegraphics[height=1.5cm]{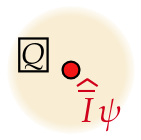}}}
\,,
\qquad
\bIo\ket{\psi}^P
=:
\ket{\bIo\psi}^Q
\end{equation} 

Topological interfaces are transparent to the energy momentum tensor, which means that they intertwine the Virasoro action  \cite{Petkova:2000ip},
 \begin{equation}\label{topdef}
 L_n^{(P)}\Io-\Io L_n^{(Q)}
 =
 0
 =
 \bL_n^{(P)}\Io-\Io \bL_n^{(Q)}
\end{equation}
For defects, this condition becomes $[L_n,\Do]=0=[\bL_n,\Do]$. Topological interfaces and defects can only connect theories of the same central charge. 

Elementary topological interfaces (also called simple or fundamental) cannot be written as \change{}{a} linear combination of other topological interfaces. We denote the set of elementary interfaces at fixed central charge between phases $P$ and $Q$ by $\TopInt{PQ}$ and its elements with lower case latin letters $a,b\in\TopInt{PQ}$. Elementary defects in phase $P$ lie in $\TopInt{PP}\equiv\TopInt{P}$.

\subsection{Fusion, Grouplike Defects and Duality Interfaces}
\label{secGrouplikeAndDuality}

A $PQ$ interface $\I_a$ can be fused with a $QR$ interface $\I_b$ into a $PR$ interface $\I_c$,
\begin{equation}\label{interfaceFusion}
\I_a\fuse_Q\I_b
    =
    \bigoplus_{c\in\TopInt{PR}}N_{ab}^c\,\I_c\,,
\end{equation}
for some integer coefficients $N_{ab}^c$. In the following, we shall drop the subscript which indicates the phase that is being \quotes{fused over}, $\fuse_Q\to\fuse$, so long as confusion cannot arise. The fusion operation $\fuse$ descends to composition of operators $\circ$, which we conveniently omit in the following, $\Io_a\circ\Io_b\to\Io_a\Io_b$.

Topological defects which fuse with their orientation inverse into the trivial defect, 
\begin{equation}\label{grouplike}
 \bD\fuse\D
 \cong
 \id
 \cong
 \D\fuse\bD,
\end{equation}
are invertible and implement (ordinary) symmetries $G$ of the theory. They are called grouplike defects $\D_g$ for $g\in G$ \cite{Frohlich_2007} and furnish representations of $G$, $\Do_g\Do_h=\Do_{g\cdot h}$. Their orientation inverse naturally implements the inverse group element, $\bD_g=\D_{g^{-1}}$. Grouplike defects are always elementary. We denote the set of grouplike defects in a single phase $P$ by $\cG_P\subseteq\TopInt{P}$.

Order-disorder dualities \cite{Frohlich_2007} are implemented by $PQ$ duality interfaces $\I_\dua$ -- we reserve the letter $\dua$ for duality interfaces in the following -- which have the defining property\footnote{\change{}{In the categorical literature, these are studied in the context of Tambara–Yamagami categories if the interfaces carry no orientation \cite{Tambara:1998vmj}. } }
\begin{equation}\label{duality}
\bI_\dua\fuse\I_\dua\cong\bigoplus_{g\in\stabR{\dua}}\D_g\,,
\end{equation}
where the right-stabilizer of the duality interface $\I_\dua$ is defined as
\begin{equation}\label{eq:rightleft_stabilizers_def}
    \stabR{\dua}:=\{g\in\cG_Q|\I_\dua\fuse\D_g\cong\I_\dua\}\,,
    \qquad 
    \stabL{\dua}:=\{g\in\cG_P|\D_g\fuse\I_\dua\cong\I_\dua\}\,
\end{equation}
For future reference, we have analogously defined the left-stabilizer. In general, $\bI_\dua$ is not guaranteed to be a duality interface. If it is however, then the right- and left-stabilizers are isomorphic, $\stabR{\dua}\cong\stabL{\dua}$, and abelian \cite{Frohlich_2007}. Note that grouplike defects are trivial duality defects, whose stabilizer consists of only the group unit corresponding to the invisible defect.

In general, topological interfaces can be fused along segments. Following the normalization for junction fields of \cite{Frohlich_2007}, and restricting for convenience to models with $N_{ij}^k=0,1$\footnote{Models with non-trivial fusion multiplicities can similarly be considered, in which case all trivalent junction spaces are many-dimensional and need be assigned an extra label. See \cite{Fuchs:2002cm} for details.}, which includes A-type Virasoro minimal models and the free boson, parallel interfaces behave as follows 
\begin{equation}\label{partialFusion}
 \vcenter{\hbox{\includegraphics[height=2.5cm]{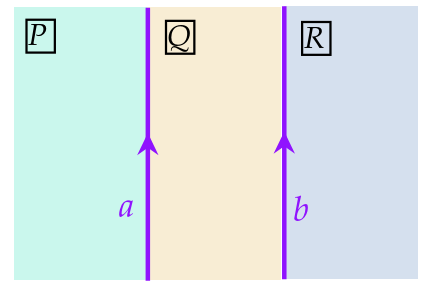}}}
 =
 \sum_{c\in a\fuse b}
 \vcenter{\hbox{\includegraphics[height=2.5cm]{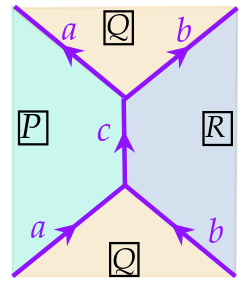}}}
\end{equation}
For antiparallel defects one has
\begin{equation}\label{antiparallelFusion}
 \vcenter{\hbox{\includegraphics[height=2.5cm]{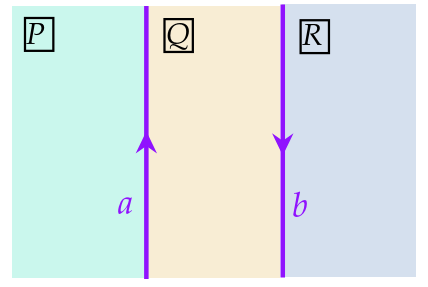}}}
 =
 \sum_{c\in a\fuse b^+}\frac{\qdim_c}{\qdim_a}
 \vcenter{\hbox{\includegraphics[height=2.5cm]{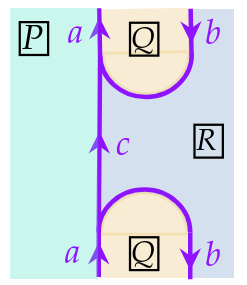}}}
\end{equation}
where $\qdim_a$ is the quantum dimension of the interface $a$. These equations apply similarly when either of the two interfaces describes a boundary, although, mathematically speaking the junctions are of different nature. 

Interfaces harbor fields. Generically, fields living on $PQ$ interfaces can change a $PQ$ interface $a$ into another $PQ$ interface $b$,
\begin{equation}\label{InterfaceField}
    \phi^{(a,b)}
    =
    \vcenter{\hbox{\includegraphics[height=1.5cm]{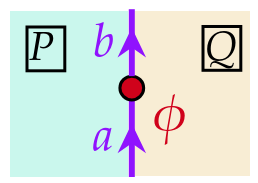}}}
\end{equation}
By the operator-state correspondence, these fields constitute the following state spaces
\begin{align}
\Hjunc{a}{b}{PQ}
=
 \bigoplus_{(i,\bi)\in\confFam\times\confFam}\njunc{a}{b}{i\bi}\,\cH_i\otimes\cH_{\bi}\,,
 \qquad
 \njunc{a}{b}{i\bi}\in\N_0,
 \qquad
 a,b\in\TopInt{PQ}
\end{align}
The bulk Hilbert space of a phase $P$ is identified with interface fields that change the trivial defect $\id$ into itself $\Hjunc{\id}{\id}{PP}\equiv\cH^P$. Twist fields terminating a line for $\I_a$ correspond to $\Hjunc{a}{\id}{PP}$, and twist fields emanating a line $\I_a$ correspond to $\Hjunc{\id}{a}{PP}$. Junctions of several interfaces similarly harbor fields with associated state spaces. The lower trivalent junction in \eqref{partialFusion} belongs to $\Hjunc{ab}{c}{PR}$ whereas the upper junction belongs to $\Hjunc{c}{ab}{PR}$; we have omitted the phase $Q$ in the notation to avoid clutter. 
\subsection{Topological Interfaces and Boundaries}
Boundary conditions $\a,\b$ in phase $P$ are taken from a set $\cB_P$ and implemented by a boundary state $\bket{\a}^P$. Boundary state spaces $\cH_{\ab}$ are labeled by two boundary conditions and contain boundary condition-changing operators,
\begin{equation}
 \cH_{\alpha\beta}
 =
 \bigoplus_{i\in\confFam}\niab\cH_i\,,
 \qquad
 \niab\in\N_0
\end{equation}

Boundaries furnish a module when acted upon by a $PQ$ interface
\begin{equation}\label{IntOnBdy}
 \I_a\fuse \a
 =
 \bigoplus_{\b\in\cB_P}\fusIB_{a\a}^\b \,\b\,,
 \qquad
    \Io_a\bket{\a}
    =
    \sum_{\b\in\cB_P}\fusIB_{a\a}^\b\bket{\b}\,,
    \qquad
    \a\in\cB_Q
\end{equation}
for $\fusIB_{a\a}^\b\in\N_0$. Mildly abusing notation, we use the same symbol $\fuse$ for fusion of boundaries and interfaces as for interface-interface fusion, see \eqref{interfaceFusion}.

Trivalent topological junctions of boundaries and topological interfaces,  
\begin{equation}\label{eq:TopIntBdyJunction}
\vcenter{\hbox{\includegraphics[height=2cm]{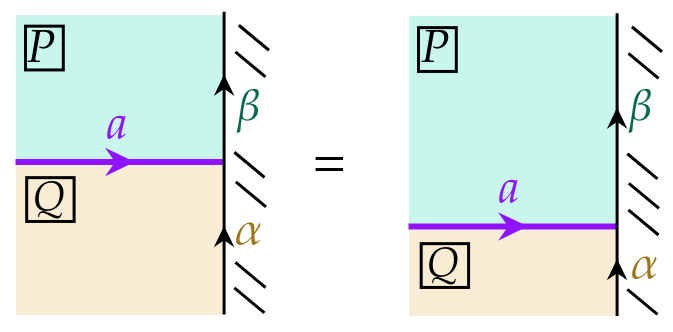}}} 
\end{equation}
are possible whenever $\b\in a\fuse \a$, i.e. $\fusIB_{a\alpha }^{\b}\neq0$. More precisely, $\fusIB_{a\a}^\b$ coincides with the dimension of the topological junction space in \eqref{eq:TopIntBdyJunction}. 
Invertibility of grouplike defects implies that their action has a single image, so that $\fusIB_{g\a}^\b=\delta_{\b,g\fuse\a}$. Junctions are topological as long as the junction field carries no energy. Hence the junction point can be moved along the boundary at will. 

A topological $PQ$ interface induces an operator $\IB^a$ from boundary state spaces with $\a,\b\in \cB_Q$ into $\a',\b'\in\cB_P$ \cite{Kojita:2016jwe}, 
\begin{align}\label{DefBdyAction}
 \IB^a: 
 \quad
 \cH_{\alpha\beta}
 \to
 \bigoplus_{\substack{\alpha'\in a\fuse \a\\\beta'\in a\fuse \b}}
 \cH_{\alpha'\beta'},
 \qquad
 \IB^a\psi_i^{\alpha\beta}
 =
 \bigoplus_{\substack{\alpha'\in\alpha\fuse a\\\beta'\in\beta\fuse a}}
 \ib{a}{\alpha\beta}{i}{\alpha'\beta'}\psi_{i}^{\alpha'\beta'},
\end{align}
It was shown in \cite{Kojita:2016jwe} that the maps $\IB_a$ follow the fusion algebra \eqref{interfaceFusion} only up to unitary equivalence, as we shall reproduce for Verlinde lines in \eqref{DefBdyCounterFusion}. Mostly, we are interested in contemplating the contribution of only a single pair $\a',\b'\in \cB_P$ of boundary conditions. We employ the notation
\begin{align}\label{IndBdyAction}
\left(\IB^a\psi_i^{\ab}\right)^{\a'\b'}
=
\ib{a}{\a'\b'}{i}{\ab}\,\psi^{\a'\b'}_i
\quad
\textrm{for}
\quad 
\a,\,\b\in\cB_P,\,\, \a',\,\b'\in\cB_Q
\end{align}
which pictorially corresponds to
\begin{equation}\label{eq:IntOnBdy}
\vcenter{\hbox{\includegraphics[height=2cm]{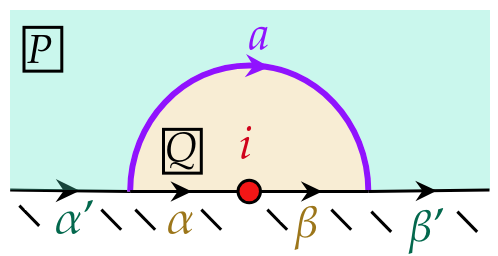}}}
=
\ib{a}{\a'\b'}{i}{\ab}\,
\vcenter{\hbox{\includegraphics[height=2cm]{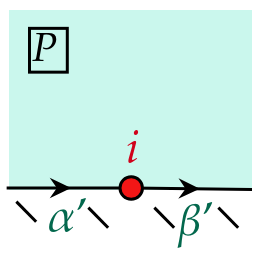}}}
\end{equation}
The trivalent junctions are topological. Interfaces oriented oppositely to boundaries induce similarly an operator $\bIB$
\begin{equation}
 \left(\bIB^a\psi_i^{\a'\b'}\right)^{\a\b}
    =
    \bib{a}{\a\b}{i}{\ab}\,\psi^{\a'\b'}_i\,,
    \qquad
    \vcenter{\hbox{\includegraphics[height=2cm]{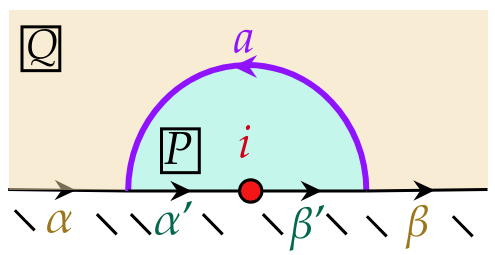}}} 
    =
    \bib{a}{\a\b}{i}{\ab}\,
    \vcenter{\hbox{\includegraphics[height=2cm]{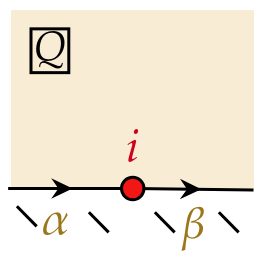}}} 
\end{equation}
At times, we manipulate interfaces without specifying a representation $i$. In this case, we write the operator as follows
\begin{equation}\label{IntBdyOps}
    \ibo{a}{\ab}{\a'\b'}
    =
    \vcenter{\hbox{\includegraphics[height=2cm]{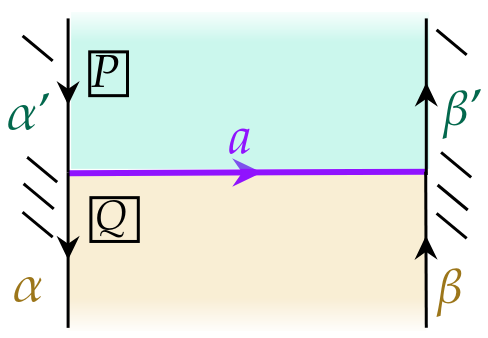}}}\,,
    \qquad
    \bibo{a}{\ab}{\a'\b'}
    =
    \vcenter{\hbox{\includegraphics[height=2cm]{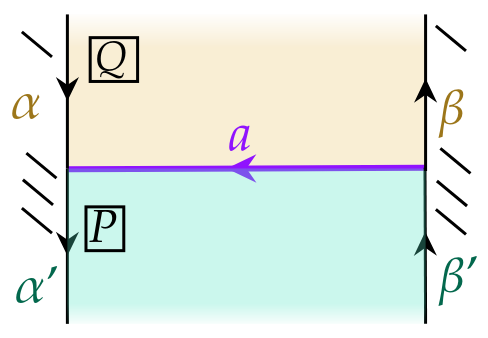}}}\,
\end{equation}
This diagram is read bottom to top. For defects we write $\DB$ instead of $\IB$. Note that in this notation $\ibo{a}{\ab}{\a'\b'}$ is an operator, while $\ib{a}{\a'\b'}{i}{\ab}$ is a number.

\subsection{Verlinde Lines}
The standard example of interfaces are the Verlinde lines \cite{Petkova:2000ip}, which are in fact defects. They map phases with diagonal modular invariant, which we refer to as $P=\id$\footnote{This \change{corresonponds}{corresponds} to the trivial Frobenius algebra object $\id$ in the categorical language of \cite{Fuchs:2002cm}.}, into itself
\begin{equation}\label{diagModInv}
 \cH^{\id}\equiv\cH=\bigoplus_{i\in\confFam}\cH_i\otimes\bar{\cH}_{i}
\end{equation}
where $\confFam$ is the set of chiral representations at fixed central charge. The set of elementary topological defects from a diagonal phase into itself is $\TopInt{P}=\confFam$. Hence, every element in $\confFam$ gives rise to one Verlinde defect \cite{Petkova:2000ip}, 
\begin{equation}\label{VerlindeLines}
 \Do_a=\sum_{i\in\confFam}\frac{\modS_{ai}}{\modS_{0i}}\proj{i},
 \qquad
 a\in\confFam
\end{equation}
where $\proj{i}:\cH_i\otimes\bar{\cH}_i\to\cH_i\otimes\bar{\cH}_i$ are projectors. Verlinde lines resemble Cardy boundary states \cite{Cardy:1989ir}
\begin{equation}\label{CardyStates}
 \bket{\alpha}
 =
 \sum_{i\in\confFam}\frac{\modS_{ai}}{\sqrt{\modS_{0i}}}\iket{i},
 \qquad
 \alpha\in\confFam
\end{equation}
Boundary state spaces are determined by the fusion ring, which is controlled by the Verlinde formula \cite{Cardy:1989ir, Verlinde:1988sn}
\begin{equation}\label{CardyStateSpaces}
 \cH_{\alpha\beta}=\bigoplus_{i\in\confFam}\fus_{\ab}^i\cH_i\,
 \qquad
  \fus_{ab}^i
    =
    \sum_{l\in\confFam}\frac{\modS_{al}\modS_{bl}\modS^*_{il}}{\modS_{0l}}\,.
\end{equation}
We remind the reader that we restrict to the cases where the fusion rules are in the range $\fus_{ab}^c\in\{0,1\}$. This includes A-type Virasoro minimal models and the free boson. 

The defect algebra \eqref{interfaceFusion} for these defects and their action \eqref{IntOnBdy} on Cardy boundary states both follow the fusion ring of chiral representations
\begin{subequations}
    \begin{align}
      \Do_a\Do_b
&=
 \sum_{c\in\confFam}\fus_{ab}^c\Do_c,\label{DefectFusion}
 \\
 \Do_a\bket{\alpha}
 &=
 \sum_{\beta\in\confFam}\fus_{a\alpha}^\beta\bket{\beta}\label{DefectFusionBdy}
\end{align}
\end{subequations}

The Ising CFT provides an example with grouplike defects \eqref{grouplike} and duality defects \eqref{duality}. It has three conformal families $\confFam=\{0,\,\nicefrac{1}{2},\nicefrac{1}{16}\}$. Defects $\D_g$ with $g=0,\nicefrac{1}{2}$ are grouplike and are in fact their own inverse, reflecting the $\Z_2$ group they generate. The celebrated Kramers-Wannier duality of the Ising model is generated by $\D_{1/16}$. It is as well its own orientation reversal so that $\Do_{\nicefrac{1}{16}}\Do_{\nicefrac{1}{16}}=\Do_0+\Do_{\nicefrac{1}{2}}$ amounts to the property \eqref{duality}.

An important ingredient in the manipulation of defect networks are the fusing matrices $\mathsf{F}$ and its inverse $\mathsf{G}$. We use the notation and conventions of \cite{Fuchs:2002cm}\footnote{Moore's and Seiberg's conventions \cite{Moore:1989vd} are reached by $$\mathsf{F}_{pq}
\begin{bmatrix}
j & k \\
i & l
\end{bmatrix}
=
\fusing{jkl}{i}{pq}$$},
\begin{subequations}\label{Fusing}
\begin{align}
 &
 \vcenter{\hbox{\includegraphics[height=2cm]{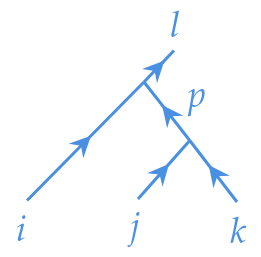}}}
 =
 \sum_{q\in\confFam}
 \fusing{ijk}{l}{pq}
 \vcenter{\hbox{\includegraphics[height=2cm]{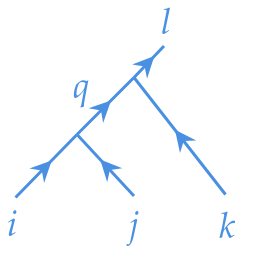}}}\,,
 &
  \vcenter{\hbox{\includegraphics[height=2cm]{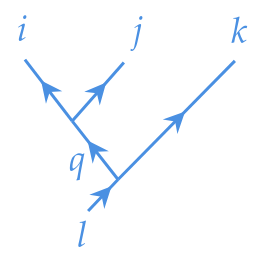}}}
 =
 \sum_{q\in\confFam}
 \fusing{ijk}{l}{qp}
 \vcenter{\hbox{\includegraphics[height=2cm]{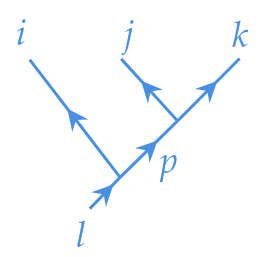}}}\,,
 \label{fusingF}\\
 &
 \vcenter{\hbox{\includegraphics[height=2cm]{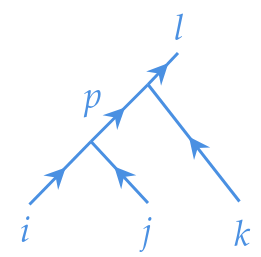}}}
 =
 \sum_{q\in\confFam}
 \fusingG{ijk}{l}{pq}
 \vcenter{\hbox{\includegraphics[height=2cm]{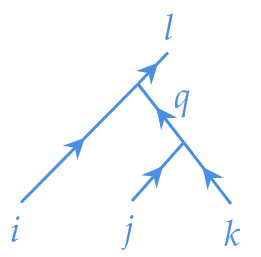}}}\,,
 &
  \vcenter{\hbox{\includegraphics[height=2cm]{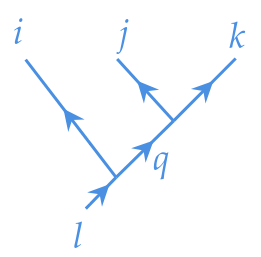}}}
 =
 \sum_{q\in\confFam}
 \fusingG{ijk}{l}{pq}
 \vcenter{\hbox{\includegraphics[height=2cm]{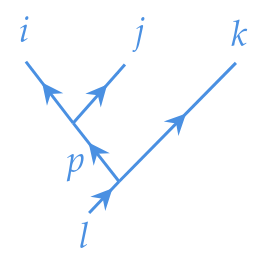}}}\,
 \label{fusingG}
\end{align}
\end{subequations}
While each line is associated with one representation in $\confFam$, they a priori do not represent interfaces or defects, hence we draw them in a different color. By virtue of our simplifying assumption $\fus_{ij}^k\in\{0,1\}$, all non-trivial trivalent junctions are associated with a one-dimensional vector space, and no multiplicity label must be assigned. These are normalized in the following way
\begin{align}\label{junctionNormalization}
\vcenter{\hbox{\includegraphics[height=3cm]{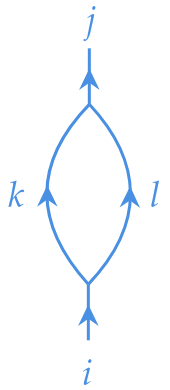}}}
=
\delta_{ij}\fus_{kl}^i\,
\vcenter{\hbox{\includegraphics[height=3cm]{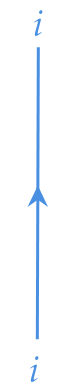}}}
\end{align}
Each diagram in \eqref{Fusing} contains concatenations of two different types of trivalent junction fields. This can be regarded as a choice of basis for conformal blocks, and the fusing matrices execute a change of basis. More details can be found in section 5 of \cite{Fuchs:2004xi} and a compressed version thereof is found in appendix A of \cite{Runkel:2007wd}. 

Because Verlinde lines and Cardy states are labeled by $\confFam$, the manipulations \eqref{Fusing} apply similarly to their defect networks inside correlators. This can be argued rigorously in the TFT approach \cite{Frohlich_2007}; see also \cite{Runkel:2007wd} for the specific example of Verlinde lines.

Turning to the defect action on boundary fields \eqref{IntBdyOps}, we have the following fusion properties
\begin{subequations}
\begin{align}
    \dbo{b}{\ab}{\a'\b'}\,\dbo{a}{\a'\b'}{\a''\b''}
    &=
    \sum_{c\in a\fuse b} \fusingG{ba\a''}{\a}{c\a'}
    \,\dbo{c}{\ab}{\a''\b''}\,
    \fusing{ba\b''}{\b}{\b'c}
    \equiv
    \sum_{c\in a\fuse b}\fusMa{ba}{c}{\ab}{\a'\b'}{\a''\b''}\dbo{c}{\ab}{\a''\b''}
    \label{BdyDefFusion}\\
    \dbo{b}{\ab}{\a'\b'}\,\bdbo{a}{\a'\b'}{\a''\b''}
    &=
    \sum_{c\in a^+\fuse b}
    \frac{\qdim_c}{\qdim_b}
    \fusing{ca\a'}{\a}{\a''b}
    \,\dbo{c}{\ab}{\a''\b''}\,
    \fusingG{ca\b'}{\b}{b\b''}
    \equiv
    \sum_{c\in a^+\fuse b}\bfusMa{ba}{c}{\ab}{\a'\b'}{\a''\b''}\dbo{c}{\ab}{\a''\b''}
    \label{BdyDefFusionCounter}
\end{align}
\end{subequations}
The relation for parallel fusion appeared before in \cite{Kojita:2016jwe} with, however, different normalizations for the junction fields than the ones used here. We exemplarily demonstrate the counteroriented fusion, since it becomes important below, 
\begin{align}\label{DefBdyCounterFusion}
 \vcenter{\hbox{\includegraphics[height=3cm]{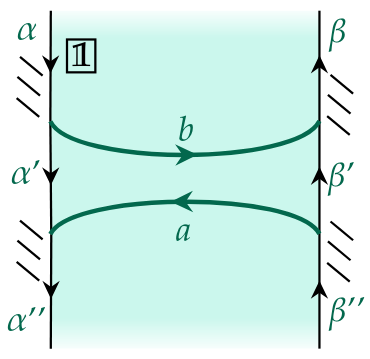}}}
 &=
 \sum_{e\in a^+\fuse b}\frac{\qdim_e}{\qdim_b}\,
 \vcenter{\hbox{\includegraphics[height=3cm]{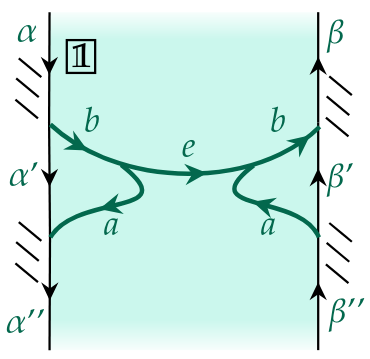}}}\notag\\
 &=
 \sum_{e\in a^+\fuse b}\frac{\qdim_e}{\qdim_b}
 \sum_{\gamma,\delta\in\confFam}
 \fusing{ea\a'}{\a}{\gamma b}\,
 \vcenter{\hbox{\includegraphics[height=3cm]{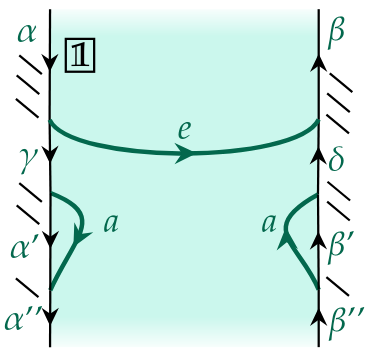}}}\,\fusingG{ea\b'}{\b}{b\delta}\notag\\
 &=
 \sum_{e\in a^+\fuse b}\frac{\qdim_e}{\qdim_b}
 \fusing{ea\a'}{\a}{\a'' b}\,
 \vcenter{\hbox{\includegraphics[height=3cm]{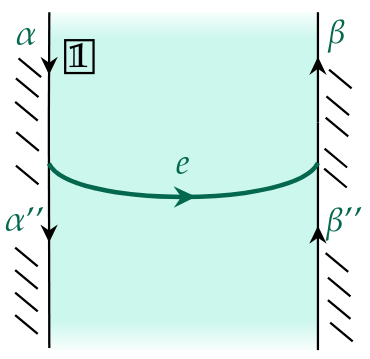}}}
 \,\fusingG{ea\b'}{\b}{b\b''}
\end{align}
The first equality uses \eqref{antiparallelFusion}, the second the fusing \change{matries}{matrices} \eqref{Fusing} and the last line the junction normalization \eqref{junctionNormalization}.

To close this overview out, one can use elementary methods \cite{Fuchs:2002cm, Frohlich_2007} to work out the boundary action \eqref{IndBdyAction},
\begin{equation}\label{VerlindeBdyAction}
    \left(\DB^a\psi_i^{\ab}\right)^{\a'\b'}
=
\db{a}{\a'\b'}{i}{\ab}\,\psi^{\a'\b'}_i
=
\frac{\eta_\ab}{\eta_{\a'\b'}}
\fusing{a\a i}{\b'}{\b\a'}\,\psi^{\a'\b'}_i
\end{equation}
where the $\eta_\ab$ are normalization constants \cite{Frohlich_2007}, which are immaterial for our purposes, because we shall only need $   \left(\DB^a\psi_i^{\ab}\right)^{\ab}
=
\db{a}{\ab}{i}{\ab}\,\psi^{\ab}_i
=
\fusing{a\a i}{\b}{\b\a}\,\psi^{\ab}_i$.
\subsection{The free boson theory}
The final class of interfaces we consider belongs to the compact free boson CFT with action
\begin{equation}\label{FBaction}
 S
 =
 \frac{1}{2\pi\a'}\int_\Sigma\mathrm{d}^2z(\p\varphi)(\bp\varphi)
 \qquad 
 \varphi\simeq\varphi+2\pi R
\end{equation}
The compactification radius $R$ is the only modulus of the theory. For convenience, we shall set the constant $\a'=1/2$. $\a'$-dependence is restored below by rescaling all instances of the compactification radius according to $R\to R/\sqrt{2\a'}$.

The modular invariant spectrum for the Hilbert space $\cH_R$,
\begin{equation}
    Z(q,\bq)
    =
    \frac{1}{|\eta(q)|^2}\sum_{(p,\bar p)\in\Lambda(R)}q^{\frac{p^2}{2}}\bq^{\frac{\bar p^2}{2}}
\end{equation}
is characterized by a two-dimensional charge lattice $\Lambda(R)$ 
\begin{equation}\label{FBstates}
    \begin{pmatrix}
        p,
        \bar p
    \end{pmatrix}
    =
     \begin{pmatrix}
        \frac{m}{2R} + Rw ,
        \frac{m}{2R} - Rw
    \end{pmatrix}
    \,\longrightarrow\,
    \ket{p,\bar p}=\ket{m,w}\in\cH^R 
\end{equation}
labeled by integer momentum and winding, $(m,w)\in\Z^2$. Note that the theory at radius $1/(2R)$ has precisely the same spectrum after swapping $(m,w)\to(w,m)$. This change is implemented by $\T$-duality. This transformation's fixed point lies at the self-dual radius $R_*=1/\sqrt{2}$.

Given the decomposition of the fundamental field $\varphi(z,\bz)=\phi(z)+\bphi(\bz)$, fields associated with states \eqref{FBstates} are written as plane wave operators 
\begin{equation}\label{VertexOp}
    V_{p,\bar p}(z,\bz)=:\exp\left(\iu p\,\phi(z)+\iu\,\bar p\bphi(\bz)\right):\,,
    \qquad
    h_p=p^2/2\,
    \quad 
    h_{\bar p}=\bar p^2/2
\end{equation}
The action \eqref{FBaction} has furthermore a $U(1)\times U(1)$ symmetry, which is generated by currents $J(z)=\sum_{n\in\Z}J_nz^{-n-1},\,\bJ=\sum_n\bJ_n\bz^{-n-1}$. Their zero modes extract the momenta, $J_0\ket{p,\bar p}=p\ket{p,\bar p}$, $\bJ_0\ket{p,\bar p}=\bar p\ket{p,\bar p}$. The $n$-point correlators on the sphere of plane wave operators, also for states outside of $\Lambda(R)$, are given by
\begin{equation}
 \corr{ \prod_{i=1}^n V_{p,\bar p} (z_i,\bar z_i )}  =\prod_{i<j}^{n}(z_i-z_j)^{\frac{p_ip_j}{4}}(\bar{z}_i-\bar{z}_j)^\frac{\bar p_i\bar p_j}{4}.
\end{equation}

The relevant $U(1)$ symmetric conformal boundary states are the well-known Cardy consistent Dirichlet and Neumann boundary states 
\begin{align}\label{DirichletNeumann}
    &\bket{\dir(\varphi_0)}^{R}
    =
    \gfd{R}\sum_{m\in \mathbb{Z}}e^{\iu\frac{m}{R}\varphi_0} \iket{m,0}^{R},
    &
    \gfd{R}=\frac{1}{\sqrt{2R}}\\
    &\bket{\neu(\varphi_0)}^{R} 
    =
    \gfn{R}\sum_{w\in \mathbb{Z}}e^{-2\iu wR\varphi_0} \iket{0,w}^{R}\,,
    & 
    \gfn{R} = \sqrt{R}\, .
\end{align}
The superscripts on the kets will only be used when two theories need distinguishing, otherwise they will be dropped. The Ishibashi states are maximally $U(1)$ symmetric and are reviewed in \eqref{FBishibashi}.

\subsubsection{U(1)-symmetric Topological Interfaces}
We are interested in interfaces between two theories at radii $R_1, R_2$ which, beyond being topological \eqref{topdef}, also intertwine the $U(1)\times U(1)$ symmetry of the free boson theory up to an automorphism \cite{Fuchs:2007tx},
 \begin{equation}\label{U1gluing}
 J_n^{(R_2)}\Io=\epsilon\,\Io J_n^{(R_1)},
 \qquad
 \bJ_n^{(R_2)}\Io=\bar\epsilon\,\Io \bJ_n^{(R_1)}\,,
 \qquad 
 \epsilon=\pm\,,\bar \epsilon=\pm
\end{equation}
These automorphisms are familiar from Dirichlet and Neumann boundary conditions. We restrict to the special case that the two moduli are related by
\begin{equation}
    R_1/\hat{R}_2 = N/M\,,
\end{equation}
where $M,N$, importantly, are coprime positive integers and
\begin{equation}
    \hat{R} := \begin{cases}
        R \hspace{1cm}\text{if}\hspace{1cm} \epsilon = \bar \epsilon\hspace{0.2cm}, \\
        \frac{1}{2R}\hspace{0.8cm}\text{if}\hspace{1cm}\epsilon=-\bar\epsilon\hspace{0.2cm}.
    \end{cases}
\end{equation}
Such interfaces have been constructed in \cite{Fuchs:2007tx} and are given by
\begin{equation} \label{duality_defect_operators_FB}
    \Io^{\epsilon,\bar\epsilon}_{R_2R_1}(x,y)
    =
    \sqrt{MN}
    \sum_{(p,\bar p) \in \Lambda_\cap }(\epsilon\bar\epsilon)^{s_{p\bar p}} e^{2\pi i(xp-y\bar p)} \pr{p & \bar p}{\epsilon p & \bar\epsilon \bar p}
    \,\to\,
    \vcenter{\hbox{\includegraphics[width=0.2\linewidth]{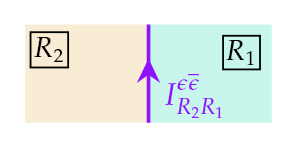}}}\,,
\end{equation}
where $s_{p\bar p}=h_p-h_{\bar p}$ is the spin. The projectors implement the gluing \eqref{U1gluing},
\begin{equation}
    \pr{p & \bar p}{\epsilon p & \bar \epsilon\bar p}\ket{q,\bq}^{R_1}
    =
    \delta_{p,q}\delta_{\bar p, \bq}\ket{\epsilon p,\bar\epsilon\bar p}^{R_2}\,,
    \qquad
    \ket{\cdot}^{R_i}\in\cH^{R_i}
\end{equation}
Since, generally, $\Lambda(R_2)\neq\Lambda(R_1)$, the interfaces can only be non-zero on their intersection $\Lambda_\cap:=\Lambda(R_2)\cap\Lambda(R_1)$. The interfaces are controlled by two parameters $(x,y)\in\R^2$\footnote{A priori, $(x,y)\in\C^2$ are allowed, but when these moduli are not real, the interface changing fields can have complex conformal weights. }. However, two interfaces can be related, 
\begin{equation}
    \Io^{\epsilon\bar\epsilon}_{R_2R_1}(x,y)
    =
    \Io^{\epsilon\bar\epsilon}_{R_2R_1}(x',y')
    \quad \textrm{iff}\quad 
    (x-x',y-y')\in\Lambda_\cap^*
\end{equation}
where $\Lambda_\cap^*$ is the dual lattice consisting of all points $(x,y)\in\R^2$ such that $xp-y\bar p\in\Z$ for any $(p,\bar p)\in\Lambda_\cap$. 

\subsubsection{Grouplike Defects}\label{secFBgrouplike}
Grouplike defects appear only for $M=N=1$, and, for $R$ different from the self-dual radius\footnote{For the purposes of our work, the self-dual radius is not too interesting so that readers are referred to \cite{Fuchs:2007tx} for more details.}, for $\bar \epsilon=\epsilon$.  We abbreviate them by 
\begin{equation}
    \go{R}^\epsilon(x,y):=\Io^{\epsilon\epsilon}_{RR}(x,y)
\end{equation}
They satisfy the following multiplication rule
\begin{equation}
    \go{R}^\epsilon(x,y)\go{R}^\nu(u,v)
    =
    \go{R}^{\epsilon\nu}(\nu x+u,\nu y+v)
\end{equation}
and form the group 
\begin{equation}\label{Piccard}
    \cG_R=(\C^2/\Lambda(R))\rtimes\Z_2
\end{equation}
The $\Z_2$ element is $\go{R}^-(0,0)$. For grouplike defects with $\epsilon=+$ we drop the superscript $\epsilon$ and write $\go{R}(x,y)$. Twist fields emanating \change{}{from} one such $g_R(x,y)$ defect are of the type \eqref{VertexOp}, but have $U(1)$ charges in the shifted (twisted) lattice
\begin{equation}\label{twistedLattice}
    \Lambda(R|x,y):=\{(p+x,\bar p+y)|(p,\bar p)\in\Lambda(R)\,\,\&\,\,(x,y)\in\R^2\backslash\Lambda(R)\}
\end{equation}
as can be easily confirmed with the techniques in \cite{Petkova:2000ip}, see also \cite{Northe:2024tnm}.

A grouplike defect acts on the boundary states \eqref{DirichletNeumann} as follows,
\begin{subequations}\label{GrouplikeOnDN}
    \begin{align}
    &\go{R}(x,y)\bket{\dir(\varphi_0)}
    =
    \bket{\dir(\varphi_0-\pi(x-y))}
    \\
    &\go{R}(x,y)\bket{\neu(\varphi_0)}
    =
    \bket{\neu(\varphi_0-\pi(x+y))}
\end{align}
\end{subequations}

Note that $\go{R}(x,x)$ stabilizes $\dir$ boundaries, while $\go{R}(x,-x)$ stabilizes $\neu$ boundaries. 
\subsubsection{$\T$-Duality Interface}\label{secFBduality}
All interfaces \eqref{duality_defect_operators_FB} can be separated into one grouplike defect and a remainder which implements the map between theories
\begin{align}
    \Io^{\epsilon\bar\epsilon}_{R_2R_1}(x,y)
    =
    \Io^{\epsilon\bar\epsilon}_{R_2R_1}(0,0)\go{R_1}(x,y)
    =
    \go{R_2}(\epsilon x,\bar \epsilon y)\Io^{\epsilon\bar\epsilon}_{R_2R_1}(0,0)
\end{align}
In \appref{appFBdualityNeumann}, we argue that $\Io_{R_1R_1}^{+-}(0,0)$ implements $\T$-duality. In a slight abuse of notation, we define
\begin{equation}\label{T}
    \T:=\I_{R_2 R_1}^{+-} (0,0)\,,
    \qquad
    \bar{\T}:=\I_{R_1 R_2}^{+-} (0,0)
\end{equation} 
for arbitrary $R_2$. One can check that it is dualizing \cite{Fuchs:2007tx}, see \eqref{duality},
\begin{equation}\label{Tduality}
    \bTo \fuse \To 
    =
    \sum_{m=0}^{M-1}\sum_{w=0}^{N-1} 
    \go{R_1}\left(\frac{m}{2MR_1}+\frac{wR_1}{N}, \frac{m}{2MR_1}-\frac{wR_1}{N}\right)
\end{equation}
The right stabilizer appearing here forms a subgroup $\stabR{\T}=\Z_{MN}\subset \cG_{R_1}$. The left-stabilizer $\stabL{\T}=\Z_{MN}\subset \cG_{R_2}$ appears in
\begin{equation}
    \To \fuse \bTo 
    =
    \sum_{m=0}^{M-1}\sum_{w=0}^{N-1} \go{R_2}\left(\frac{w}{2N\hat R_2}+\frac{m\hat R_2}{M}, \frac{w}{2N\hat R_2}-\frac{m\hat R_2}{M}\right)
\end{equation}
Hence both, $\T$ and $\bar\T$ are duality interfaces. We clearly see that their stabilizers are abelian and isomorphic as groups \cite{Frohlich_2007}. 

In \eqref{appdefect_sum_FB_full}, we show how to $\T$-dualize a Neumann boundary state. The result is organized into a sum of simple Dirichlet boundaries with equidistant shift over the circle of radius $R_2$,
\begin{subequations}\label{defect_sum_FB_full}
\begin{align}
    \To \bket{\neu(\varphi_0)}^{R_1}
    & =
    \sum_{l=0}^{N-1} \bket{\dir(\varphi_0 -\frac{2\pi R_2}{N}l)}^{R_2}\label{defect_sum_FB_a}\\
    &
    = \sum_{l=0}^{N-1} \go{R_2 } \left(R_2\frac{l}{N},-R_2 \frac{l}{N}\right) \bket{\dir(\varphi_0)}^{R_2} \label{defect_sum_FB}.
    \end{align}
\end{subequations}
 As expected from general theory \cite{Fuchs:1999xn}, $\T$-duality does not act as a one-to-one map on boundary states. Importantly, each summand has multiplicity 1, cf. \eqref{IntOnBdy}, which coincides with the dimensionality of the trivalent junction space between boundary and interface. In the last line, this sum is expressed by the action of group elements forming a $\Z_N\in\cG_{R_2}$. Note that the final expression can be written as the action of the projector onto the trivial $\Z_N$ representation $N \mathcal{P}_{\id}^{\mathbb{Z}_N} \bket{\dir(\varphi_0)}^{R_2}$.

\section{Defect-dressed Reduced Density Matrices I -- Twists}\label{secDisorderStates}
In this section, we extend the factorization prescription introduced in \cite{Ohmori_2015} and reviewed in
\eqref{iota_def}, \eqref{RDM} to arbitrary defect states in 1+1 Rational CFT.
We start in \secref{secTwistedStateRDMs} with a general description of RDMs in presence of a single topological line defect piercing the spatial slice. This can be used to study EE in pure global twisted states. 
We then follow with an application of our framework to the free boson theory. A further example can be found in \cite{Northe:2025zmv}, where twist field RDMs were studied in the critical Ising model.
Further generalization to multiple line defects and topological interfaces will be the focus of the next section.

\subsection{Twisted state RDMs}\label{secTwistedStateRDMs}
The RDM construction \eqref{RDM} can be generalized to various distinct physical situations by dressing it with a defect network. 

Consider the scenario in which the circular spatial slice is pierced by a topological defect $\D_a$ for $a\in\TopInt{PP}$. Then the global 
pure state of the system is specified by an element of the twisted sector $\ket{\phi^{(a)}}\in{\Hjunc{a}{\id}{PP}}$. We continue restricting our attention to primary states for definiteness.
In the space-time picture, the defect line labeled by $a$ emanates from the associated twist field $\phi^{(a)}(z,\bar z)$ preparing the state in the distant past: 
\begin{center}
    \includegraphics[height=3cm]{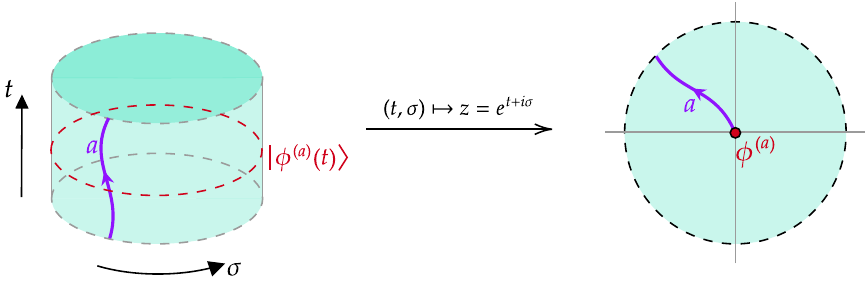}
\end{center}

\begin{figure}[ht]
    \centering
    \includegraphics[height=3cm]{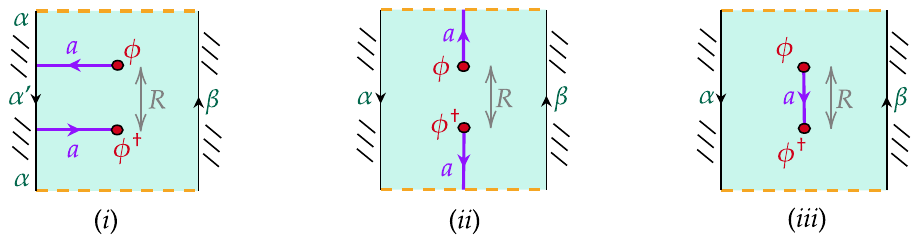}
    \caption{Twist field RDMs obtained by qualitatively different choices of defect networks, in the strip frame \eqref{wCoordn}. Here we assumed minimal fusion rules $N_{ab}^c\in\{0,1\}$, otherwise one should add extra labels for the $(\alpha a \alpha')$ junctions in $(i)$.
    As in \eqref{RDM}, these should be normalized to ensure ${\rm Tr} \rho^{\phi,a}_{\ab}=1$. The normalization factor is the correlator obtained by gluing the ends of the strip. The parameter $R$ accounts for the (dimensionless) length of the interval, see \appref{app:EE_bulkRDMs}.}
    \label{fig:TwistedRDMs}
\end{figure}

Since the defect is topological, it does not matter where it pierces the spatial slice exactly. However, once we choose a subregion $A$, the relative position of $A$ and the defect is relevant. In the setup under consideration, we have three distinct cases: $a$ crosses \change{(i) within the interval $A$, (ii) at one of its boundary points,}{(i) at one of the boundary points of the interval $A$, (ii) within $A$} or (iii) outside $A$:
\begin{center}
    \includegraphics[height=3cm]{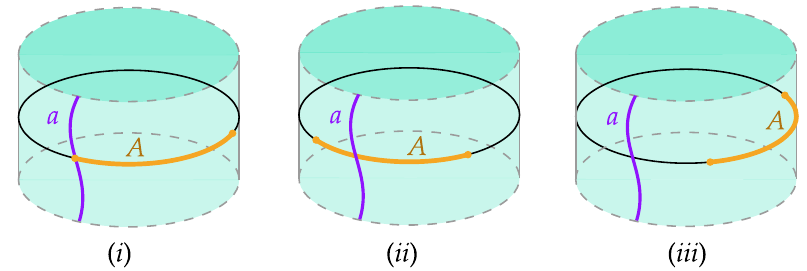}
\end{center}
The density matrix associated to the field $\phi^{(a)}$, the factorization homomorphism and the resulting RDM can be constructed following the same steps \eqref{eq:globalDMfactorized}, \eqref{cutSphere}, \eqref{RDM} with the important subtlety that in case (i) the line defect $a$ attaches to one of the two boundaries (say $\a$). This requires us to introduce another boundary condition $\a'$ compatible with the fusion rules, i.e. $\a'\in \a \fuse a$. For non-minimal fusion rules, there can be many trivalent junctions between $a,\a,\a'$ so in that case one needs to introduce an extra label $n$. We will not consider examples with non-minimal fusion rules in this work and for this reason we avoid introducing this extra label to keep the notation simpler, although our setup could easily accommodate for that. We moreover assume this junction to be topological, hence its precise location does not matter. We avoid introducing non-topological junctions at this point since their physical interpretation would be unclear to us.
The resulting RDMs for the three cases are reported in Fig. \ref{fig:TwistedRDMs}.

Replica manifolds for the cases \change{(iii) and (ii)}{(ii) and (iii)} are mapped into each other by replacing $R\mapsto 1-R$, so it is enough to analyze cases (i) and (ii). 
The \change{former}{latter} corresponds to an impurity-doped subsystem $A$ and was dubbed \quotes{symmetric entropy} in \cite{Roy_2022, Northe:2025zmv}. The EE for (i), instead, provides a measure of quantum correlations across the defect and was called \quotes{interface entropy} in \cite{Roy_2022,Northe:2025zmv}.

A few structural remarks are in order:
\begin{itemize}
    \item The twist field $\phi^{(a)}$ is not part of the modular invariant describing the bulk spectrum $\cal H$. In fact, looking at entanglement through topological defect requires considering $\cal H$ together with all twisted sectors.
    \item $\phi^{(a)}$ cannot be the bulk vacuum of the theory when $a\neq\id$. Therefore we are required to look at entanglement in excited states.
    \item The BCFT approach allows us to have technical control over the RDM in presence of any kind of defect network. In some cases it even grants control over the entanglement spectrum, as we shall see below. We remark that the conventional (replica) twist field picture of entanglement \cite{Calabrese_2009} becomes technically challenging when treating the \quotes{interface} case (ii) since several defects connect to the entangling edge.
    \item The present method also requires us to make some choices that fully specify these defect networks (such as the relative position between the defect $a$ and the interval $A$ as well as the intermediate boundary condition $\a'$). 
    In fact, the RDMs in Fig. \ref{fig:TwistedRDMs} all define EE in the same state $\phi^{(a)}$, but differ only by a choice of defect network. 
    Generically, these lead to physical consequences on the predicted value of EE \cite{Northe:2025zmv}. It is therefore of prime importance to investigate the impact that the different defect network has on EE.
\end{itemize}

\subsection{Free boson grouplike twist field}\label{secFBtwistField}
To demonstrate the theory laid out in \secref{secTwistedStateRDMs}, we now describe entanglement in the presence of a grouplike defect. Such a calculation has been described for the Ising model before in \cite{Northe:2025zmv}. Here we extend the portfolio of examples by considering grouplike defects for the free boson, as reviewed in \secref{secFBgrouplike}.

We focus on the symmetric entropy, i.e. the case that the grouplike defect $g_R(x,y)$ pierces the entanglement interval. This defect emanates in the distant past from a twist field and terminates in the distant future in a conjugate twist field. We are led to the following reduced density matrix
\begin{align}
    \rho_{\ab}^{(x,y)}
    &=
    \frac{1}{Z^{(x,y)}_{\ab}(q)}
    \vcenter{\hbox{\includegraphics[width=0.2\linewidth]{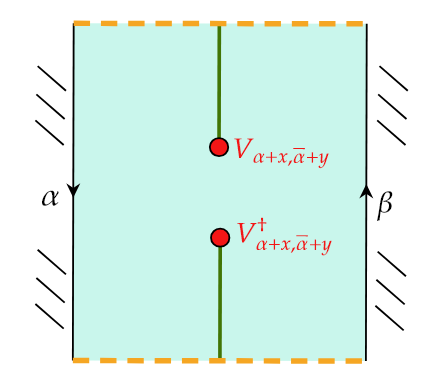}}}\\
    Z^{(x,y)}_{\ab}(q)
    &=
    \tr_{\ab}\left[q^{H_{\ab}} V_{p+x,\bar p+y}(w(0))V_{-p-x,-\bar p-y}(w(\infty))
    \right],\quad (p,\bar p)\in\Lambda(R)\notag
\end{align}
where by superscript $(x,y)$ we abbreviate the field twisted by $g_R(x,y)$. The R\'enyi entropy $S_n\left(\rho_{\ab}^{(x,y)}\right)$ is computed from the $n^\th$ moment
\begin{align}\label{n_partition_function_twistedFB}
        \tr\left[\left(\rho_{\ab}^{(x,y)}\right)^n\right]
        & = 
        \frac{1}{{(Z_{\ab}^{(x,y)}(q))}^n}
        \tr\left[q^{n H_{\ab}}
        \prod_{m=1}^{n}V_{p+x,\bar p+y}(w(0))V_{-p-x,-\bar p-y}(w(\infty))
        \right] \\
        & =
        \frac{1}{{(Z_{\ab}^{(x,y)}(q))}^n}
        \bbra{\a}
        \tq^{\frac{H}{4n}}
        \prod_{m=1}^{n}V_{p+x,\bar p+y}(\tilde{w}(0_m)) V_{-p-x,-\bar p-y} (\tilde{w}(\infty_m)) \tq^\frac{H}{4n}
        \bket{\b} \notag \\
        & = 
        \frac{1}{{(Z_{\ab}^{(x,y)}(q))}^n}
        \sum_{k,l} 
        \bbra{\a} \tq^{\frac{H}{4n}} \ketbra{k}{k} \notag\\
        & \qquad \qquad
        \prod_{m=1}^{n}V_{p+x,\bar p+y}(\tilde{w}(0_m)) V_{-p-x,-\bar p-y} (\tilde{w}(\infty_m)) \ketbra{l}{l}\tq^{\frac{H}{4n}}\bket{\b}\notag
\end{align}
where the anti-holomorphic coordinate labels are implicit. Here, $H_{\a\b}$ is the open string Hamiltonian and $H=L_0 + \bar{L}_0 -\frac{1}{12}$ is the bulk Hamiltonian.

The insertion points $\tilde{w} (\infty_m), \hspace{0.1cm}\tilde{w}(0_m)$ are the ones of \eqref{InsertionsTwN}. In the $\tq \rightarrow 0 $ limit, the boundaries are pushed to infinity and only the lowest weight states in the sums over $k,l$ survive. We get
\begin{equation}
    \begin{split}
         &\hspace{.5cm} \sum_{k,l} \bbra{\a} \tq^{\frac{H}{4n}} \ket{k}\bra{k} \prod_{m=1}^n V_{p+x,\bar p+y}(\tilde{w}(0_m)) V_{-p-x,-\bar p-y}(\tilde{w}(\infty_m)) \ket{l}\bra{l}\tq^{\frac{H}{4n}}\bket{\b} \\
         & \longrightarrow \tq^{-\frac{1}{24n}} \gf_{\a} \gf_{\b}\corr{ \prod_{m=1}^n\,V_{p+x,\bar p+y}(\tilde w (0_m)) V_{-p-x,-\bar p-y}(\tilde w (\infty_m))}_\cyl
    \end{split}
\end{equation}
We now apply to the correlators for both numerator and the denominator the conformal map
\begin{equation}
    f(z) = 2\pi \left( z + \frac{R+1}{2n}\right).
\end{equation}
This produces a conformal Jacobian that is the same for numerator and denominator and thus cancels,
\begin{align}
     \tr\left[\left(\rho_{\ab}^{(x,y)}\right)^n\right]
     &=
     \frac{\tq^{-\frac{1}{24n}}}{\tq^{-\frac{n}{24}}}
     \frac{\gf_{\a} \gf_{\b}\, 
     \langle \prod_{m=1}^n V_{-p-x,-\bar p-y} (\frac{2\pi m}{n}) V_{p+x,\bar p+y}(\frac{2\pi(R+m)}{n})\rangle_\cyl}{(\gf_\a \gf_\b)^n \biggl( \langle V_{-p-x,-\bar p-y}(0) V_{p+x,\bar p+y} (\frac{2\pi R}{n})\rangle_{\cyl} \biggr)^n}\notag\\
     &=
     \tq^{\frac{1}{12}(n-\frac{1}{n})}(\gf_\a \gf_\b )^{1-n}\cF^{(p+x,\bar p+y)}_n(R).
\end{align}
where we defined the expression
\begin{equation}
    \cF^{(x+p,\bar p+y)}_n(R) := \frac{\langle  \prod_{m=1}^{n}V_{-p-x,-\bar p-y} (\frac{2\pi m}{n}) V_{p+x,\bar p+y}(\frac{2\pi(R+m)}{n})\rangle_\cyl}{\biggl( \langle V_{-p-x,-\bar p-y}(0) V_{p+x,\bar p+y} (\frac{2\pi R}{n})\rangle_{\cyl} \biggr)^n}
\end{equation}
Note the unfortunate clash of notation: The parameter $R$ is not the compactification radius of the boson, but the size of the entangling interval. This takes exactly the form presented in the seminal work \cite{Ibáñez_Berganza_2012}, where expressions of this type had been considered for bulk fields. Here we show that entanglement entropies for twist fields result in an expression of the same type. The only difference is that here we do not consider bulk but twist fields. Because the grouplike twist fields here are also of the type \eqref{VertexOp}, the bulk field result $\cF^{(p+x,\bar p+y)}_n(R)=1$ \cite{Ibáñez_Berganza_2012} directly translates to the grouplike twist fields.

Using $\tq = e^{-2W}$ (see appendix \ref{app:EE_bulkRDMs}) we can compute the R\'enyi entropy
\begin{equation}\label{REgrouplikeFB}
    S_n\left(\rho_{\ab}^{(x,y)}\right)
    =
    \frac{1}{1-n}\log\text{Tr}\left[\left(\rho_{\ab}^{(x,y)}\right)^n\right]
    =
    \frac{W}{12}\frac{n+1}{n} +  \log(\gf_\a\gf_b) +\dots
\end{equation}
where dots stand for subleading terms in the UV cutoff.

The entanglement entropy for the grouplike twist fields takes the same form as that of bulk field primaries, which in turn take the same form as that of the vacuum state. This happens because grouplike defects introduce a simple deformation of the lattice $\Lambda$ and therefore just a shift in the conformal weight of operators. However, \change{all the dependence from the conformal weight that of the R\'enyi entropy is within}{all the dependence of the R\'enyi entropy on the conformal weights is contained in}  $\cF^{(a,b)}_n(R)$, which is trivial for any pair of holomorphic and anti-holomorphic momenta $(a,b)$.

We briefly comment on the interface entropy and the symmetric entropy for interval $B$, i.e. cases $(i)$ and \change{$(iii)$}{$(ii)$} in \figref{fig:TwistedRDMs}, respectively. Because the defect is grouplike, it can be unfused from the boundary in case $(i)$, so that we obtain case $(iii)$. When constructing the replica manifold, it becomes clear that the defects lie shifted with respect to case $(ii)$ analyzed above. The calculation however follows the same pattern and ultimately leads to the same result. It follows that the defect network is in fact irrelevant in this particular case.

The physical picture is that free boson grouplike defects do not quantitatively affect quantum correlations, only qualitatively. Indeed, a grouplike defect is only expected to reshuffle degrees of freedom, and hence quantum correlations can neither diminish nor increase. This expectation has similarly been borne out in the Ising model's grouplike defect \cite{Northe:2025zmv}. Crucially, for non-invertible defects this behavior is not expected to occur. An example of this is easily found in the Ising model's Kramers-Wannier defect which has also been analyzed in \cite{Northe:2025zmv}.

\section{Defect-dressed Reduced Density Matrices II -- Interfaces
}\label{secDualityIntRDM}
In this section, we generalize the factorization prescription discussed above to topological interfaces between different 1+1-dimensional Rational CFTs.
We begin with a general description of RDMs in the presence of multiple topological lines or interfaces piercing the spatial slice in \secref{secInterfaceRDMs}.
In \secref{secRDMduality}, we then illustrate this framework by specializing to duality interfaces, which provide a particularly interesting example with a rich algebraic structure.
A central result of this analysis is that RDMs in the presence of duality interfaces naturally implement projectors of the type appearing in symmetry resolved entanglement entropy \cite{di2023boundary, Kusuki:2023bsp, northe2023entanglement}. In this case, resolution is made with respect to the group-like symmetry \eqref{eq:rightleft_stabilizers_def} associated with the duality. The information theory of these RDMs is relegated to \secref{secInfoThyDuality}. Finally, we consider RDMs of thermal states in \secref{secThermalStates}.

\subsection{Defect and Interface RDMs}\label{secInterfaceRDMs}
Let us start by generalizing the RDM construction to any number of topological defects piercing the spatial slice. With $n$ defect lines $a_1 ,\cdots ,a_n\in\TopInt{PP}$, the state $\ket{\phi^{(a_1 ,\cdots,a_n)}}$ of the system belongs to the $n$-junction space $\Hjunc{a_1a_2\dots a_{n-1}}{a_n}{PP}$. The associated field $\phi^{(a_1 ,\cdots,a_n)}(z,\bar z)$ preparing the state in the distant past lives at the junction of the $n$ defects and is a defect junction field.

In the construction of the RDM, we need to specify where each defect ends relative to the position of the interval $A$, and eventually also specify intermediate boundary conditions and corresponding topological junctions. This clearly leads to a growing number of alternative RDMs computing entanglement entropy in the same defect field $\phi^{(a_1,\cdots,a_n)}$. A priori, different defect networks lead to different entanglement entropies. 

\begin{figure}[ht]
    \centering
    \includegraphics[height=3.5cm]{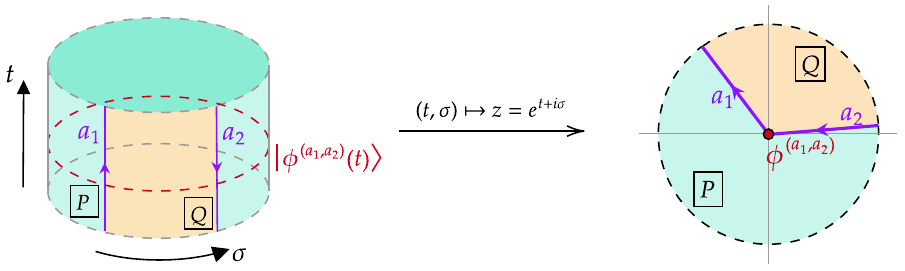}
    \caption{System in the presence of two interface lines $a_1,a_2$ separating phases $P$ and $Q$. In radial quantization, the state $\ket{\phi^{(a_1,a_2)}}$ corresponding to the interface field $\phi^{(a_1,a_2)}$ sits at the junction between $a_1$ and $a_2$.}
    \label{fig:DefectInterfaceState}
\end{figure}

Let us consider $n=2$ lines for definiteness. When $a_2=\id$ is the trivial defect, $a_1$ is necessarily a defect, so that our analysis of \secref{secTwistedStateRDMs} applies. In this section, we take $a_1$ and $a_2$ to be generic topological $PQ$ interfaces, see \figref{fig:DefectInterfaceState}. Of course we may also straightforwardly generalize to $n\geq2$, but these setups are not detailed here. 

Similarly to \secref{secTwistedStateRDMs}, we can apply the Ohmori-Tachikawa factorization homomorphism, see \eqref{eq:globalDMfactorized}, to produce a state in a bipartite Hilbert space. 
Again, depending on the relative position between the interval $A$ and the interfaces, this produces multiple possible states, differing only by a choice of defect network. Two examples are
\begin{center}
    \includegraphics[height=3cm]{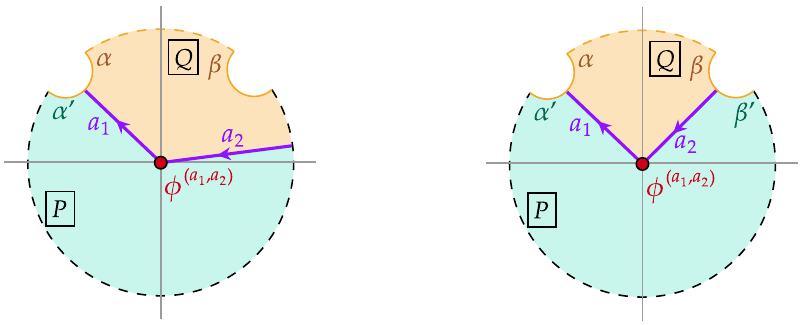}
\end{center}
Following the steps \eqref{eq:globalDMfactorized}, \eqref{cutSphere}, \eqref{RDM}, we then construct the RDM associated to each of these alternatives. Some of the resulting RDMs are equivalent or related upon considering their replica manifolds (as for the \quotes{symmetric} cases (ii) and (iii) in \secref{secTwistedStateRDMs}). Keeping this in mind, the inequivalent RDMs are, up to normalization,
\begin{equation}\label{eq:defectFieldRDMs}
\rho_{\ab}^{\phi,A}\to\quad \vcenter{\hbox{\includegraphics[height=4.5cm]{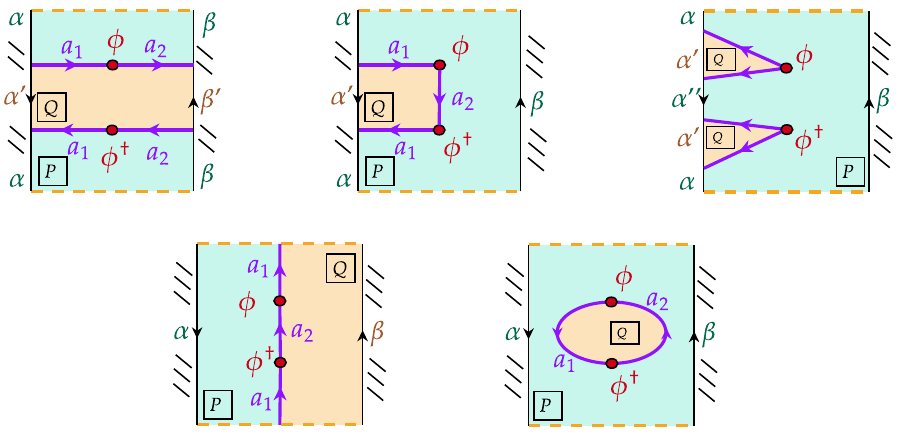}}}
\end{equation}
Here we simplified the notation $\phi^{(a_1,a_2)}\to\phi$ to avoid cluttering.

Notice that, for $a_1=a_2\equiv a$, it is always possible to pick the defect vacuum field $\phi=\id_{a}$ carrying zero conformal dimension. This situation is in sharp contrast with the case of one defect line, where we are forced to consider fields with non-zero conformal dimension. 
In more detail,
the second and third cases in \eqref{eq:defectFieldRDMs} reduce to the RDM of the vacuum.\footnote{This is due to our normalizations, in which a bubble attached to a boundary without field insertion acts as the identity. But even for other conventions where one can pick up a proportionality factor, it cancels away against the normalization, leading to \change{vacuum state's}{the vacuum-state} RDM.} The fifth RDM picks up a quantum dimension $\qdim_a$, which cancels out against the normalization. The first and fourth RDMs are still non-trivial.

The computation of replica manifolds proceeds as usual by taking $n$ copies and gluing these cyclically. We thus arrive at replica geometries of the same size but with different interface networks. Generically, these now depend highly on the specifics of the interfaces and boundaries involved. To elucidate this and to make further formal progress, in the rest of the section we take a detailed look at the first option of \eqref{eq:defectFieldRDMs} for the case where $a\in\TopInt{PQ}$ is a duality interface $\I_\dua$ (see  \eqref{duality}) and the field $\phi=\id_{\dua}$ is its identity field. 
\subsection{Entanglement through duality interfaces}\label{secRDMduality}
We fix a constant time slice as in \figref{fig:DefectInterfaceState} with $a_1=a_2=\dua$ and consider an entangling interval $A$ stretching from interface to interface in phase $P$; that is, all of phase $Q$ belongs to the region $B$. A consistent completion of this interface network on the sphere is achieved by connecting the interfaces in the distant past with a field in the spectrum $\phi_{i\bi}\in\Hjunc{\dua}{\dua}{PQ}$. Similarly the network is completed in the distant future by the conjugate field $\phi_{i^+\bi^+}\in\Hjunc{\dua}{\dua}{PQ}$. 

We restrict to the ground state of this configuration. The lightest state in $\Hjunc{\dua}{\dua}{PQ}$ is the topological field $\id_\dua$ and has $h=0=\bh$. As with the bulk identity field, we do not draw it. The RDM describing the subsystem is hence 
\begin{align}\label{RDMtwoDualities}
   \rho^{\id_\dua\, \a'\b'}_\ab 
   \propto
\vcenter{\hbox{\includegraphics[height=2.5cm]{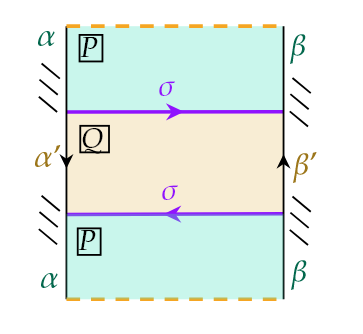}}}
=
q^{\frac{1}{3}H_\ab^P}\,
   \ibo{\dua}{\ab}{\a'\b'}\,
   q^{\frac{1}{3}H_{\a'\b'}^Q}
   \bibo{\dua}{\a'\b'}{\ab}
   q^{\frac{1}{3}H_\ab^P}
   =
   q^{H_\ab^P}\,
   \ibo{\dua}{\ab}{\a'\b'}\,
   \bibo{\dua}{\a'\b'}{\ab}
\end{align}
Besides the boundaries $\a,\b\in\cB_P$, the factorization \eqref{iota_def} has two intermediate boundary conditions $\a',\,\b'\in\cB_Q$, i.e. $\underline{\a}=(\a,\b,\a',\b')$. The RDM's proportionality is up to a normalization factor ensuring $\tr_\ab[\rho_\ab^\dua]=1$ and $H_\ab^P$ is the strip Hamiltonian with boundary conditions $\a,\b$ in phase $P$. Note that the incoming and outgoing boundary conditions must agree in the RDM for $\rho^{\id_\dua\, \a'\b'}_\ab\in\End(\cH_\ab^A)$ to apply; otherwise this operator would not be an RDM. 

We now show in three different cases, that this RDM acts as a projector on BCFT Hilbert spaces $\cH_{\ab}^A$, which in \secref{secInfoThyDuality} are interpreted as entanglement spectra. The first are Verlinde line defects in \secref{secDualityVerlinde}, which apply to diagonal theories. Then we drop the assumption of a diagonal modular invariant and turn to duality interfaces, whose orientation reversal is also a duality interface in \secref{secDualityProjectorGeneral}. Finally, in \secref{secDualityProjFB}, we consider an example with extended symmetry, namely $\T$-duality in the free boson, in a mildly modified setup, resulting in a modified projector. As we describe in more detail in \secref{secInfoThyDuality}, this means that the RDM's entanglement spectrum is diluted to the subspace which is projected onto.

\subsubsection{Diagonal Theories and Verlinde Duality defects}\label{secDualityVerlinde}
Verlinde defects \eqref{VerlindeLines} connect a diagonal bulk theory \eqref{diagModInv} to itself, so that $P=Q\cong\bf 1$, and we drop the phase label unless necessary. Cardy boundary conditions \eqref{CardyStates} are employed to fix the boundary spectra $\cH_\ab^A$. The counteroriented partner $\bD_\dua$ of a duality Verlinde defect $\D_\dua$ is also dualizing. Therefore their left- and right-stabilizers agree, $\stab{\dua}\equiv\stabR{\dua}\cong\stabL{\dua}$ and form an abelian group \cite{Frohlich_2007}. 

Fusing the two duality defects via \eqref{BdyDefFusionCounter} leads exclusively to grouplike defects, due to \eqref{duality}, with fusion coefficients
\begin{align}
    \bfusMa{\dua\dua}{g}{\ab}{\a'\b'}{\ab}
    =
    \frac{\qdim_g}{\qdim_\dua}
    \fusing{g\dua\a'}{\a}{\a\dua}
    \,
    \fusingG{g\dua\b'}{\b}{\dua\b}
\end{align}
Grouplike Verlinde lines have unit quantum dimensions, $\qdim_g=1$. Furthermore, the intermediate boundaries $\a',\b'\in\cB_Q$ cannot be arbitrary: The fusing matrix $\fusing{g\dua\a'}{\a}{\a\dua}$ effectuates the change
\begin{align}
\vcenter{\hbox{\includegraphics[height=2cm]{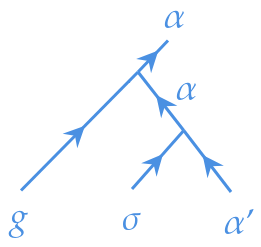}}}
=
\fusing{g\dua\a'}{\a}{\a\dua}
\vcenter{\hbox{\includegraphics[height=2cm]{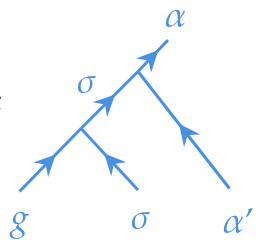}}}
\end{align}
No sum appears on the right-hand side because $g\in\stab{\dua}$, so that all other channels are forbidden. As seen from the left-hand side, this fusing element is non-zero if 1) $\a\in\dua\fuse\a'$ and 2) $\a\in g\fuse \a$. Because grouplike families map simple objects in $\confFam$ into simple objects, requirement 2) becomes $\a=g\fuse\a$. The relation 1) then turns into $g\fuse \a\in\dua\fuse\a'$, which is satisfied when $\a=\dua$ and $\a'\in\stab{\dua}$ are chosen. A similar argument for the fusing matrix $\fusingG{g\dua\b'}{\b}{\dua\b}$ establishes $\b=\dua$ and $\b'\in\stab{\dua}$. We are thus interested in the RDM 
\begin{equation}\label{RDMduality}
    \rho_{\dua\dua}^{\id_\dua\, \ab}
    \propto
    q^{H_{\dua\dua}}\,
   \dbo{\dua}{\dua\dua}{\ab}\,
   \bdbo{\dua}{\ab}{\dua\dua}\,,
   \qquad
   \a,\b\in\stab{\dua}
\end{equation}
Note that we have dropped the primes on the intermediate boundary labels to reduce clutter.

The following relations can be established for $\a,\b\in\stab{\dua}$ by the pentagon identity:
\begin{align}\label{reprProperty}
    \fusing{g\dua\a}{\dua}{\dua\dua}\fusing{h\dua\a}{\dua}{\dua\dua}
    &=
    \fusing{g\cdot h\,\dua\a}{\dua}{\dua\dua}\\
    \fusingG{g\dua\b}{\dua}{\dua\dua}\fusingG{h\dua\b}{\dua}{\dua\dua}
    &=
    \fusingG{g\cdot h\,\dua\b}{\dua}{\dua\dua}
\end{align}
The first of these relations for instance is seen via
\begin{equation}\label{PentagonDuality}
    \vcenter{\hbox{\includegraphics[height=5cm]{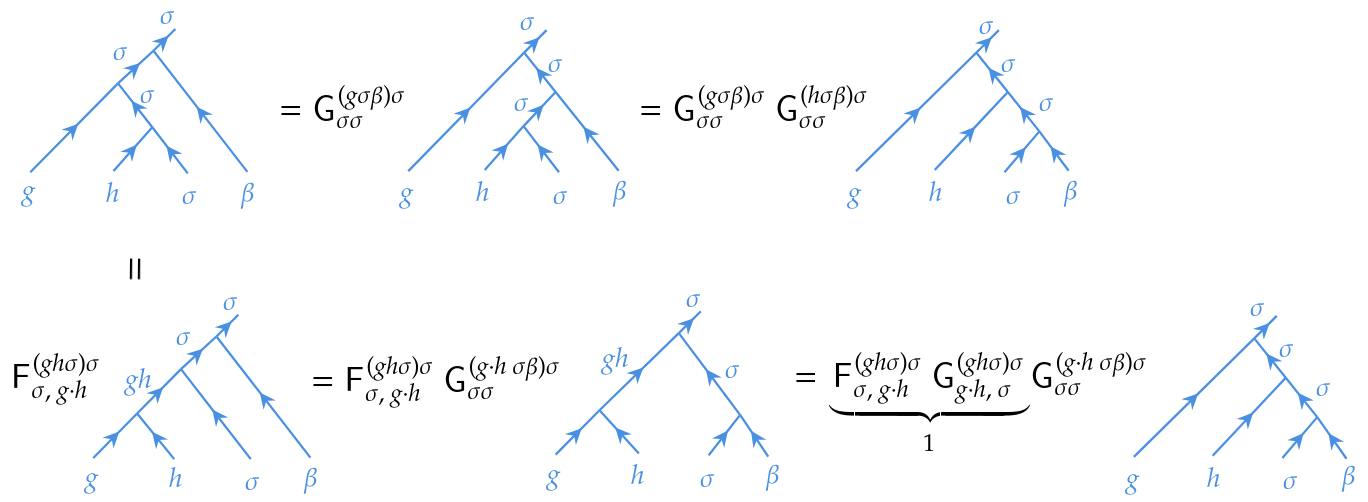}}}
\end{equation}
Together with $\fusing{\id\dua\a}{\dua}{\dua\dua}=1=\fusingG{\id\dua\a}{\dua}{\dua\dua}$, these $\mathsf{F}$ and $\mathsf{G}$ fusing elements clearly furnish a representation of the stabilizer $\stab{\dua}$ by themselves. 

Because the stabilizer $\stab{\dua}$ is abelian, the fusing matrix entries at hand are trivially class functions, so that they must be characters $\varrho$ of $\stab{\dua}$. Because of $\fusing{g\dua\a}{\dua}{\dua\dua}\fusingG{g\dua\a}{\dua}{\dua\dua}=1$, we fix
\begin{align}\label{characterF}
    \varrho_\a(g)
    =
    \fusing{g\dua\a}{\dua}{\dua\dua}
    =
    \left(\fusingG{g\dua\a}{\dua}{\dua\dua}\right)^{-1}
\end{align}

By construction, the cardinality of the $\a$s is $|\stab{\dua}|$, but this does not yet guarantee that every single character is in fact realized. Similarly to \eqref{PentagonDuality}, one can demonstrate that also the third entry in parentheses of the fusing matrices furnishes a representation of the stabilizer $\stab{\dua}$,
\begin{align}
    \fusing{g\dua\a}{\dua}{\dua\dua}\fusing{g\dua\b}{\dua}{\dua\dua}
    &=
    \fusing{g\dua\,\a\cdot\b}{\dua}{\dua\dua}\\
    \fusingG{g\dua\a}{\dua}{\dua\dua}\fusingG{g\dua\b}{\dua}{\dua\dua}
    &=
    \fusingG{g\dua\,\a\cdot\b}{\dua}{\dua\dua}\\
    \fusing{g\dua\id}{\dua}{\dua\dua}
    =
    1
    &=
    \fusingG{g\dua\id}{\dua}{\dua\dua}
\end{align}
Hence the numbers $\fusing{g\dua\a}{\dua}{\dua\dua}$ and $\fusingG{g\dua\a}{\dua}{\dua\dua}$ are in fact bihomomorphisms $\stab{\dua}\times\stab{\dua}\to\C^\times$. For the characters this implies, as expected,
\begin{align}
    \varrho_\a(g)\varrho_\b(g)=\varrho_{\a\cdot\b}(g)
\end{align}
Because characters of finite abelian groups, such as $\stab{\dua},$ are distributed on the unit circle, we have $(\varrho_\a(g))^{-1}=\varrho_{\a^{+}}(g)$. 

Combining everything, and writing $\check{g}^\dua:=\dbo{g}{\dua\dua}{\dua\dua}$ to reduce clutter, the RDM \eqref{RDMduality} is expressed through
\begin{align}\label{RDMDualityProj}
    \rho_{\dua\dua}^{\id_\dua\, \ab}
    \propto
    q^{H_{\dua\dua}}\sum_{g\in\stab{\dua}}
    \frac{1}{\qdim_\dua}
    \varrho_\a(g)\varrho_{\b^{+}}(g)\,
    \check{g}^\dua
    =
    \frac{|\stab{\dua}|}{\qdim_\dua}q^{H_{\dua\dua}}\cP_{\a^{+}\otimes\b}
\end{align}
where $\cP_{\a^{+}\otimes\b}$ is the projector onto the representation $\a^{+}\otimes\b$. It satisfies $\cP_{\a}\cP_{\b}=\delta_{\a,\b}\cP_\a$. The prefactor $\frac{|\stab{\dua}|}{\qdim_\dua}$ cancels out due to the normalization $\tr_{\dua\dua}[\rho_{\dua\dua}^{\id_\dua\, \ab}]=1$.

Because the representations at hand are all one-dimensional, essentially only one label is needed, i.e., one can set $\a=\id$. This can be seen by the following move
\begin{equation}\label{MoveRDMDuality}
    \vcenter{\hbox{\includegraphics[height=3cm]{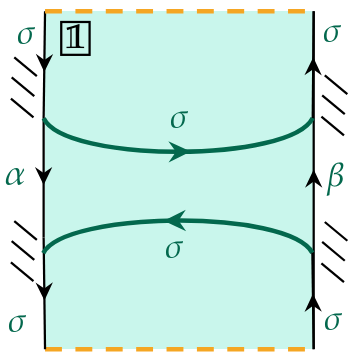}}}
    =
    \vcenter{\hbox{\includegraphics[height=3cm]{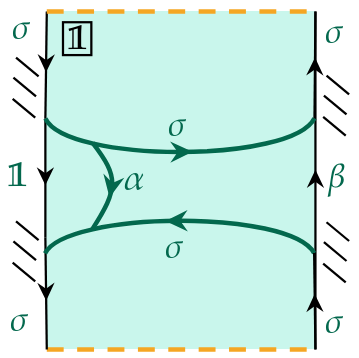}}}
    =
    \vcenter{\hbox{\includegraphics[height=3cm]{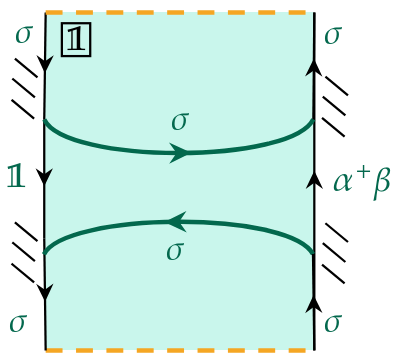}}}
\end{equation}
Since $\a,\,\b$ are both grouplike, their fusion results in a simple label for the boundary on the right. 

It remains to be seen whether the representations at hand are actually regular or projective. To that end one must consider the parallel fusion of two boundary-anchored defects. The fusion rule \eqref{BdyDefFusion} for grouplike defects anchored to boundaries labeled by $\dua$ furnishes a regular representation
\begin{equation}
    \check{h}^\dua\check{g}^\dua
    =
    \underbrace{\fusingG{hg\dua}{\dua}{g\cdot h\,\dua}\,
    \fusing{hg\dua}{\dua}{\dua\,g\cdot h}}_{1}\,
    \check{(gh)}^\dua
\end{equation}

The result \eqref{RDMDualityProj} is remarkable. While duality defects naturally produce the projector onto the trivial representation of the \change{stablilizer}{stabilizer} $\stab{\dua}$ when acting on the bulk CFT Hilbert space, as seen in \eqref{duality}, RDMs dressed with two parallel boundary-anchored duality interfaces yield projectors for any representation contained in the boundary Hilbert space. These representations are selected by the intermediate boundary conditions $\a,\,\b\in\stab{\dua}\subset\cG_\id$.

Finally, it is useful to have the action of this projecting RDM on boundary fields. For that we need the action of $\check{g}^\dua=\dbo{g}{\dua\dua}{\dua\dua}$ on the boundary fields $\psi_i^\ab$ in $\cH^A_{\dua\dua}$. This is read out from \eqref{VerlindeBdyAction} to be 
\begin{equation}
    \check{g}^\dua_i\psi_i^\ab
    \equiv
    \db{g}{\dua\dua}{i}{\dua\dua}\psi_i^\ab
    =
    \fusing{g\dua i}{\dua}{\dua\dua}\psi_i^\ab
\end{equation}
and therefore \eqref{RDMDualityProj} acts as 
\begin{align}\label{RDMDualityProjPsi}
    \rho_{\dua\dua}^{\id_\dua\, \a0}\psi_i^{\dua\dua}
    \propto
    \frac{q^{H_{\dua\dua}}}{|\stab{\dua}|}
    \sum_{g\in\stab{\dua}}
    \varrho_\a(g)\,
    \check{g}^\dua\,
    \psi_i^\ab
    =
    \frac{q^{H_{\dua\dua}}}{|\stab{\dua}|}
    \sum_{g\in\stab{\dua}}
    \fusing{g\dua\a}{\dua}{\dua\dua}
    \fusing{g\dua i}{\dua}{\dua\dua}
    \psi_i^\ab
\end{align}
As discussed in detail in \secref{secInfoThyDuality}, one can interpret this result in two ways: 
\begin{itemize}
    \item[1)] The RDM \eqref{RDMDualityProj} naturally implements symmetry resolution on the entanglement spectrum of the RDM corresponding to the global vacuum state $\rho_{\dua\dua}^\id$. Note the similarity with the projectors constructed in \cite{Heymann_2025}, modulo conventions. It is important to point out that their construction applies also to non-invertible symmetries, while ours is valid strictly for grouplike symmetries, concretely duality defect stabilizers $\stabR{\dua}$.
    \item[2)] The entanglement spectrum for the global state corresponding to the identity field on $\D_\dua$, i.e. $\id_\dua$, is naturally a subsector of $\cH_{\dua\dua}^A$, which forms the entanglement spectrum of the bulk identity field $\id$. 
\end{itemize}

\subsection*{Example: Ising model}
Consider first the RDM $\rho^\id_{\dua\dua}$ for the bulk vacuum, i.e. in absence of defects. The BCFT Hilbert space, which represents the entanglement spectrum of the vacuum state $\ket{0}\in\cH$ for a single interval, is according to \eqref{CardyStateSpaces}
\begin{equation}
    \cH_{\dua\dua}^A=\cH_0\oplus\cH_{\nicefrac{1}{2}}
\end{equation}
The first summand carries bosonic fields and the second fermionic ones. These are the Ising model's boundary spin flip $\Z_2$ representations. 

The Ising model's duality defect implements Kramers-Wannier duality and corresponds to the label $\sigma=\nicefrac{1}{16}$. For the global state $\id_\dua$ two RDMs $\rho_{\dua\dua}^{\id_\dua\, \a 0}$ are possible, controlled by $\a=0,\nicefrac{1}{2}$. While $\rho_{\dua\dua}^{\id_\dua\, 00}$ projects onto $\cH_0$, $\rho_{\dua\dua}^{\id_\dua\, \nicefrac{1}{2}0}$ projects onto $\cH_{\nicefrac{1}{2}}$. This is easily confirmed via the fusing symbols
\begin{equation}
    \fusing{0\dua\a}{\dua}{\dua\dua}
    =
    \fusing{\nicefrac{1}{2}\,\dua\, 0}{\dua}{\dua\dua}
    =
    -\fusing{\nicefrac{1}{2}\,\dua\, \nicefrac{1}{2}}{\dua}{\dua\dua}
    =
    1
\end{equation}
which indeed are $\Z_2$ characters as argued in \eqref{characterF}. 
Accordingly, the action \eqref{RDMDualityProjPsi} is
\begin{equation}
    \rho_{\dua\dua}^{\id_\dua\, 00}\psi_0^{\dua\dua}
    \propto
    \psi_0^{\dua\dua}\,,
    \quad
    \rho_{\dua\dua}^{\id_\dua\, 00}\psi_{\nicefrac{1}{2}}^{\dua\dua}
    =
    0\,,
    \qquad
    \rho_{\dua\dua}^{\id_\dua\, \nicefrac{1}{2}0}\psi_0^{\dua\dua}
    =0\,,
    \quad
    \rho_{\dua\dua}^{\id_\dua\, \nicefrac{1}{2}0}\psi_{\nicefrac{1}{2}}^{\dua\dua}
    \propto
   \psi_{\nicefrac{1}{2}}^{\dua\dua}\,,
\end{equation}
which are the promised projectors onto the state spaces $\cH_0$ and $\cH_{\nicefrac{1}{2}}$.

\subsubsection{Non-diagonal Theories and Duality interfaces with abelian stabilizer}\label{secDualityProjectorGeneral}
We now lift the assumption of diagonal bulk spectra and turn to $PQ$ duality interfaces $\I_\dua$. The phases $P$ and $Q$ are a priori unconstrained, but we demand that $\bI_\dua$ also be dualizing. The standard example is $\T$-duality \cite{Fuchs:2007tx} in compact free boson theories at non-self-dual radius to which we however turn in-depth in the next subsubsection. 

Our analysis in the case of the Verlinde lines profited from the known fusing matrices \eqref{Fusing} which allow to manipulate defect networks. Note that the fusing matrices pertain to the category of representations $\confFam$, for which we use the same symbol as for the set of chiral families. For diagonal theories, this set $\confFam$ determines the bulk spectrum. Verlinde lines form a bimodule category associated with $\confFam$, and inherit its label set as well as its fusing matrices.  

For non-diagonal modular invariants in distinct phases $P$ and $Q$, the bimodule category $\confFam_{P|Q}$ describing interfaces is generically distinct from $\confFam$ and fusing matrices must first be constructed. As seen above, it suffices to know the elements $\fusing{g\dua\a}{\dua}{\dua\dua},\, \fusingG{g\dua\b}{\dua}{\dua\dua}$. These are examples of the following bihomomorphic fusing matrix elements $\zeta_\dua(g,h): \stabL{\dua}\times\stabR{\dua}\to\C^\times$ associated specifically with duality interfaces \cite{Frohlich_2007} in $\confFam_{P|Q}$,
\begin{equation}
    \vcenter{\hbox{\includegraphics[height=2.5cm]{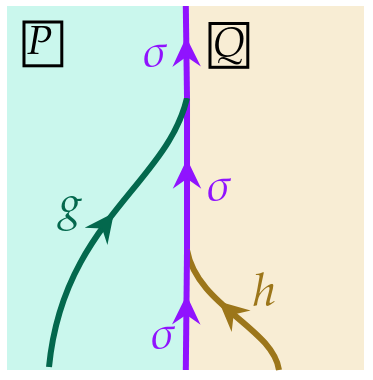}}}
    =
    \zeta_\dua(g,h)
     \vcenter{\hbox{\includegraphics[height=2.5cm]{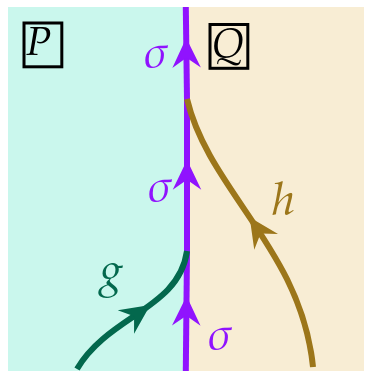}}}
\end{equation}
where $g\in\stabL{\dua}$ and $h\in\stabR{\dua}$. Each junction space involved here is one-dimensional given that the grouplike defects $\D_g$ and $\D_h$ sit in stabilizers. Glancing back at the left-hand side of \eqref{fusingF}, $\zeta_\dua$ is identified as a generalization of $\fusing{g\dua h}{\dua}{\dua\dua}$ specifically. As in the right-hand side of \eqref{fusingF}, there is also the move
\begin{equation}
    \vcenter{\hbox{\includegraphics[height=2.5cm]{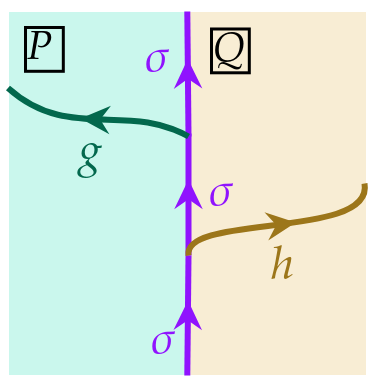}}}
    =
    \zeta_\dua(g,h)
     \vcenter{\hbox{\includegraphics[height=2.5cm]{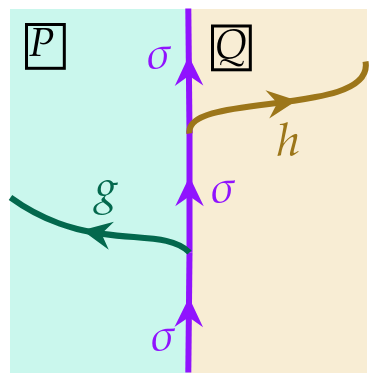}}}
\end{equation}
which allows one to commute the two grouplike defects $\D_g$ and $\D_h$ attached \change{}{to} the $PQ$ duality interface $\I_\dua$. 

As proven in \cite{Frohlich_2007}, the map $\zeta_\dua(g,h)$ has the following two properties: 
\begin{itemize}
    \item[1)] It is a bihomomorphism in its two arguments, $\zeta_\dua(g,h)\zeta_\dua(g',h)=\zeta_\dua(gg',h)$ and $\zeta_\dua(g,h)\zeta_\dua(g,h')=\zeta_\dua(g,hh')$. This is proven by repeating the moves in \eqref{reprProperty} and requires only that $g\in\stabL{\dua}$ and $h\in\stabR{\dua}$, i.e. $\I_\dua$ need a priori not be a duality interface. The unit property $\zeta_\dua(e,h)=1=\zeta_\dua(g,e)$ follows immediately.
    \item[2)]  Requiring that $\I_\dua$ be a duality interface, $\zeta_\dua$ is seen to be non-degenerate in its first argument, i.e. if $\zeta_\dua(g,h)=\zeta_\dua(g',h)$ for some $g,\,g'\in\stabL{\dua}$ and all $h\in\stabR{\dua}$ then $g'=g$. Moreover, $\sum_{h\in\stabR{\dua}}\zeta_\dua(g,h)=\delta_{g,e}\,|\stabR{\dua}|$, which resembles the orthogonality relation for characters of $\stabR{\dua}$.
\end{itemize}
If $\bI_\dua$ is also a duality interface, then
\begin{itemize}
    \item[3)] The left and right stabilizers are abelian and isomorphic $\stabL{\dua}\cong\stabR{\dua}$. 
    \item [4)] Complementary to 2), $\zeta_\dua(g,h)$ is seen to be non-degenerate in its second argument and $\sum_{g\in\stabL{\dua}}\zeta_\dua(g,h)=\delta_{h,e}|\,\stabL{\dua}|$, which resembles the orthogonality relation for characters of $\stabL{\dua}$.
\end{itemize}
Altogether, these properties allow us to identify the bihomomorphism $\zeta_\dua$ as characters for either group, and since $|\stabR{\dua}|=|\stabL{\dua}|$, all characters are realized once. For our purposes picking characters of $\stabL{\dua}$ is appropriate, $\zeta_\dua(g,h)=\varrho_h(g)$. 

Besides the Verlinde lines considered above, \change{}{we} can consider the example of the free boson in passing. Their $U(1)$ preserving duality interfaces have stabilizers given by the group $Z_{MN}$. Let it be generated by $g$ so that any group element is $g^m$. Then the character $\zeta_d(g^m,\alpha)=e^{2\pi\iu \frac{\alpha m}{MN}}$, where $m=0,\dots,MN-1$ and $\alpha=0,\dots,MN-1$.

We now let both interfaces $\I_\dua$ and $\bI_\dua$ in  \eqref{RDMtwoDualities} be duality defects. Manipulating their interface network similarly to \eqref{DefBdyCounterFusion} leads to
\begin{equation} \label{GeneralProjector}
    \vcenter{\hbox{\includegraphics[height=4cm]{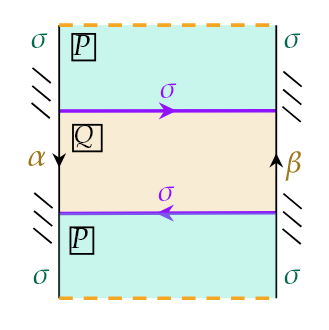}}}
    =
    \sum_{g\in\stabL{\dua}}
    \frac{\qdim_g}{\qdim_\dua}
    \zeta_\dua(g,\a)
    \vcenter{\hbox{\includegraphics[height=4cm]{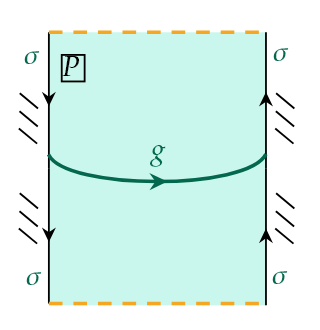}}}
    (\zeta_\dua(g,\b))^{-1}
\end{equation}
where $\a,\b\in\stabR{\dua}$ has been chosen. Note that the sum runs over the left stabilizer, because these duality interfaces appear ordered as $\I_\dua\fuse\bI_\dua$, which is reversed with respect to \eqref{duality}. Grouplike defects of a phase $P$, which lie in a stabilizer of some other interface, have fixed quantum dimension $\qdim_g=\qdim_P$ \cite{Frohlich_2007}, where, as customary, we use the same symbol $P$ to denote a phase and its corresponding Frobenius algebra object in $\confFam$. Replacing $\zeta_\dua(g,\a)=\varrho_\a(g)$ yields 
\begin{align}\label{RDMDualityIntProj}
    \rho_{\dua\dua}^{\id_\dua\, \ab}
   &\propto
   q^{H_{\dua\dua}}\,
   \ibo{\dua}{\dua\dua}{\ab}\,
   \bibo{\dua}{\ab}{\dua\dua}\notag\\
   &=
   q^{H_{\dua\dua}}
   \frac{\qdim_P}{\qdim_\dua}
   \sum_{g\in\stabL{\dua}}
    \varrho_\a(g)\varrho_{\b^{+}}(g)
    \check{g}^\dua
    =
    |\stabL{\dua}|\frac{\qdim_P}{\qdim_\dua}
    \,
    q^{H_{\dua\dua}^P}\cP_{\a^{+}\otimes\b}\,,
\end{align}
which generalizes the result \eqref{RDMDualityProj} for duality Verlinde lines to $PQ$ duality interfaces $\I_\dua$, whose orientation reversal $\bI_\dua$ is also a duality interface. We remind the reader of our shorthand notation $\check{g}^\dua=\dbo{g}{\dua\dua}{\dua\dua}$ for the boundary state space action of a grouplike defect $g\in\stabL{\dua}$. The prefactor $\frac{|\stabL{\dua}|\qdim_P}{\qdim_\dua}$ cancels out due to the normalization $\tr_{\dua\dua}[\rho_{\dua\dua}^{\id_\dua\, \ab}]=1$. The entanglement Hamiltonian is easily read off to be proportional to $H_{\dua\dua}^P|_{\a^+\otimes\b}$, where the superscript indicates the phase $P$, not the projector $\cP$.

We can perform the same move as in \eqref{MoveRDMDuality} and the fact that all irreducible representations of abelian finite groups are one-dimensional to drop one label on the RDM \eqref{RDMDualityIntProj}
\begin{equation}\label{RDMdualityProjSimple}
    \rho_{\dua\dua}^{\id_\dua\, \b}
   =
   \frac{
   q^{H_{\dua\dua}^P}\cP_{\b}}{\chi_\b(q)},
   \qquad
   \chi_\b(q)
   =
   \tr_{\dua\dua}\left[q^{H_{\dua\dua}^P}\cP_{\b}\right]
   \qquad
   \b\in\stabR{\dua}\,,
\end{equation}
where $\a^+\otimes\b\to\b$. The denominator secures $\tr\rho_{\dua\dua}^{\id_\dua\, \b}=1$. This result applies similarly to Verlinde lines and diagonal bulk invariants, i.e. $P=\id$. This form of the RDM is one of the main results of this paper. 

\subsubsection{Free boson duality interface RDM}\label{secDualityProjFB}
We now turn to the free boson. We connect two distinct phases, i.e. two compact bosons at radii $R_1$ and $\hat R_2$, via the $\T$ duality interface \eqref{T}. In contrast to the previous section, we do not pick an incoming boundary condition $\dua$ for the factorization, but a standard Neumann boundary \eqref{DirichletNeumann}, which leads to mild, yet interesting modifications. The RDMs of interest are proportional to the following operator. 
\begin{equation} \label{FB_transfmatrix}
    \vcenter{\hbox{\includegraphics[width=0.44\linewidth]{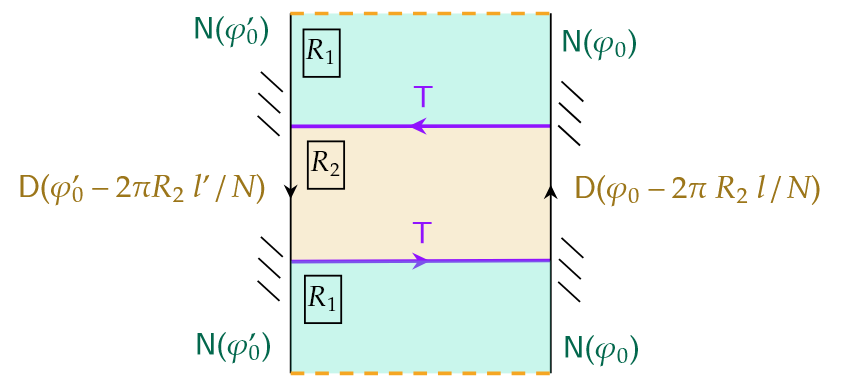}}}
    =
    \sum_{g\in \stabR{\T}}\frac{\qdim_g}{\qdim_{\T}}\hspace{0.2cm}\vcenter{\hbox{\includegraphics[width=0.46\linewidth]{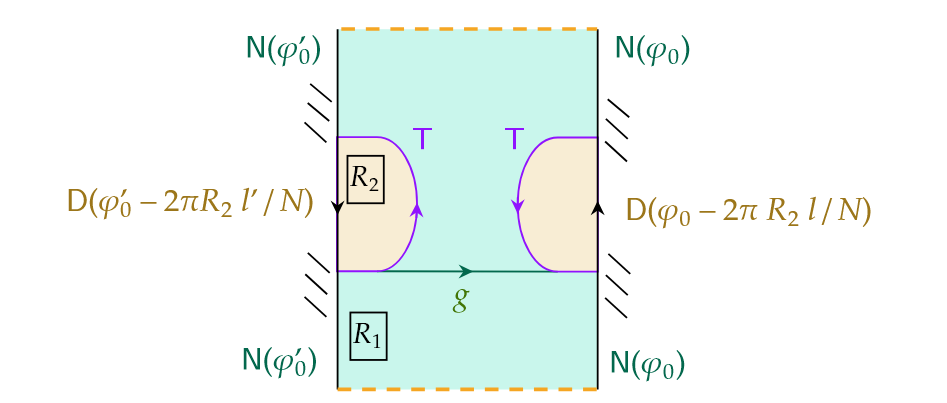}}}
\end{equation}
where \eqref{antiparallelFusion} has been used. Note that the orientations of $\T$ are opposite to the case \eqref{GeneralProjector}, so that they are compatible with the Neumann boundaries. The intermediary boundaries, i.e. those of phase $R_2$, are allowed because by virtue of \eqref{defect_sum_FB_a}
\begin{equation}
    \dir\left(\varphi_0'-\frac{2\pi R_2 l'}{N}\right)\in \T\fuse \neu(\varphi_0'),
    \qquad
    \dir\left(\varphi_0-\frac{2\pi R_2 l}{N}\right)\in \T\fuse \neu(\varphi_0),\qquad l,l'=0,\dots,N-1.
\end{equation}
These boundary labels appear frequently below as \change{index}{indices}, where we abbreviate them as
\begin{equation}\label{DNabbreviations}
    \dir\left(\varphi_0'-\frac{2\pi R_2 l'}{N}\right)\to\dir' \,, 
    \quad 
    \neu(\varphi_0')\to\neu'\,,
    \quad
    \dir\left(\varphi_0-\frac{2\pi R_2 l}{N}\right)\to\dir\,,
    \quad
    \neu(\varphi_0)\to\neu.
\end{equation}
 Zooming in on the red dotted regions \change{of}{on the} right-hand side of \eqref{FB_transfmatrix}
\begin{equation}
    \vcenter{\hbox{\includegraphics[width=0.3\linewidth]{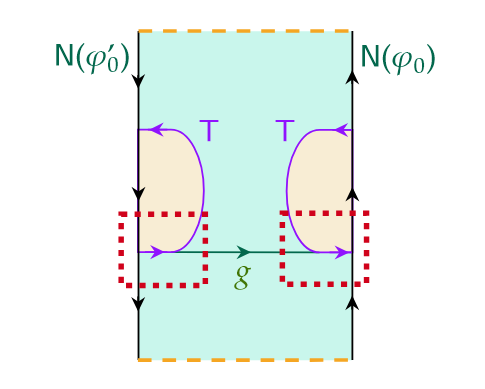}}}
\end{equation}
we can perform the following moves:
\begin{equation}
\vcenter{\hbox{\includegraphics[width=0.13\linewidth]{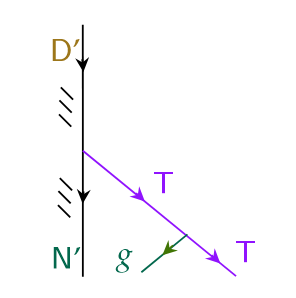}}}
      =
      \fusingG{\T g\neu}{\dir}{\T,g\fuse \neu}\hspace{0.4cm}\vcenter{\hbox{\includegraphics[width=0.13\linewidth]{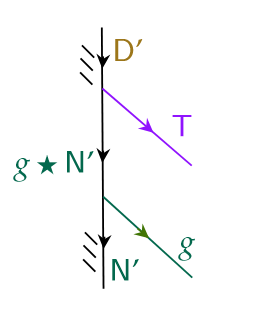}}}\,,
    \qquad
      \vcenter{\hbox{\includegraphics[width=0.15\linewidth]{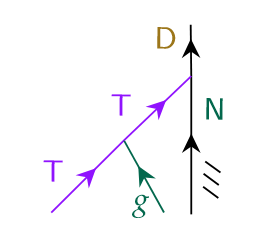}}}
     =
     \fusing{\T g\neu'}{\dir'}{g\fuse \neu',\T}\hspace{0.4cm}\vcenter{\hbox{\includegraphics[width=0.16\linewidth]{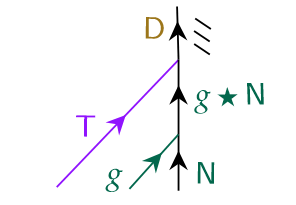}}}.
\end{equation}
Note that there is no sum in these fusing moves since the moved line is grouplike. Moreover, $g\in\stabR{\T}$ so that $\T \fuse (g\fuse \neu) = \T\fuse \neu$. Hence, the only allowed channels are the indicated ones. Recalling \eqref{GrouplikeOnDN} and the form of $\stabR{\T}$, see the right-hand side of \eqref{Tduality}, we in particular have,
\begin{equation} \label{B_boundaryconditionFB}
    \go{R_1}\left(\frac{m}{2MR_1}+\frac{wR_1}{N}, \frac{m}{2MR_1}-\frac{wR_1}{N}\right)\bket{\neu(\varphi_0)}^{R_1}
    =
    \bket{\neu(\varphi_0 -2\pi \hat{R}_1\frac{m}{M} )}^{R_1} .
\end{equation}
After these moves, \eqref{FB_transfmatrix} becomes
\begin{equation}\label{FB_transfermatrix2}
    \sum_{g\in \stabR{\T}}\frac{\qdim_g}{\qdim_\T}
    \fusing{\T g\neu'}{\dir' }{g\fuse \neu',\T}
    \vcenter{\hbox{
        \includegraphics[width=0.4\linewidth]{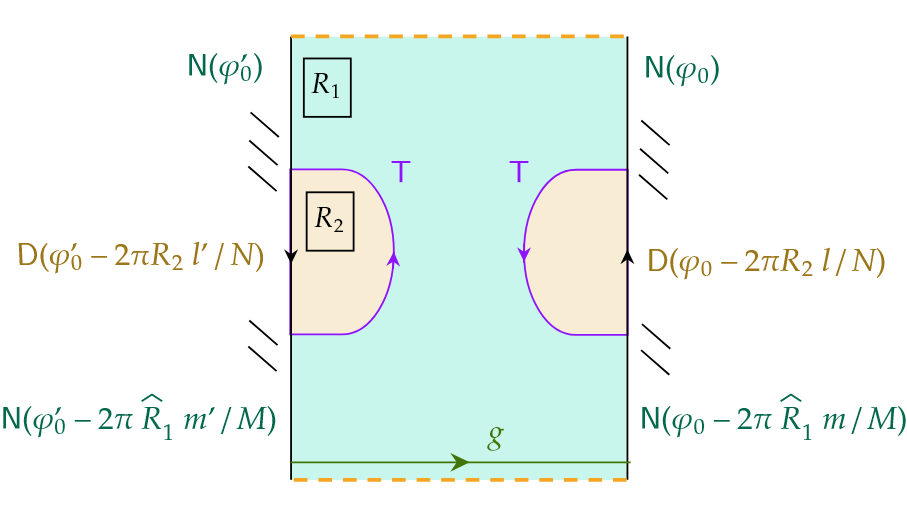}}} 
        \fusingG{\T g\neu}{\dir}{\T,g\fuse \neu}
\end{equation}
We can make use of the following inner product on junction spaces
\begin{align} \label{FB_delta_condition}
    \vcenter{\hbox{
        \includegraphics[width=0.3\linewidth]{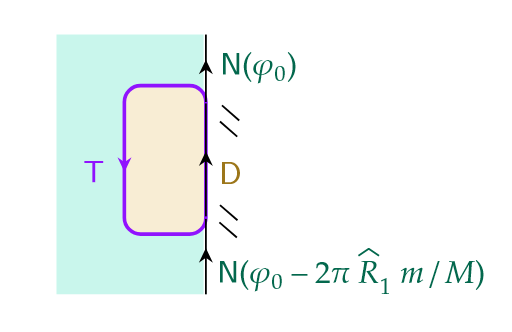}}} 
        =
        \begin{cases}
        0\,, & m \neq 0\mod M\,,\, w=0,\cdots,N-1 \\
        \lambda(\dir,\neu)\vcenter{\hbox{\includegraphics[width=0.15\linewidth]{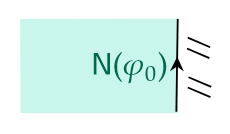}}}\,,& m=0 \mod M\,,\, w=0,\cdots,N-1
    \end{cases}
\end{align}
due to the fact that the trivalent interface-boundary junction spaces are 1-dimensional, see \eqref{defect_sum_FB_a}. The $\delta_{m ,0\text{ mod } M}$ does not only remove the momentum $m$ dependence in \eqref{B_boundaryconditionFB}, but more importantly reduces the sum in \eqref{FB_transfermatrix2} over the right stabilizer $\stabR\T$ to its zero-momentum $\Z_N$ subgroup. This stands in contrast to the discussion of the previous section and emphasizes that distinct defect networks -- distinguished here by means of the \change{incomming}{incoming} Neumann boundaries $\neu,\neu'$, which cannot be labelled by $\dua=\T$ -- easily lead to distinct answers. We are left with operators
\begin{equation} \label{subgrouplike_defects}
   \vcenter{\hbox{
        \includegraphics[width=0.3\linewidth]{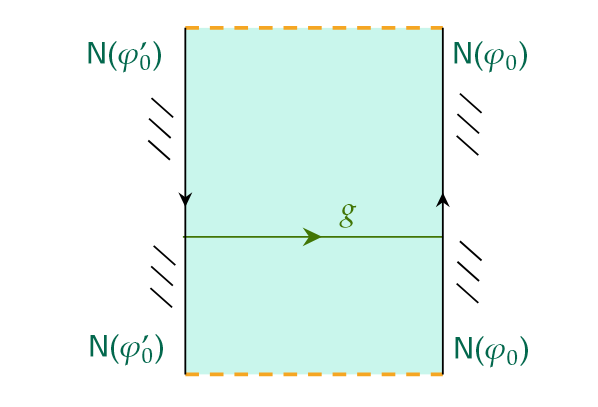}}} 
        \qquad
        g\equiv g_{R_1}\left(R_1\frac{w}{N},-R_1\frac{w}{N}\right)\in\Z_N\subset\stabR{\T}\subset\cG_{R_1}.
\end{equation}
In \appref{appBosonStrip}, we show that \eqref{FB_delta_condition} allows to build consistent $U(1)$ preserving twisted boundary states for this geometry. 

Keeping track of all factors, and using $\qdim_g=\qdim_{R_1}$\footnote{This is the equivalent of $\qdim_g=\qdim_P$, valid for $g\in\cG_P$ which stabilize some other $PQ$ interface \cite{Frohlich_2007}. If $g$ does not stabilize any other interface, one only finds $\qdim_g=\pm\qdim_P$.}, \eqref{FB_transfermatrix2} becomes
\begin{equation}
\begin{split}
    & \hspace{0.5cm}\sum_{g \in \stabR{\T}}
    \frac{\qdim_g}{\qdim_\T} 
    \fusing{\T g\neu'}{\dir' }{g\fuse \neu',\T}
    \vcenter{\hbox{\includegraphics[width=0.4\linewidth]{Pics/FB_3.png}}}
    \fusingG{\T g\neu}{\dir}{\T,g\fuse \neu} \\ 
    & =
    \frac{\qdim_{R_1}}{\qdim_\T}
    \lambda(\dir',\neu')\lambda(\dir,\neu)\sum_{g\in \Z_N \subset \stabR{\T}}
    \fusing{\T g\neu'}{\dir' }{g\fuse \neu',\T}
    \vcenter{\hbox{
        \includegraphics[width=0.3\linewidth]{Pics/FB_4.png}}} 
        \fusingG{\T g\neu}{\dir}{\T,g\fuse \neu}
\end{split}
\end{equation}
By introducing the $\Z_N$ generator $g_N$, all other elements can be written as $(g_N)^w$ with $w=0,\dots,N-1$. To avoid clutter, we write its action \change{}{on} the boundary Hilbert space for the Neumann boundaries $\neu(\varphi_0')$ and $\neu(\varphi_0)$ as $\check{g_N}$ and avoid appending extra labels.

Recall the labels \eqref{DNabbreviations}. For simplicity, we now restrict to the simplest case, $\varphi_0'=\varphi_0$ and $l'=l$, i.e. $\dir'=\dir$. The general case should follow the same pattern, as evidenced by our analysis in \secref{secDualityProjectorGeneral}. Because $\mathsf{F}$ and $\mathsf{G}$ are inverses and there is only one channel for the symbols above, we have
\begin{equation}
    \fusing{\T g\neu}{\dir}{g\fuse\neu,\T}\fusingG{\T g\neu}{\dir}{\T, g\fuse\neu}=1
\end{equation}
We arrive at
\begin{equation}
\begin{split}
   \vcenter{\hbox{
        \includegraphics[width=0.45\linewidth]{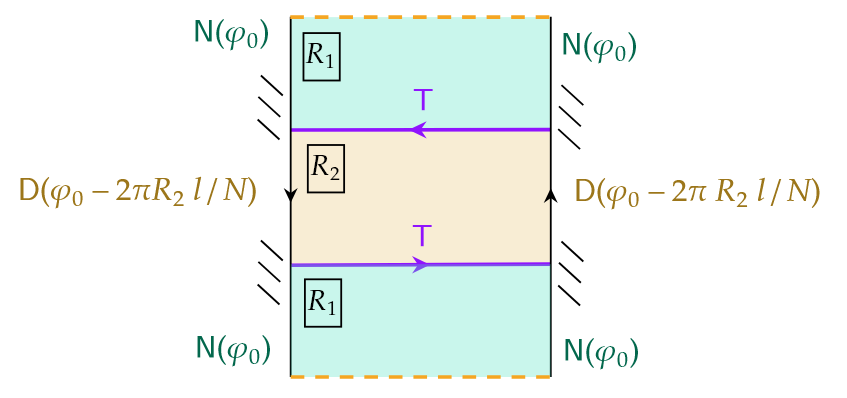}}} & 
        =
        \frac{\qdim_{R_1}}{\qdim_{\T}}\lambda(\dir,\neu)^2
        q^{H_{\neu\neu}}
        \sum_{w=0}^{N-1} (\check{g}_N)^w \\
        & =
        \frac{\qdim_{R_1}}{\qdim_{\T}}\lambda(\dir,\neu)^2
        q^{H_{\neu\neu}}
        \, N \cP_{0}^{\mathbb{Z}_N}.
\end{split}
\end{equation}
where we abbreviated the strip Hamiltonian $H_{\neu\neu}:=H_{\neu(\varphi_0)\neu(\varphi_0)}$. When constructing an RDM out of this operator, the prefactors cancel due to the normalization, so that we shall not evaluate them explicitly. In the last line we have recognized a projector $\cP^{\Z_N}_{0}$ onto the trivial representation of $\Z_N\subset\Z_{MN}=\stab{\T}$. This falls in line with our general discussion. Although we have not checked explicitly, we suspect that for $l'\neq l$, but still $\varphi_0'=\varphi_0$, the fusing symbols supply the $\Z_N$ character for the representation $l'-l=0,\dots,N-1$,
\begin{equation}
    \fusing{\T g\neu}{\dir}{g\fuse\neu,\T}\fusingG{\T g\neu}{\dir}{\T, g\fuse\neu}
    =
    \exp\left(-2\pi\iu\frac{w(l'-l)}{N}\right)
    =
    \rho_{l'-l}^{\mathbb{Z}_N}(w)^{-1}
\end{equation}
where $g$ supplies the $w$. This then would replace the projector $\cP_0^{\Z_N}$ by $\cP_{l'-l}^{\Z_N}$.

For the case of $\varphi_0'\neq\varphi_0$, we expect projective representations of $\Z_N$ to appear. This is backed by the explicit action of $\check{g}_N$ on the strip, which we evaluate in \eqref{gNstrip} of the appendix. Because projective representations of $\Z_N$ are equivalent to regular representations, we do not treat this case here separately.

We conclude that in the case of the free boson, the RDMs $\rho^{\id_\T\,\dir\dir}_{\neu\neu}$ built from \eqref{FB_transfmatrix} follow the structure \eqref{RDMdualityProjSimple} with the important difference that only representations of a subgroup of the stabilizer $\stab{\T}$ appear,
\begin{equation}\label{RDMTdualityProjSimple}
    \rho_{\dua\dua}^{\id_\T\, l}
   =
   \frac{
   q^{H_{\neu\neu}^{R_1}}\cP^{\Z_N}_{l}}{\chi^{\Z_N}_l(q)},
   \qquad
   \chi^{\Z_N}_l(q)
   =
   \tr_{\neu\neu}\left[q^{H_{\neu\neu}^{R_1}}\cP_{l}\right]
   \qquad
   l=0,\dots,N-1,
\end{equation}
where $\chi^{\Z_N}_l(q)$ is the multiplicity space of the $l$ representation of $\Z_N$.

\subsection{Comparison with thermal states}\label{secThermalStates}
Many physical phenomena do not pertain to ground or pure state, but thermal states. These may as well be dressed with interface networks and in this section we construct their RDMs. The reader is reminded that R\'enyi entropies are no useful measures to extract entanglement in thermal states, since they cannot separate thermal from quantum correlations. Other measures can instead be consulted, see \cite{LiuVertexStates} for some options. To that end it is important to understand how to construct their RDMs, which are fed into these information measures. They contain the full information of the subsystem after all. This presents furthermore a useful exercise in interface manipulations and sheds light on the zero-temperature limits considered above.

Let us consider the thermal density matrix $\rho(\tau)=e^{-\beta_0 H+i\ell P}=e^{2\pi i(\tau L_{0}-\bar{\tau}\bar{L}_{0})}$
with modular parameter $\tau:=(\ell+i\beta_0)/2\pi$, acting on the twisted Hilbert space $\Hjunc{a'}{a}{P}$ of defect fields $\phi^{(aa')}$ sitting at the junction between the lines $a$ and $a'$. The inverse temperature $\b_0$ is not to be confused with a boundary condition $\b\in\cB_P$. Pictorially\footnote{Upon taking the trace, this reproduces the twisted partition function $Z_{aa'}(\tau,\bar{\tau})={\rm Tr}_{\Hjunc{a}{a'}{P}}\rho(\tau)$} 
\begin{equation}
\rho(\tau)|_{{\cal H}_{(aa')}} = \vcenter{\hbox{\includegraphics[height=2.8cm]{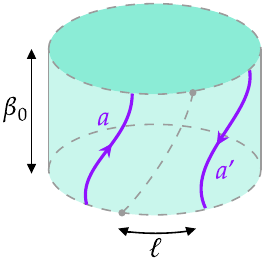}}} 
\end{equation}
Employing the Ohmori-Tachikawa homomorphism, we face the usual choice of relative position between the defects endpoints and the subregion. We consider only one example for simplicity here, where the defects are joined to the boundary conditions symmetrically at the opposite ends of the cylinder (analogous to (i) in figure \ref{fig:TwistedRDMs}). Tracing over the outside region we get the RDM: 
\begin{equation}\label{eq:thermal_RDM}
\iota_{\alpha\beta}\circ\rho(\tau)|_{\Hjunc{a}{a'}{P}}\circ\iota_{\ab}^{\dagger} 
=
\vcenter{\hbox{\includegraphics[height=2.5cm]{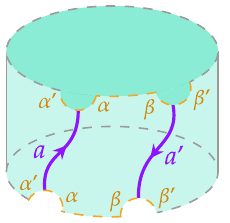}}} \qquad \Rightarrow \quad \rho_{\a\b}(\tau)  = \vcenter{\hbox{\includegraphics[height=2.5cm]{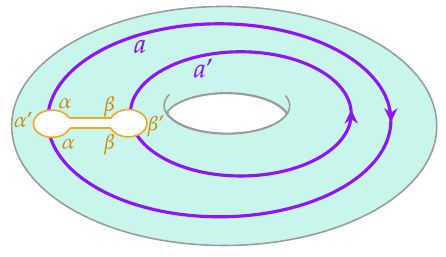}}}
\end{equation}

\begin{figure}
\begin{equation*}\begin{aligned}
\vcenter{\hbox{\includegraphics[height=2.5cm]{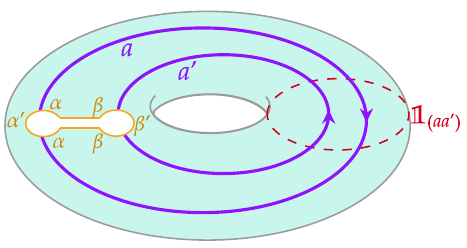}}} 
\cong 
\sum_{\phi\in {\cal H}_{(aa')}} \vcenter{\hbox{\includegraphics[height=2.5cm]{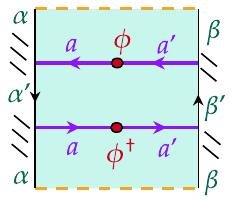}}} \xrightarrow[a = a']{\beta\to \infty} \vcenter{\hbox{\includegraphics[height=2.5cm]{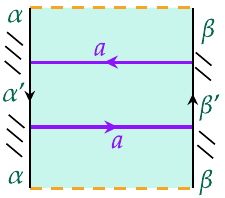}}}
\end{aligned}\end{equation*}
\caption{Steps to simplify the RDM for the thermal state. $\cong$ means a relation through an appropriate conformal map. The $\b_0 \to \infty$ limit is meaningful only for $a=a'$ and selects the vacuum state on the defect, reproducing a special case of \eqref{RDMtwoDualities} for $P=Q$.} \label{fig:thermal_RDM}
\end{figure}

It is useful to resolve the handle by inserting an identity decomposition along one of the two cycles of the torus. In the current frame, cutting transversally to the defects we insert $\id|_{\Hjunc{a}{a'}{P}}=\sum_{\phi \in {\cal H}_{aa'}}\ketbra{\phi}{\phi}$, which reduces the genus introducing two punctures on the sphere. After a suitable conformal transformation, the RDM is mapped to an infinite sum of terms in the form of the first option of \eqref{eq:defectFieldRDMs}, here for equal phases $P=Q$; notice however that the sum is over primaries \textit{and} descendant  fields. Sending $\beta_0\to\infty$ selects the vacuum state $|0^{(aa')}\rangle$, which is present only when $a=a'$, thus
reducing to the simpler case of $\phi^{(aa)}=\id_a$.
The above steps are graphically displayed in Fig. \ref{fig:thermal_RDM}.
Thus we obtain that applying the homomorphism \eqref{iota_def} to the thermal state in the defect Hilbert space $\Hjunc{a}{a'}{P}$, recovers RDMs of the type \eqref{RDMtwoDualities}. 

\begin{figure}
\begin{equation*}
\vcenter{\hbox{\includegraphics[height=2.5cm]{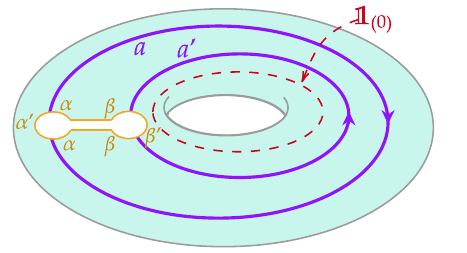}}} \cong  \sum_{\phi\in {\cal H}} \vcenter{\hbox{\includegraphics[height=2.5cm]{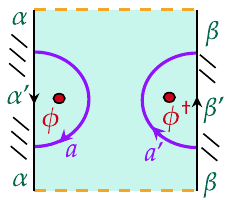}}} \xrightarrow{\ell\to \infty} \vcenter{\hbox{\includegraphics[height=2.5cm]{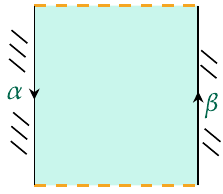}}}
\end{equation*}
\caption{Another simplification of the RDM for the thermal state. Using the S-transformed frame, the identity decomposition is in terms of bulk fields. The $\ell \to \infty$ limit selects the vacuum state, after which the defect network can be simplified and one recovers the bulk vacuum RDM.} \label{fig:thermal_RDM2}
\end{figure}

If instead we start from \eqref{eq:thermal_RDM} and perform an $S$ transformation, swapping ``time'' and ``space'' directions on the torus, we insert an identity decomposition $\id|_{{\cal H}}$ in the bulk (untwisted) Hilbert space cutting along the cycle that does not intersect any defect line.
After a suitable conformal transformation, the RDM is mapped to a
strip with the defect lines surrounding two bulk punctures and ending
respectively on the opposite boundaries. Upon taking the limit $\ell\to\infty$
we select the vacuum state $\ket{0}$ in the bulk Hilbert space. In this limit, the defect lines can be simplified by means of \eqref{eq:IntOnBdy}, which acts trivially here because there is no boundary field insertion\footnote{This is in fact a consequence of our choice of normalization for the boundary to defect junction fields. We follow the conventions of \cite{Fuchs:2002cm, Frohlich_2007}. See for instance \cite{Kojita:2016jwe, Konechny:2019wff} to see other conventions for these normalizations.}. We then recover the RDM for the vacuum state in the untwisted bulk Hilbert space. These steps are displayed graphically in \figref{fig:thermal_RDM2}.

\section{Information Theory in Presence of Duality Interfaces}\label{secInfoThyDuality}
In the previous section, we have constructed RDMs for the identity field $\id_\dua$ on duality interfaces, whose orientation reversal is also a duality interface. Hence their left and right stabilizers are isomorphic as groups and abelian, $\stabL{\dua}\cong\stabR{\dua}\equiv\stab{\dua}$. We now turn to the quantum information theory of these RDMs. It is best understood in relation to the information theory of the bulk vacuum field $\id$ corresponding to $\ket{0}\in\cH$. In \secref{secESBulkVac} we discuss the entanglement spectrum of the vacuum field $\id$ in the phase $P$  and partially also for $\id_\dua$. In \secref{secESDualityVac} we  turn explicitly to the entanglement spectrum for the duality interface identity field $\id_\dua$ and in \secref{secRelEnt} we compare the RDMs for $\id$ and $\id_\dua$ by means of their relative entropy. Finally, we construct entanglement spectra for the free boson in \secref{secFBentSpec}.
\subsection{Entanglement spectrum of the bulk vacuum state}\label{secESBulkVac}
It is instructive to work out the entanglement spectrum of the bulk vacuum state $\id$  of phase $P$ with RDM $\rho_{\dua\dua}^\id\in\End(\cH_{\dua\dua}^P)$, since the entanglement spectrum in presence of the duality interface, i.e. for the RDM $\rho^{\id_\dua\,\b}_{\dua\dua}\in\End(\cH_{\dua\dua}^P)$, is best understood in relation to it. 

The boundary states in phase $P$ for a factorization \eqref{iota_def} with $\a=\b=\dua$ and no interface junctions are constructed from any boundary state $\bket{h}^Q$ of phase $Q$ that can be labeled by $h\in\stabR{\dua}$ via
\begin{equation}\label{duaBdyState}
    \bket{\dua}^P:=\Io_\dua\bket{h}^Q,
    \qquad
    \Do_g\bket{\dua}^P=\bket{\dua}^P 
    \quad 
    \textrm{for }g\in\stabL{\dua}
\end{equation}
All boundary states $\bket{h}^Q$ lie in the $\stabR{\dua}$ orbit of the distinguished boundary state labeled by the unit element\footnote{In the categorical language, the appropriate label is in fact the Frobenius algebra $Q\in\confFam$ itself. Since the algebra $Q$ forms a module over itself, it behaves like the unit under fusion in its own phase. In particular, for any $PQ$ interface $\I$ the defect with label $Q$ behaves according to $\I\fuse_Q\D_Q\cong\I$. We hence write $\D_Q=\id^Q$.}, i.e. $\bket{h}^Q=\Do_h\bket{\id}^Q$. Without loss of generality we therefore restrict to the case $h=\id$. In the case of the Ising model, the duality interface is the Kramers-Wannier defect $\Do_\sigma$ and the boundary states $\bket{h}$ can be either $\bket{\epsilon}$ or $\bket{\id}$.

The property on the right of \eqref{duaBdyState} means that $\bket{\dua}$ is strongly symmetric under $\stabL{\dua}$, and hence $[H^P_{\dua\dua},\check{g}^\dua]=0$, where $\check{g}^\dua$ are the strip operators on the right-hand side of \eqref{GeneralProjector}. In consequence, the entanglement spectrum of the vacuum state $\id$ in phase $P$, i.e. the eigenspectrum of the RDM 
\begin{equation}\label{RDMdualityBdies}
    \rho_{\dua\dua}^\id
    =
    \frac{q^{H_{\dua\dua}^P}}{Z^P_{\dua\dua}(q)}
\end{equation}
decomposes into irreducible representations $r_\a\in\rep(\stabL{\dua})$, where we use $\a\in\stabR{\dua}$ to label the representations\footnote{This distinction may in fact be dropped in our case and we could simply write $r_\a\in\rep(\stab{\dua})$ with $\a\in\stab{\dua}$, but carrying the labels around helps to highlight the origin of all involved components when turning to duality interface-dressed RDMs.},
\begin{align}\label{ESvacuum}
    \cH_{\dua\dua}^P
    =
    \bigoplus_{\a\in\stabR{\dua}}r_\a\otimes\cH_\a,
    \qquad
    Z_{\dua\dua}^P(q)=\tr_{\dua\dua}^P\left[q^{H_{\dua\dua}^P} \right]=\sum_{\a\in\stabR{\dua}}\chi_\a(q)\,.
\end{align}
$\cH_\a$ is an infinite-dimensional multiplicity space with character $\chi_\a$. They appear with multiplicity $\dim(r_\a)=1$ because $\stabL{\dua}$ is abelian. Note that this has the form \eqref{SymmetrySplitting} and thus lends itself to symmetry resolution. To that end, it is useful to have the first charged moment for the element $g\in\stabL{\dua}$
\begin{align}
    \tr_{\dua\dua}^P\left[\check{g}^\dua q^{H_{\dua\dua}^P} \right]
    =
    \sum_{\a\in\stabR{\dua}}\tr_\a(g) \tr_{\a}\left[q^{H_{\dua\dua}^P} \right]
    \equiv
    \sum_{\a\in\stabR{\dua}}\varrho_\a(g) \chi_\a(q) 
\end{align}
where $\varrho_\a(g)$ is the group character for representation $r_\a$ and $\chi_\a(q)$ is its multiplicity space.

There is an additional insightful way to obtain $Z^P_{\dua\dua}$ via the phase $Q$ with bulk Hamiltonian $H^Q$. It employs the fundamental property of duality interfaces \eqref{duality}, 
\begin{align}\label{ESvacuumClosed}
    Z^P_{\dua\dua}(q)
    =
    {}^{P}\bbra{\dua}\tq^{H^P}\bket{\dua}^P
    &=
    {}^Q\bbra{\id}\tq^{H^Q}\bIo_\dua\Io_\dua\bket{\id}^Q\notag\\
    &=
    \sum_{h\in\stabR{\dua}}{}^Q\bbra{\id}\tq^{H^Q}\Do_h\bket{\id}^Q\notag\\
    &=
    \sum_{h\in\stabR{\dua}}{}^Q\bbra{\id}\tq^{H^Q}\bket{h}^Q
    =
    \sum_{h\in\stabR{\dua}}Z_{\id h}^Q(q)
\end{align}
We shall require the $\gf$-factors contained in $Z^P_{\dua\dua}$, which are extracted easily from the right-hand side at leading order in $\tq$,
\begin{subequations}\label{Zleading}
\begin{align}
 Z^Q_{\id h}(q)
 &=
 ^Q\bbra{\id}\tq^{H^Q}\bket{h}^Q
 \overset{\tq\to0}{\longrightarrow}
 \tq^{-\frac{\cc}{24}}
    {}^Q\bbra{\id}\Do_h\ket{0}^Q{}^Q\langle0\bket{\id}^Q
  =  
  \tq^{-\frac{\cc}{24}}
    \left(\gf_{\id}^Q\right)^2\label{ZleadingA}\\
    Z^P_{\dua\dua}(q)
    &\overset{\tq\to0}{\longrightarrow}
    \tq^{-\frac{\cc}{24}}
    |\stabR{\dua}|
    \left(\gf_{\id}^Q\right)^2
\end{align}
\end{subequations}
which used $\Do_h\ket{0}^Q=\frac{\qdim_h}{\qdim_Q}\ket{0}^Q=\ket{0}^Q$ \cite{Frohlich_2007} and $\gf_\id^Q=\langle0\bket{\id}^Q$.
The R\'enyi entropies are then
\begin{equation}\label{RenyiDualityBdies}
    S_n(\rho_{\dua\dua}^\id)
    =
    \frac{\cc \wi}{12}\frac{n+1}{n}+\log\left[|\stabR{\dua}|\left(\gf_\id^Q\right)^2\right]+\dots
\end{equation}
The fact that the summation indices on the right-hand sides of \eqref{ESvacuum} and \eqref{ESvacuumClosed} agree suggests $Z_{\id \a}^Q=\chi_\a$. This can be confirmed using the RDM \eqref{RDMdualityProjSimple} since it projects onto the representation $\b$ in \eqref{ESvacuum}. To similarly create a useful expression in the bulk channel of the strip, we unfuse the duality interfaces from the boundaries
\begin{equation}
    \vcenter{\hbox{\includegraphics[height=3cm]{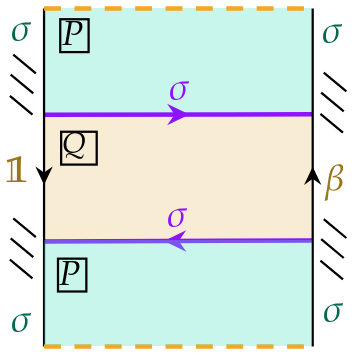}}}
    =
    \vcenter{\hbox{\includegraphics[height=3cm]{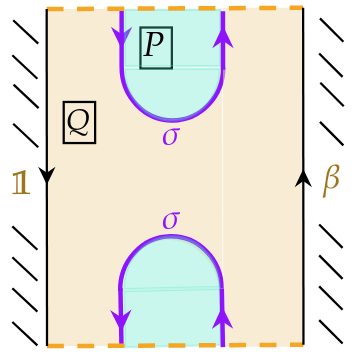}}}
\end{equation}
Note that one cannot unfuse the duality interfaces such that one directly obtains a loop of \change{}{the} duality interface $\I_\dua$. If this were the case, the effect of the interface on the entanglement spectrum would be trivial. 

Tracing this relation gives a non-trivial constraint. On the left-hand side, it means to trace $q^{H_{\dua\dua}}\,
   \ibo{\dua}{\dua\dua}{\id\b}\,
   \bibo{\dua}{\id\b}{\dua\dua}
=|\stabL{\dua}|\frac{\qdim_P}{\qdim_\dua}
    \, q^{H_{\dua\dua}^P}\cP_{\b}$, see \eqref{RDMDualityIntProj}, and leads to $|\stabL{\dua}|\frac{\qdim_P}{\qdim_\dua}
    \,\chi_\b(q)$. On the right-hand side one uses the standard relation \cite{Frohlich_2007} 
\begin{equation}
    \vcenter{\hbox{\includegraphics[height=2cm]{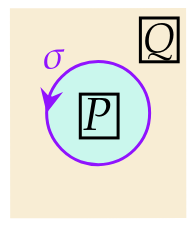}}}
    =
    \frac{\qdim_\dua}{\qdim_Q}
    \vcenter{\hbox{\includegraphics[height=2cm]{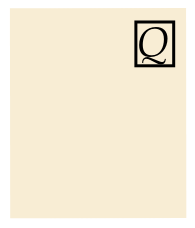}}}
\end{equation}
which leads to
\begin{align}
    |\stabL{\dua}|\frac{\qdim_P}{\qdim_\dua}
    \,\chi_\b(q)
    =
    \frac{\qdim_\dua}{\qdim_Q}\,{}^Q\bbra{\id}\tq^{H^Q}\bket{\b}^Q
    =
    \frac{\qdim_\dua}{\qdim_Q}{}Z_{\id\b}^Q(q)\,.
\end{align}
Finally, one employs that $|\stabL{\dua}|=\frac{\qdim_\dua^2}{\qdim_P\qdim_Q}$, which is found after tracing the defining relation \eqref{duality} \cite{Frohlich_2007}. One obtains the claim,
\begin{equation}\label{chiZ}
    \chi_\b(q)
    =
    Z_{\id\b}^Q(q)\,.
\end{equation}
\subsection{Entanglement spectrum of the duality interface vacuum field}\label{secESDualityVac}
The RDM $\rho_{\dua\dua}^{\id_\dua\,\b}\in\End{\cH^P_{\dua\dua}}$ in \eqref{RDMdualityProjSimple} projects onto the representation $\b$ in \eqref{ESvacuum}. Therefore, its entanglement Hamiltonian is proportional to $H_{\dua\dua}^P$ and its entanglement spectrum is encoded in $\tr_{\dua\dua}^P\left[\cP_\beta q^{H_{\dua\dua}^P} \right]=\chi_\b(q)$. This is one of the main results of this paper and falls into our lap given our detailed analysis in \secref{secRDMduality}. 

The overall construction clearly mirrors the formalism of symmetry resolution of entanglement \cite{Goldstein:2017bua, di2023boundary, northe2023entanglement}, which we briefly recalled around \eqref{SymmetrySplitting}. The main difference is that in symmetry resolution the symmetry projectors are inserted by hand in order to extract a desired charge or representation. Here, the projection is a consequence of the presence of duality interfaces. In other words, the duality interfaces act like a filter on the entanglement spectrum and therefore select particular quantum correlations for a fixed representation $\b$. The information loss through the interface is quantified by all representations that are lost upon projection. 

In the following, we shall convince ourselves that this information loss is, at least asymptotically in the UV cutoff $\varepsilon$, in fact independent of the specific choice of representation $\b$ in \eqref{RDMdualityProjSimple} and is ultimately dictated entirely by the phase $Q$ \quotes{behind the looking glass} $\I_\dua$ as viewed from the phase $P$. 

By introducing the modular $\modS_\ab$ matrix for the characters $\chi_\a$, we can easily write down the R\'enyi entropies for the RDM \eqref{RDMdualityProjSimple},
\begin{equation}\label{RenyiDuality}
S_n(\rho_{\dua\dua}^{\id_\dua,\b})
=
\frac{1}{1-n}\log\left(\frac{\chi_\b(q^n)}{(\chi_\b(q))^n}\right)
=
\frac{\cc\wi}{12}\frac{n+1}{n}+\log\modS_{\b0}+\dots
\end{equation}
where we assumed unitarity, imposing that the bulk channel character with smallest energy is $\chi_0(\tq)$. Details for this calculation are found in \cite{northe2023entanglement}. For Verlinde lines, the modular $\modS$ matrix can be related to the quantum dimension $\qdim_a=\frac{\modS_{a0}}{\modS_{00}}$. Because for grouplike Verlinde lines $\b\in\cG_\id$, we have $\qdim_\b=1$, so that $\modS_{\beta 0}=\modS_{00}$. In consequence, all $\rho_{\dua\dua}^{\id_\dua,\b}$ are asymptotically equipartitioned, as claimed.

For the general case, we resort to \eqref{chiZ}, which allows to write the subleading term in \eqref{RenyiDuality} as $2\log(\gf^Q_\id)$ by virtue of \eqref{ZleadingA}. We arrive at 
\begin{equation}
S_n(\rho_{\dua\dua}^{\id_\dua,\b})
=
\frac{1}{1-n}\log\left(\frac{\chi_\b(q^n)}{(\chi_\b(q))^n}\right)
=
\frac{\cc\wi}{12}\frac{n+1}{n}+2\log\gf_{\id}^Q+\dots
\end{equation}
which is independent of the intermediary boundary condition $\b$. It depends only on the $\gf$-factor of $\bket{\id}^Q$, which is in turn fixed by the phase $Q$, delivering the claim also for $PQ$ duality interfaces whose orientation reversal is also a duality interface. 

Finally, comparing with \eqref{RenyiDualityBdies} we find 
\begin{equation}\label{REreduction}
S_n(\rho_{\dua\dua}^\id)-S_n(\rho_{\dua\dua}^{\id_\dua,\b})
=
\log|\stabR{\dua}|+\dots,
\end{equation}
where the dots are terms which vanish when the UV cutoff is taken to zero. Hence, there is more uncertainty in the bulk vacuum field $\id$ than in the ground state field of the duality interface $\id_\dua$. This is clear because the entanglement spectrum of the latter is much smaller than that of the former. \change{Therefore quantum correlations between $A$ and $B$ are reduced, and one may think that information is reflected back into A by the duality interface.}{Given that the entanglement spectrum can be thought of as the entangling degrees of freedom living at the entangling edge, a reduced entanglement spectrum leads to reduced quantum correlations between $A$ and $B$. One may think of this as information being reflected back into $A$ by the duality interface. }

\subsection{Relative entropy}\label{secRelEnt}

We can now investigate how much duality interfaces modify a density matrix for the particular case that both RDMs lie in $\End(\cH_{\dua\dua}^P)$ by means of the relative entropy \eqref{RelEntropy}. When choosing $\rho=\rho_{\dua\dua}^{\id_\dua\,\a}$ as in \eqref{RDMdualityProjSimple} and $\sigma=\rho_{\dua\dua}^\id$ as in \eqref{RDMdualityBdies}, we observe that
\begin{align}
    \tr_{\dua\dua}\left[\rho_{\dua\dua}^{\id_\dua\,\a}\left(\rho_{\dua\dua}^{\id}\right)^{n-1}\right]
    =
    \frac{\chi_\a(q^n)}{\chi_\a(q)(Z^P_{\dua\dua}(q))^{n-1}}
    =
    \tr_{\dua\dua}\left[\left(\rho_{\dua\dua}^{\id_\dua\,\a}\right)^n\right]
    \left(\frac{\chi_\a(q)}{Z_{\dua\dua}^P(q)}\right)^{n-1}
\end{align}
Therefore,
\begin{align}
    S(\rho_{\dua\dua}^{\id_\dua\,\a}||\rho_{\dua\dua}^\id)
    =
    \lim_{n\to1}\frac{1}{1-n}\log\left[\left(\frac{\chi_\a(q)}{Z_{\dua\dua}^P(q)}\right)^{n-1}\right]
    =
    \log\left(\frac{Z_{\dua\dua}^P(q)}{\chi_\a(q)}\right)
    \overset{\tq\to0}{\longrightarrow}
    \log|\stabR{\dua}|
\end{align}
which used \eqref{Zleading} and \eqref{chiZ}. Note that this is independent of the specific choice $\a\in\stabR{\dua}$. The RDM $\rho_{\dua\dua}^{\id_\dua\,\a}$ can thus be approximated by $\rho_{\dua\dua}^\id$ at the cost of a finite error $\log|\stabR{\dua}|$. \change{}{We have thus found that the entanglement reduction in \eqref{REreduction} also properly quantifies the error made when approximating  $\rho=\rho_{\dua\dua}^{\id_\dua\,\a}$ by $\sigma=\rho_{\dua\dua}^\id$.} Indeed, the RDM $\rho_{\dua\dua}^{\id}$ has the information accounting for all representations of $\stabR{\dua}$ instead of only the single one in $\rho_{\dua\dua}^{\id_\dua\,\a}$. Note that when the stabilizer is the trivial group, i.e. the duality interface is a grouplike defect, the two RDMs encode the same probability distribution.

When choosing $\rho=\rho_{\dua\dua}^\id$ and $\sigma=\rho_{\dua\dua}^{\id_\dua\,\a}$, on the other hand, we observe that
\begin{align}
    \tr_{\dua\dua}\left[\rho_{\dua\dua}^{\id}\left(\rho_{\dua\dua}^{\id_\dua\,\a}\right)^{n-1}\right]
    =
    \frac{\chi_\a(q^n)}{Z^P_{\dua\dua}(q)(\chi_\a(q))^{n-1}}
    =
    \tr_{\dua\dua}\left[\left(\rho_{\dua\dua}^{\id_\dua\,\a}\right)^n\right]
    \frac{\chi_\a(q)}{Z_{\dua\dua}^P(q)}
\end{align}
Therefore,
\begin{align}
    S(\rho_{\dua\dua}^\id||\rho_{\dua\dua}^{\id_\dua\,\a})
    &=
    \lim_{n\to1}\frac{1}{1-n}\log\left[
    \frac{\tr_{\dua\dua}\left[\left(\rho_{\dua\dua}^{\id_\dua\,\a}\right)^n\right]}{\tr_{\dua\dua}\left[\left(\rho_{\dua\dua}^{\id}\right)^n\right]}
    \frac{\chi_\a(q)}{Z_{\dua\dua}^P(q)}\right]\notag\\
    &=
    S_1(\rho_{\dua\dua}^{\id_\dua\,\a})
    -
    S_1(\rho_{\dua\dua}^{\id})
    +
    \lim_{n\to1}\frac{1}{1-n}\log\left(\frac{Z_{\dua\dua}^P(q)}{\chi_\a(q)}\right)
\end{align}
This limit does not exist and $\rho^{\id}_{\dua\dua}$ cannot be approximated by $\rho_{\dua\dua}^{\id_\dua\,\a}$. This is intuitively clear, because the former RDM has a block decomposition over several representations while the latter contains only the block $\a$ and is zero on the other blocks. Mathematically, this is seen by recalling that $\rho^{\id}_{\dua\dua}$ has larger support than $\rho_{\dua\dua}^{\id_\dua\,\a}$. Physically, $\rho_{\dua\dua}^{\id\dua\,\a}$ is too \quotes{narrow} to approximate $\rho_{\dua\dua}^{\id}$.
\subsection{Free Boson Entanglement Spectra}\label{secFBentSpec}
Turning to the free boson, we similarly wish to compare the entanglement spectra for the fields $\id$ and $\id_\T$ in the phase $R_1$. Recall that we picked a factorization above, which places the Neumann boundary conditions \eqref{DNabbreviations} on the entangling edges. We directly work with the simple case that $\varphi_0'=\varphi_0$.

The entanglement spectrum of the RDM $\rho_{\neu\neu}^\id$, i.e. the empty strip with $\neu\neu$ boundaries, is captured by the strip partition function
\begin{equation}\label{NeumannSpec}
    Z_{\neu\neu}^\id(q)\equiv Z_{\neu\neu}(q)
    =
    \sum_{m\in\Z}\chi^{U(1)}_m(q)
    =
    \frac{1}{\eta(q)}\sum_{m\in\Z}q^{\frac{1}{2}\left(\frac{m}{R}\right)^2}
\end{equation}
where $\eta(q)$ is the Dedekind eta function. It allows to compute the R\'enyi entropies of the vacuum state
\begin{equation}\label{RenyiFBvac}
    S_n(\rho_{\neu\neu}^\id)
    =
    \frac{1}{1-n}\log\left[\frac{Z_{\neu\neu}(q^n)}{(Z_{\neu\neu}(q))^n}\right]
    =
    \frac{ \wi}{12}\frac{n+1}{n}+2\log\left[\gfn{R_1}\right]+\dots
\end{equation}

The spectrum \eqref{NeumannSpec} admits a $\Z_N$ action and decomposes accordingly,
\begin{equation}
    \cH_{\neu\neu}
    =
    \bigoplus_{m\in\Z}\cH_m^{U(1)}
    =
    \bigoplus_{l=0}^{N-1}r_l\otimes\cH_l^{\Z_N},
    \qquad
    Z_{\neu\neu}(q)=\sum_{l=0}^{N-1}\chi_l^{\Z_N}(q)
\end{equation}
The multiplicity space $\cH_l^{\Z_N}$ of each $\Z_N$ representation $r_l$ is extracted via the strip projector
\begin{align}
    \cP_{l}=\frac{1}{N}\sum_{w=0}^{N-1}e^{-2\pi\iu\frac{wl}{N}}(\check{g_N})^w
\end{align}
as easily seen by
\begin{align}\label{ZNcharacter}
   \chi_l^{\Z_N}(q)
   :=
     \tr_{\neu\neu}\left[\cP_l^{\Z_N}q^{H^{R_1}_{\neu\neu}}\right]
    &=
    \frac{1}{N}\sum_{w=0}^{N-1}\sum_{m\in\Z}e^{2\pi\iu\frac{w(m-l)}{N}}\chi^{U(1)}_m(q)\notag\\
    &=
    \sum_{m\in\Z}\delta_{m,l \text{ mod } N}\,\chi^{U(1)}_m(q)
    \qquad
    =
     \sum_{m\in N\Z+l}\chi^{U(1)}_m(q)
\end{align}
These are precisely the entanglement spectra of the RDMs \eqref{RDMTdualityProjSimple}. 

To obtain their R\'enyi entropies, it is useful to express these characters in the bulk channel $\tq$
\begin{align}
    \chi_l^{\Z_N}(q)
    &=
    \frac{1}{N}\sum_{w=0}^{N-1}e^{-2\pi\iu\frac{wl}{N}}
    \bbra{\neu_w(\varphi_0)}\tq^{H/2}\bket{\neu_w(\varphi_0)}\notag\\
    &=
    \frac{1}{N}\sum_{w=0}^{N-1}e^{-2\pi\iu\frac{wl}{N}}
    \sum_{k_w}
    \tq^{\frac{\Delta_w(k)}{2}-\frac{1}{24}}\bbra{\neu_w(\varphi_0)}k_w\rangle\langle k_w\bket{\neu_w(\varphi_0)}\notag\\
    &\approx 
    \frac{1}{N}\tq^{-\frac{1}{24}}(\gfn{R_1})^2
\end{align}
In the second line, we have expressed the action of $(\check{g_N})^w$ through the twisted boundary states $\bket{\neu_w(\varphi_0)}$. We explicitly construct these in \eqref{twistedNeumann}, but their concrete form is not necessary here. In the third line we have resolved the identity operator in the $w$-twisted sector by means of a basis $\ket{k_w}$. In the third line we have only kept the leading contribution in the regime $\tq\to0$, which minimizes the energy $\Delta_w(k)\to0$, which leads to $k=0$ and $w=0$.

The R\'enyi entropies are now read off,
\begin{equation}\label{RenyiFBT}
    S_n(\rho_{\neu\neu}^{\id_\T,l})
    =
    \frac{1}{1-n}\log\left[\frac{\chi_l^{\Z_N}(q^n)}{(\chi_l^{\Z_N}(q))^n}\right]
    =
    \frac{ \wi}{12}\frac{n+1}{n}+2\log\left[\gfn{R_1}\right]-\log N+\dots
\end{equation}
This is the expected symmetry resolution result \cite{Kusuki:2023bsp}. As in \secref{secESDualityVac}, the entropy is independent of the particular representation $l$ fixed by the intermediate boundary condition to subleading order in the UV cutoff. 

The uncertainty in this state is reduced with respect to that in the bulk vacuum state,
\begin{equation}
S_n(\rho_{\neu\neu}^\id)-S_n(\rho_{\neu\neu}^{\id_\T,l})
=
\log|\Z_N|+\dots,
\end{equation}
Therefore, the interval $A$ in the state $\id_\T$ is less correlated with the complement $B$ than in the state $\id$. This may again be interpreted as information being reflected at the interface $\T$. Relative entropy may be studied in complete analogy to the previous section.

\section{Discussion and Outlook}\label{sec:conclusion}
The purpose of this paper was to further develop the theory presented first in \cite{Northe:2025zmv} on entanglement in the presence of interfaces. Given that the prior theory for the quantification of entanglement through interfaces  \cite{Sakai_2008, Brehm:2015lja, Gutperle:2017enx, Brehm_2016, Gutperle_2016, gutperle2024note,capizzi2023domain} was violated in numerical simulations \cite{Roy_2022,Roy:2023wer}, the prior work \cite{Northe:2025zmv} proposed a new approach based on techniques which had already proven useful in the exploration of entanglement spectra \cite{Ohmori_2015, Lauchli:2013jga, Cardy_2016, northe2023entanglement}. 

The main insight was that entanglement through topological defects could be described by entanglement in twisted excitations, and that it depends on the defect network describing the physical setup. In the present paper, we have worked out the example of a grouplike twist field whose defect pierces the entangling interval in the free boson in \secref{secFBtwistField}. As with the grouplike twist in the Ising model \cite{Northe:2025zmv},\footnote{Note that also the theory presented in \cite{Brehm_2016} yields the correct result for grouplike twists, but fails for a Kramers-Wannier twist \cite{Roy_2022}.} the entanglement reduces to \change{the entanglement}{that} of the vacuum state as seen in \eqref{REgrouplikeFB}. This follows the expectation that grouplike defects only shuffle degrees of freedom \change{. Hence defects}{, hence they} do not affect the amount of quantum correlation between two subsystems. It would be interesting to prove this expectation for grouplike twists in generality or find counterexamples.  

The main aim of this paper was to extend this analysis to the situation of interfaces between two possibly distinct CFTs $P$ and $Q$ with various interface networks. This was done in \secref{secDualityIntRDM}. In this case, twist fields are not allowed and we are naturally \change{lead}{led} to the study of entanglement for interface fields. We discussed the construction of RDMs for all possible physical situations leading to distinct defect networks. We have restricted ourselves to the case of only two interfaces piercing the constant time slice harboring the entangling interval. Extensions to three or more defects are easily implemented by considering junction fields between several interfaces.

We then focus on the important case that the entangling interval $A$ is separated at its two edges by a duality interface $\I_\dua$\change{}{, as defined in \eqref{duality},} from its complement $B$; see \figref{fig:DefectInterfaceState} for $a_1=a_2=\dua$. \change{We showed in three different cases, namely diagonal theories, non-diagonal theories and the free boson, that the RDMs describing the physics on the subregion $A$ are projectors.}{We showed in three different cases that the RDMs describing the physics on the subregion $A$ are projectors: 1) Beginning with diagonal minimal models, the canonical interfaces are Verlinde lines, see \eqref{VerlindeLines}. They are in fact defects and carry no orientation, which means in particular that duality defects fuse with themselves according to \eqref{duality}. Verlinde lines can represent at most groups which are products of $\Z_2$, hence the projectors are onto representations of this group. As we show explicitly in the case of the Ising model, this group is the spin flip symmetry. 2) Generalizing to the more general case of proper interfaces, i.e. that distinct CFTs are glued, we reinstate orientations, but keep the requirement that the orientation inverse be also a duality interface. This requires that both phases have an abelian symmetry group, which stabilize the duality interface, and that these are isomorphic. The RDMs then project onto irreducible representations of this group. 3) In both previous cases the entangling edges were chosen such that they are compatible with the duality interface, see \eqref{duaBdyState}. While natural in both preceding cases, in the free boson, it is more natural to work with Neumann boundary conditions at the entangling edges, following the maximal entangling edge entropy principle \cite{Roy_2025}. The duality of interest is $\T$-duality here between boson theories of radius $R_1$ and $1/(2R_2)$, such that $2R_1R_2=N/M$ for coprime integers $M,N$. While its stabilizer group is $\stab{\T}=\Z_{M}\times\Z_N$, the entanglement spectrum only supports a $\Z_N$ symmetry. The construction of the RDM naturally reflects this, so that only projectors onto irreducibles of this smaller group are realized.} 

In all three cases, these RDMs may also be viewed as corresponding to the interface identity field $\id_\dua$. The entanglement spectrum of this field is in consequence only a subsector of the entanglement spectrum of the bulk identity field $\id$. \change{}{It is amusing to note that the formalism of symmetry resolution \cite{di2023boundary,northe2023entanglement,Kusuki:2023bsp} for the stabilizer symmetry $\stab{\dua}$ arises naturally when studying entanglement in the state $\id_\dua$. The physical setup must be clearly distinguished however. In symmetry resolution, one uses the fact that an entanglement spectrum carries representations of a symmetry $G$, and one implements projectors by hand to study the uncertainty contained in these sectors. In contrast, when studying entanglement through duality interfaces, the projectors emerge from the physical setup and implement the reflection of information at the interface.}

\change{We not}{}

Because the entanglement spectrum is diluted in presence of $\I_\dua$, the uncertainty\change{}{, quantified by the entanglement entropy,} reduces. Therefore quantum correlations between $A$ and its complement $B$ \quotes{behind the looking glass} $\I_\dua$ shrink, which can be seen as information being reflected at the interface back into $A$. The degree to which the uncertainty reduces is controlled by the order of the stabilizer of the duality interface $|\stab{\dua}|$, which is a natural symmetry associated with the interface. 

\change{In the case of the free boson we in fact only found a subgroup $\Z_N$ of the stabilizer $\stab{\T}=\Z_{MN}$, which is compatible with the spectrum of a $\neu\neu$ (Neumann-Neumann) strip. It would be interesting to understand in more detail, why and when only subgroups of the stabilizer arise.  }{}

\change{It is amusing to note that the formalism of symmetry resolution \cite{di2023boundary,northe2023entanglement,Kusuki:2023bsp} for the symmetry $\stab{\dua}$ arises naturally when studying entanglement in the state $\id_\dua$. The physical setup must be clearly distinguished however. In symmetry resolution, one uses the fact that an entanglement spectrum carries representations of a symmetry $G$, and one implements projectors by hand to study the uncertainty contained in these sectors. In contrast, when studying entanglement through duality interfaces, the projectors emerge from the physical setup and implement the reflection of information at the interface.}{}

\change{}{Finally, we have also compared how the changed entanglement spectra affect the distance between the two RDMs $\rho=\rho_{\dua\dua}^{\id_\dua\,\a}$ and $\sigma=\rho_{\dua\dua}^\id$. We found that the error made in approximating the former by the latter accurately captures the uncertainty difference between the two. On the other hand, approximating the latter by the former leads to an infinite error since the smaller entanglement spectrum of $\rho=\rho_{\dua\dua}^{\id_\dua\,\a}$ is too narrow to account for the quantum correlations mediated in the entanglement spectrum of $\sigma=\rho_{\dua\dua}^\id$.}

\subsection*{Outlook}
Moving forward, a comparison with numeric simulations is warranted. The interface RDMs presented here have so far not been simulated, but can presumably be implemented at least for the Ising model via the techniques of \cite{Roy_2022}. Similarly, it is interesting to construct the RDMs corresponding to the subsystems studied in \cite{Roy:2023wer} for Luttinger liquids and in \cite{Sinha:2023hum} for the three-state Potts model and compare field-theoretic with numeric results as for the Ising model in \cite{Northe:2025zmv}.

\change{}{Extending the study of projectors, it would be interesting to consider the case that the orientation reversal $\bI_\dua$ is not itself a duality interface. In this case the stabilizer groups are generically non-abelian. The most prominent case is the duality interface between the tetracritical Ising model and the three-state Potts model. It would be interesting to study other classes of non-invertible interfaces, for instance Fibonacci categories or Haagerup lines as used in \cite{Huang:2026lqf}.}

It would furthermore be interesting to study RDMs in presence of conformal interfaces. This presents many technical challenges, since interface-boundary junctions must be set up carefully and all deformations of topological interfaces are disallowed. This would provide an additional tool to test bounds on effective central charges \cite{Karch:2023evr, Karch:2024udk}. 

It would also be useful to study the behavior of entanglement spectra under defect RG flows. This would allow to investigate the variation of a subsystem's information content with energy scales. 

Furthermore, a theoretical discussion on the choice of boundary conditions -- even in the absence of interfaces and defects -- is necessary. Numerical observations for vacuum states of critical states indicate that the simple boundary conditions emerging in the factorization \eqref{iota_def} tend to maximize the boundary entropy \cite{Roy_2025}. This is at least the case for $A$-series minimal models, but exceptions are easily found outside of this class of theories, for instance in the three-state Potts model. First theoretic advances in the presence of defects are proposed in \cite{Northe:2025zmv}, and it would be useful to develop these further. 

Finally, an important related development concerns entanglement asymmetry, which has been conceived in order to study the restoration of a subsystem's symmetries after perturbations \cite{Ares:2022koq,Kusuki:2024gss, benini2025entanglementasymmetryhighernoninvertible, Fossati_2024, Yamashika:2024hpr, Benini:2025hbj}. These inherently utilize the representation of symmetries on the subsystem, as was done here. It would be interesting to elaborate the connection between defect-dressed RDMs and entanglement asymmetry clearly.

\acknowledgments
It is a pleasure to thank José Barbón, Saskia Demulder, Jared Heymann, Monica Kang, Thomas Quella, Ananda Roy, Martin Schnabl and Germán Sierra for useful discussions.
CN's work is funded by the European Union’s Horizon Europe Research and Innovation Programme under the Marie Sklodowska-Curie Actions COFUND, Physics for Future, grant agreement No 101081515. The research of RP was supported by the Grant Agency of the Czech Republic under the grant project 26-23188S. The research of PR was co-funded by the European Union and supported by the Czech Ministry of Education, Youth and Sports  (Project FORTE CZ.02.01.01/00/22\_008/0004632).


\begin{appendix}

\section{Bulk State RDMs and Primary State R\'enyi Entropies}\label{app:EE_bulkRDMs}
In this appendix, we demonstrate the use of the Hilbert space factorization map \eqref{iota_def} for the simple case of a bulk primary state, i.e. in the absence of interfaces, and compute its R\'enyi entropies.

In 1+1-dimensions, ${\cal H}$ is typically taken
to be the Hilbert space on the circle, following radial quantization in CFT.
Throughout this work the subregion $A$ is a single interval on the circle and $B=\bar{A}$, so the entangling edge consists of two points\footnote{Generalizing to disconnected intervals would be straightforward in principle, although it would cause interesting technical challenges which are not the focus of this note. 
}, so that $\underline{\a}=\{\a,\b\}$. To obtain $\iota_{\ab}$ we excise two small half-disks of radius $\varepsilon$ centered at the entangling points and impose boundary conditions\footnote{Introductions to BCFT can be found in \cite{Recknagel:2013uja, Northe:2024tnm}.} $\a,\, \b$, which we always take to be conformal.\footnote{As explained in \cite{Ohmori_2015}, non-conformal boundary conditions can be considered. However, in the limit of a large number of replicas, these would eventually flow to conformal ones.}
$\varepsilon$ is a UV-regulator and eventually eventually shrinks away, $\varepsilon\to0$. In pictures, for a pure global state $\ket{\phi}\in \cal H$ we have 
\begin{equation}\label{eq:globalDMfactorized}
 \ket{\phi}
 =
 \vcenter{\hbox{\includegraphics[height=2.5cm]{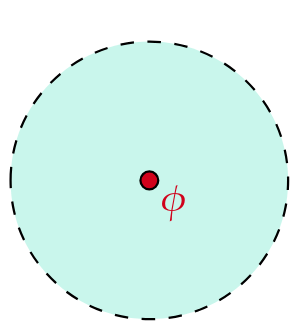}}}
 \quad
 \mapsto
 \quad
 \iota_\ab\ket{\phi}
 =
 \vcenter{\hbox{\includegraphics[height=2.5cm]{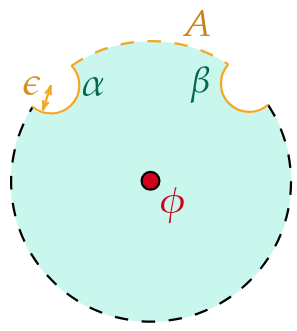}}}
\end{equation}
We place the interval $A$ on a radial arc with unit radius in the complex plane, parameterized by $z$, and fix endpoints $a=e^{-\iu \pi R}$ and $b=e^{\iu \pi R}$ with $R\in[0,1/2]$. For simplicity, we restrict ourselves to primary states $\ket{\phi}=\lim_{z,\bz\to0}\phi(z,\bz)\ket{0}\in\cH$ with conformal weights $(h_i,h_{\bi})$, although the construction here is straightforwardly extended to descendants. The labels $i,\bi$ are taken from the set of chiral representations $\confFam$ for fixed central charge. The conjugate state is as usual $\bra{\phi}=\lim_{z,\bz\to\infty}z^{2h}\bz^{2\bh}\bra{0}\phi^\dagger(z,\bz)$ and corresponds to the conjugate families $(i^+,\bi^{\,+})\in\confFam\times\confFam$.

One copy of $\iota_\ab\ket{\phi}$ and its adjoint are glued together along the domain $B$ by tracing over $\cH_{\ab}^B$. One is left with a sphere which is cut along $A$,
\begin{equation}\label{cutSphere}
 \rho_\ab^{A,\phi}
 =
 \tr_{B}[\iota_{\alpha\beta}\ketbra{\phi}{\phi}\iota_{\alpha\beta}^\dagger]
 =
 \frac{1}{N}\,
 \raisebox{-.45\height}{\includegraphics[scale=.22]{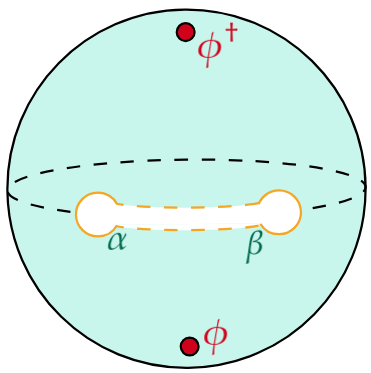}}
\end{equation}
with $N=\tr_A[\rho_\ab^{A,\phi}]$. We now recapitulate how to construct a replica annulus from the cut sphere and the conformal mapping of \eqref{cutSphere} to the strip \eqref{RDM} \cite{Ibáñez_Berganza_2012, Northe:2025qcv}.

The cut sphere is mapped onto a cut annulus via
\begin{equation}
 \xi(z)
 =
 e^{-\iu \pi R}\frac{e^{\iu\pi R}-z}{z-e^{-\iu \pi R}}
\end{equation}
The point $b$ is mapped to the origin and a boundary state $\bket{\beta}$ is imposed on its surrounding disk, while the point $a$ is mapped to infinity on the Riemann sphere and the boundary state $\bbra{\alpha}$ is imposed on its surrounding disk. The field insertions are mapped to $\xi(\infty)=e^{\iu\pi(1-R)}$, $\xi(0)=e^{\iu\pi(1+R)}$. The phase $e^{-\iu R}$ ensures that the interval $A$ is mapped onto the positive real axis of the $\xi$-plane. 


In order to compute $\tr_A[(\rho_\ab^{A,\phi})^n]$, $n$ cyclically glued copies of the sphere \eqref{cutSphere}, each parameterized by $z_m$ with $m=1,\dots,n$, are now mapped to an annulus via the uniformization map $\xi\to\xi^{1/n}$. The resulting replica annulus is mapped onto a cylinder slab of circumference 1 via
\begin{equation}\label{twCoordn}
 \tw(z_m)
 =
 \frac{1}{2\pi n \iu}
 \log\xi(z_m)
 =
 \frac{1}{2\pi n\iu}
 \log\xi(z)+\frac{m-1}{n}
\end{equation}
with modular nome $\tq^{1/n}=e^{2\pi \iu\ttau/n}=e^{-2\wi/n}$ and parameter $\ttau=\iu\frac{\wi}{\pi}$, which are expressed in terms of the (single-replica) annulus width $\tq^{\,-1/2}=e^{\wi}$ with
\begin{equation}\label{width}
 \wi
 =
 2\pi\biggl|\tw(b(1-\iu\pi\epsilon))-\tw(a(1+\iu\pi\epsilon))\biggr|
 =
 2\log\biggl(\frac{2}{\pi\epsilon}\sin(\pi R)\biggr)\,.
\end{equation}
The field insertion points are 
\begin{equation}\label{InsertionsTwN}
  \tw(\infty_m)
 =
 \frac{1-R}{2 n}+\frac{m-1}{n},
 \quad
 \tw(0_m)
 =
 \frac{1+R}{2 n}+\frac{m-1}{n}
\end{equation}
As boundary states are constructed from elements in $\cH$\footnote{Boundary states are non-normalizable in $\cH$ and thus strictly speaking not elements of $\cH$.}, the description is fully in terms of bulk degrees of freedom. It is useful to perform furthermore a modular $\modS$ transformation, since the entanglement spectrum is in one-to-one correspondence with the boundary spectrum \cite{Cardy_2016, di2023boundary, northe2023entanglement}. $\modS$ is implemented by $\tau=-1/\ttau=\iu\frac{\pi}{\wi}$ ($q=e^{-2\pi^2/\wi}$) and a rescaling,
\begin{align}\label{wCoordn}
 w(z_m)
 =
 n\tau\,\tw(z_m)
 =
 \frac{1}{2\wi}\log\xi(z_m)
\end{align}
The modular nome is now $q^n$ and the field insertions are mapped to
\begin{equation}\label{wCoordnInsertions}
 w(\infty_m)
 =
 \iu\frac{\pi}{2\wi}(1-R+2(m-1))\,,
 \qquad
 w(0_m)
 =
 \iu\frac{\pi}{2\wi}(1+R+2(m-1))\,.
\end{equation} 
While the $\tw$ frame is best for calculations since $\tq\to0$ when $\epsilon\to0$, the frame $w$ is best for (representation theoretic) analysis of the entanglement spectrum, see for instance \cite{Cardy_2016, di2023boundary, northe2023entanglement}.

A perk of the $w$ coordinates \eqref{wCoordn} is that the Jacobian transformation of the primary $\phi$ with holomorphic dimension $h$ drops out of the RDM \cite{Northe:2025qcv}. Hence, the RDM \eqref{cutSphere} takes the form advertised in the main text in \eqref{RDM} and repeated here for convenience:
\begin{equation}\label{appRDM}
 \rho_{\ab}^{\phi}
 =
 \frac{1}{Z^{\phi}_{\ab}(q)}
 \raisebox{-.5\height}{\includegraphics[scale=.15]{Pics/RDMphi.png}}\,,
 \qquad
 Z^{\phi}_{\ab}(q)
 =
 \tr_{\ab}\left[q^{H_{\ab}} \phi(w(0))\phi^\dagger(w(\infty))
 \right]
\end{equation}
The strip has height $\tau$, width $1/2$ and is understood to carry the evolution operator $q^{H_\ab}$, where $H_\ab$ is the Hamiltonian on the strip, and evolves from bottom to top. Hence for the vacuum, $\phi=\id$, the expected result $\rho_\ab^\id=q^{H_\ab}/Z_\ab^\id(q)$ is recovered \cite{Cardy_2016}, in which case the entanglement spectrum coincides with that of $H_\ab$, up to an additive shift. To avoid clutter the interval label $A$ has been dropped from the RDM \eqref{RDM}. 

To compute the R\'enyi entropies for \eqref{appRDM}, we employ \change{}{the} well-known relation $\tr[O\,q^{H}]=\tr[q^H]\corr{O}$ between traces and correlators and manipulate as follows:
\begin{align}\label{psiMoment}
 \tr_{\ab}\left[(\rho_{\ab}^\phi)^n\right]
 &=
 \frac{1}{(Z_\ab^\phi(q))^n}\tr_{\ab}\left[q^{nH_{\ab}}\prod_{m=1}^n\phi(w(0_m))\phi^\dagger(w(\infty_m))\right]\\
 &=\frac{1}{(Z_\ab^\phi(q))^n}Z_\ab(q^n)\,\corr{\prod_{m=1}^n\phi(w(0_m))\phi(w(\infty_m))}_\ab(q^n)\notag\\
 &=\frac{Z_\ab(q^n)}{(Z_\ab(q))^n}\,\frac{\corr{\prod_{m=1}^n\phi(w(0_m))\phi(w(\infty_m))}_\ab(q^n)}{\left(\corr{\phi(w_0)\phi^\dagger(w_\infty)}_\ab(q)\right)^n}\notag\\
 &=\tr_{\ab}\left[(\rho_{\ab}^\id)^n\right]\,\frac{\corr{\prod_{m=1}^n\phi(w(0_m))\phi(w(\infty_m))}_\ab(q^n)}{\left(\corr{\phi(w_0)\phi^\dagger(w_\infty)}_\ab(q)\right)^n}\notag
\end{align}
Writing $\X_n^\phi(w)$ for the collection of fields, the trace can also be evaluated in the $\modS$-modular frame
\begin{align}
\tr_\ab[g\,q^{nH_\ab}\X_n^\phi(w)]
=
(n\tau)^{-2nh_\phi}(n\btau)^{-2n\bh_\phi}
\bbra{\a_g}\tq^{\frac{H}{4n}}
\,\X_n^\phi(\tw)\,
\tq^{\frac{H}{4n}}\bket{\b_g}
\end{align}
Note that the fields are evaluated at the loci \eqref{twCoordn}. This frame is the one best used in actual calculations since \change{$\tq\to1$}{$\tq\to0$} at the end. In this limit, the amplitude reduces to a correlator on the sphere.

\section{More on the Free Boson}\label{app:ExtraFreeBoson}
In this appendix, we provide additional information and computations that aid in understanding the main text and filling in a few gaps. In \appref{appU1boundaries}, we provide a small review of $U(1)$ symmetric boundaries. In \appref{appFBdualityNeumann} we go on to justify our definition of the $\T$-duality interface and derive its action on a Neumann boundary state. Finally, in \appref{appBosonStrip}, we construct grouplike twisted Neumann boundary states and evaluate their overlaps.  
\subsection{$U(1)$-symmetric Boundary Conditions}\label{appU1boundaries}
Conformal boundary states satisfy the gluing condition
\begin{equation}
    (L_n-\bL_{-n})\bket{B}=0
\end{equation}
which secures $T=\bT$ at the boundary. The $U(1)\times U(1)$ symmetry of the free boson can furthermore accommodate an automorphism when gluing the currents, i.e. $J=\pm\bJ$ at the boundary. Because $T\sim JJ\,,\bT\sim \bJ\bJ$, conformal invariance of the boundary is preserved. This descends to gluing conditions on the $U(1)$ modes,
\begin{subequations}\label{IshibashiConditions}
\begin{align}
 \dir:\,\,  (a_n-\ba_{-n})\iket{(m,0)}=0\,,\qquad
 \neu:\,\,  (a_n+\ba_{-n})\iket{(0,w)}=0
\end{align}
\end{subequations}
The Ishibashi states solving these equations are built upon bulk states \eqref{FBstates},
\begin{subequations}\label{FBishibashi}
    \begin{align}
\dir:\qquad &\iket{m,0}=\exp\left(\sum_{n=1}^\infty\frac{1}{n}a_{-n}\ba_{-n}\right)\ket{m,0},
 &
 m\in\Z
 \\
 \neu:\qquad &\iket{0,w}=\exp\left(-\sum_{n=1}^\infty\frac{1}{n}a_{-n}\ba_{-n}\right)\ket{0,w},
 &
 w\in\Z
\end{align}
\end{subequations}
These are combined into the Cardy consistent Dirichlet and Neumann boundary states 
\begin{align}\label{appDirichletNeumann}
    &\bket{\dir(\varphi_0)}^{R}
    =
    \gfd{R}\sum_{m\in \mathbb{Z}}e^{\iu\frac{m}{R}\varphi_0} \iket{m,0}^{R},
    &
    \gfd{R}=\frac{1}{\sqrt{2R}}\\
    &\bket{\neu(\varphi_0)}^{R} 
    =
    \gfn{R}\sum_{w\in \mathbb{Z}}e^{-2\iu wR\varphi_0} \iket{0,w}^{R}\,,
    & 
    \gfn{R} = \sqrt{R}\, .
\end{align}
The superscripts on the kets will only be used when two theories need distinguishing, otherwise they will be dropped. 

\subsection{$\T$-Duality and Neumann Boundaries}\label{appFBdualityNeumann}
All interfaces \eqref{duality_defect_operators_FB} can be separated into one grouplike defect and a remainder which implements the map between theories
\begin{align}
    \Io^{\epsilon\bar\epsilon}_{R_2R_1}(x,y)
    =
    \Io^{\epsilon\bar\epsilon}_{R_2R_1}(0,0)\go{R_1}(x,y)
    =
    \go{R_2}(\epsilon x,\bar \epsilon y)\Io^{\epsilon\bar\epsilon}_{R_2R_1}(0,0)
\end{align}
One can easily check that
\begin{subequations}\label{DualityOnBulkState}
\begin{align}
    & \Io_{R_2 R_1}^{+ +}(0,0) \ket{p,\bar p}^{R_1} 
     =
     \sqrt{MN}\ket{p,\bar p}^{R_1},
     &
     \Io_{R_2 R_1}^{+ +}(0,0) \ket{m,w}^{R_1} 
     =
     \sqrt{MN}\ket{m,w}^{R_2},\\
     &\Io_{R_2 R_1}^{+ -}(0,0) \ket{p,\bar p}^{R_1}
     =
     \sqrt{MN}\ket{p,-\bar p}^{\hat R_2},
     &
     \Io_{R_2 R_1}^{+ -}(0,0) \ket{m,w}^{R_1} 
     =
     \sqrt{MN}\ket{w,m}^{R_2}.
\end{align}
\end{subequations}
Therefore $\Io_{R_2 R_1}^{+ +}(0,0)=\sqrt{MN}\id_\cap$. Note that there is no hat on the very last ket, which is seen as follows
\begin{equation}
    \ket{p,-\bar p}^{\hat R_2}:\quad
    p=\frac{m}{2\hat R_2}+w\hat R_2=\frac{w}{2 R_2}+m R_2
    \quad , \quad
    -\bar p=-\frac{m}{2\hat R_2}+w\hat R_2=\frac{w}{2 R_2}-m R_2.
\end{equation}
We observe that $\Io_{R_1R_1}^{+-}(0,0)$ implements $\T$-duality. For the reader's convenience, we repeat our definition of our $\T$-duality interface given in the main text,
\begin{equation}\label{appT}
    \T:=\I_{R_2 R_1}^{+-} (0,0)\,,
    \qquad
    \bar{\T}:=\I_{R_1 R_2}^{+-} (0,0).
\end{equation} 
Recall our slight abuse of notation regarding the appearance of arbitrary $R_2$. 

It is easily checked that the action of a duality interface on a bulk state \eqref{DualityOnBulkState} descends to the Ishibashi states
\begin{align}
    & \hat{I}_{R_2 R_1}^{+ +}(0,0) \iket{m,0}^{R_1} = \iket{m,0}^{R_2}\,,\qquad
    & \hat{I}_{R_2 R_1}^{+ -}(0,0) \ket{m,0}^{R_1} =\iket{0,m}^{R_2}.
\end{align}
and analogously for pure winding Ishibashi states. This allows us to $\T$-dualize a Neumann boundary state:
\begin{equation}
     \To \bket{\neu(\varphi_0)}^{R_1}
     =
     \gfn{R_1}\sqrt{MN} \sum_{w\in \Lambda_\cap} e^{2iwR_1 \varphi_0} \iket{w,0}^{R_2}
\end{equation}
Recalling the $\gfn{R}$-factors for Neumann/Dirichlet boundaries in \eqref{DirichletNeumann}, we can use the relation $MR_1=N\hat R_2=N/(2R_2)$ to rewrite $\gfn{R_1}=N\gfd{R_2}$. To continue, denote by $w$ the winding in the lattice $\Lambda(R_1)$ and by $k$ the winding in the lattice $\Lambda(R_2)$. In order for them to belong to the intersection lattice $\Lambda_\cap$, they need to satisfy
\begin{equation}
    w R_1 = k \hat{R}_2
    \qquad
    \Leftrightarrow
    \qquad
    w  =\frac{M}{N}k.
\end{equation}
Therefore, for $w$ to be integer, we require that $k \in N\mathbb{Z}$. Hence,

\begin{subequations}\label{appdefect_sum_FB_full}
\begin{align}
    \To \bket{\neu(\varphi_0)}^{R_1}
    &=
    N\gfd{R_2} \sum_{k\in N\mathbb{Z}} e^{2\iu k \hat R_2  \varphi_0}  \iket{k,0}^{R_2}\notag\\
    &=
    N\gfd{R_2}\sum_{k\in\mathbb{Z}}P_{k\hspace{0.5mm}\text{mod}N} e^{2\iu \frac{k}{R_2} \varphi_0} \iket{ k,0}^{R_2}\notag\\
    & =
    \sum_{l=0}^{N-1} \bket{\dir(\varphi_0 -\frac{2\pi R_2}{N}l)}^{R_2}\label{appdefect_sum_FB_a}\\
    &
    = \sum_{l=0}^{N-1} \go{R_2 } \left(R_2\frac{l}{N},-R_2 \frac{l}{N}\right) \bket{\dir(\varphi_0)}^{R_2} \label{appdefect_sum_FB}.
    \end{align}
\end{subequations}
By inserting the projector $P_{k\hspace{0.5mm}\text{mod}N}  = \frac{1}{N}\sum_{l=0}^{N-1} e^{-2\pi \iu \frac{kl}{N}}$ in the second line, we have extended the sum over $k$ to run over all integers. The result can be organized into a sum of simple Dirichlet boundaries with equidistant shift over the circle of radius $R_2$. Importantly, each summand has multiplicity 1, cf. \eqref{IntOnBdy}, which coincides with the dimensionality of the trivalent junction space between boundary and interface. In the last line, this sum is expressed by the action of group elements forming a $\Z_N\in\cG_{R_2}$. Note that the final expression can be written as the action of the projector onto the trivial $\Z_N$ representation $N \mathcal{P}_{\id}^{\mathbb{Z}_N} \bket{\dir(\varphi_0)}^{R_2}$. This is the result \eqref{defect_sum_FB_full} claimed in the main text.

\subsection{Free Boson Grouplike Defect Action on the Strip}\label{appBosonStrip}

In this appendix, we work out grouplike-twisted boundary Neumann boundary states which are required in dealing with strips such as in \eqref{subgrouplike_defects}. 

Start from the generic open string partition function with a defect attached to the boundaries $\a$ and $\b$. That is
\begin{equation} \label{opn_string_partition_function}
\begin{split}
        \text{tr}_{\a\b}[g \hspace{0.1cm}q^H] 
        & = \sum_{i}n_{\a\b}^i \text{tr}_{i}[g\hspace{0.1cm}q^H] \\
        & = \sum_{i} n_{\a\b}^i \check{I}_{i,\a\b}^{g,\ab} \chi_i(q)
\end{split}
\end{equation}
At the same time,
\begin{equation}
    \bbra{\a_g}\tq^{H/2}\bket{\b_g} 
    =
    \sum_{i,j}(B^{i}_{\a_g})^* B_{\b_g}^i S_{ij}\,\chi_j(q).
\end{equation}
where we defined $\bket{\a_g}=\sum_iB_{i}^{\a_g}\iket{i}$ as twisted boundary state. This gives a Cardy condition for twisted boundary states, which is in principle difficult to solve since $\niab \ib{g}{\ab}{i}{\ab}$ need not be an integer. We expect $\ib{g}{\ab}{i}{\ab}$ to be a phase entering the Cardy condition.

In order to write the twisted boundary states we are going to employ the bulk-boundary OPE of a twist field or rather the near-boundary behavior of a bulk twist field. Consider the following picture
\begin{equation}
    \vcenter{\hbox{\includegraphics[width=0.18\linewidth]{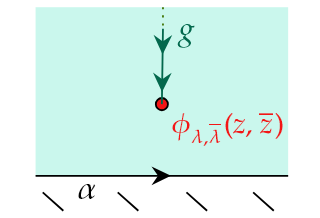}}}
\end{equation}
 By sending the twisted vertex operators on the boundary with the proper limit, we can deduce the twisted boundary state coefficient $B_{\a_g}^i$ and find an expression for $\check{I}_{i,\a\b}^{g,\a\b}$.\\
Consider a twisted vertex operator for the grouplike defects,
\begin{equation} \label{Vertex_operator_FB}
\begin{split}
        \phi_{\lambda,\bar\lambda}(z,\bar z) &=\hspace{0.2cm} :e^{i\lambda X(z) + i\bar\lambda \bar X (\bar z)}: \\
        &=\hspace{0.2cm}
        e^{i\lambda X_< (z)+i\bar \lambda \bar X_{<}(\bz)}e^{i\lambda \hat{x}_0+i\bar\lambda \hat{\bar x}_0}e^{\frac{\lambda}{4}\hat{q}\log z + \frac{\bar\lambda}{4}\hat{\bar q}\log \bz} e^{i\lambda X_>(z) + i\bar\lambda \bar X_>(\bz)}.
\end{split}
\end{equation}
for $(\lambda,\bar\lambda)\notin\Lambda(R)$, $(z,\bz)$ are coordinates on the plane, $X_> (X_<)$ contains all annihilators (creators) $a_n$ with $n>0\,(n<0)$ where the subscripts indicate positive or negative modes and $\hat{p},\hat{\bar p}$ are operators that extract the eigenvalues \eqref{FBstates}, i.e. $\hat q\ket{q,\bar q}=q\ket{q,\bar q}$, and finally $(\lambda,\bar\lambda)\in\Lambda(R|x,y)$, see \eqref{twistedLattice}. 

We are particularly interested in grouplike twists corresponding to the right-stabilizer $\stabR{\T}$, see \eqref{Tduality}. We introduce the shorthand 
\begin{equation}
    x_{m,w} = \frac{m}{2MR_1}+\frac{wR_1}{N}\hspace{0.5cm}, \hspace{2.5cm} y_{m,w}= \frac{m}{2MR_1}-\frac{wR_1}{N}.
\end{equation}
to parameterize the elements $\go{R_1}(x_{m,w},y_{m,w})\in\stabR{T}$. Hence we are looking at fields \eqref{Vertex_operator_FB} with
\begin{equation}
    (\lambda, \bar\lambda)
    =
    (p+x_{m,n}, \bar p + y_{m,n})
    \in
    \Lambda(R_1|x_{m,w},y_{m,w}).
\end{equation}
At the same time, it is useful to express $\Lambda(R_1|x_{m,w},y_{m,w})$ via 
\begin{equation}
    \lambda = \frac{u}{2R_1}+R_1v,\hspace{3.5cm}\bar\lambda=\frac{u}{2R_1}-R_1v.
\end{equation}
For our purpose, it suffices to restrict the one-dimensional zero momentum lattice defined by the $\mathbb{Z}_N$ defects of the form \eqref{subgrouplike_defects}. That is,
\begin{equation}\label{lambdav}
        \lambda = R_1v,\hspace{2.5cm}\bar\lambda=-R_1v.
\end{equation}
By standard free field theory techniques, we act with \eqref{Vertex_operator_FB} on a generic (Neumann) Ishibashi state and obtain
\begin{equation} \label{VO_on_Ishibashi}
     \phi_{\lambda,\bar\lambda}(z,\bar z)\iket{w} 
     =
     \frac{z^{\lambda R_1w/4}\bar z^{-\bar\lambda R_1w/4}}{(1-|z|^{-2})^{-\lambda\bar\lambda/4}}
     e^{i\lambda X_< (z)+i\bar \lambda \bar X_{<}(\bz)+i(\lambda X_>(z^-1) + \bar\lambda \bar X_>(\bz^-1))}\iket{w+v}.
\end{equation} 
Two remarks are in order: firstly, if we allowed a 2-dimensional lattice with non-vanishing momentum direction, the twisted Ishibashi state would pick a generically non vanishing momentum direction, therefore stopping to be $U(1)$ preserving. Secondly, the Ishibashi state, and therefore the boundary, lies at $z=\bz=1$ of the complex plane, hence the form of the singularity. 

Due to \eqref{lambdav}, we label a twisted Neumann boundary state by $v\in\Z$ and define
\begin{align}\label{twistedNeumann}
    \bket{\neu_v(\varphi_0)}^{R_1}
    &:=
    \lim_{z,\bz \rightarrow1}(1-|z|^{-2})^{\frac{\lambda^2}{4}}
    \phi_{\lambda,\bar\lambda}(z,\bar z)\bket{\neu(\varphi_0)}^{R_1}\notag\\
    &=
    \gfn{R_1}\sum_{w\in\Z}e^{-2\iu wR_1\varphi_0}
    \iket{w+v}^{R_1}.
\end{align}
The twisted annulus amplitude can now be evaluated by standard means
\begin{equation}
    \begin{split}
       &\bbra{\neu_v(\varphi_0')}\tq^{H/2}\bket{\neu_v(\varphi_0)}\\
        & =
        \left(\gfn{R_1}\right)^2 \sum_{w,w'\in \mathbb{Z}} e^{2 iw'R_1 \varphi_0'-2iw R_1\varphi_0} \ibra{w'+v}\tq^{H/2}\iket{w+v} \\
        & =
        \left(\gfn{R_1}\right)^2\sum_{w} e^{2 iw R_1 (\varphi_0'-\varphi_0)} \int_\R db\, S_{b,w+v}\chi_{b}(q)\\
        & =
        \left(\gfn{R_1}\right)^2 \int_\R db \sum_{w} e^{-2 iw R_1 \Delta \varphi_0} e^{2\pi i b(v+w)R_1} \chi_{b}(q)\\
        & =
        \left(\gfn{R_1}\right)^2 \int db \sum_{w} e^{2 iw R_1 ( b - \frac{\Delta \varphi_0}{\pi})} e^{2\pi i bv R_1} \chi_{b}(q).
    \end{split}
\end{equation}
We find that $\left(\gfn{R_1}\right)^2=R_1$.
We can now use the following identity
\begin{equation}
    \sum_w e^{2 \pi i w R_1 (b-\frac{\Delta \varphi_0}{\pi})}e^{2\pi i bv R_1}
    =
    \sum_m \delta\left[\left(bR_1 - R_1 \frac{\Delta\varphi_0}{\pi}\right)-m\right],
\end{equation} and we obtain the final expression
\begin{equation}
        \bbra{\neu_v(\varphi_0')}\tq^{H/2}\bket{\neu_v(\varphi_0)} = 
        \sum_m e^{2\pi i  v(\frac{\Delta \varphi_0}{\pi}+m)} \hspace{0.18cm} \chi_{\frac{\Delta \varphi_0}{\pi}+m}(q).
\end{equation}
Comparing with \eqref{opn_string_partition_function}, we can identify the phase
\begin{equation}
    \check{I}^{v\hspace{1cm}\varphi_0 \varphi_0'}_{{\frac{\Delta \varphi_0}{\pi}+m},\hspace{0.1cm}\varphi_0 \varphi_0'} = e^{2\pi i  v(\frac{\Delta \varphi_0}{\pi}+m)} .
\end{equation}
In particular,
\begin{align}\label{gNstrip}
    \check{I}^{\frac{w}{N}\hspace{1cm}\varphi_0 \varphi_0'}_{{\frac{\Delta \varphi_0}{\pi}+m},\hspace{0.1cm}\varphi_0 \varphi_0'} = \begin{cases}
         \rho_m^{\mathbb{Z}_N}(w) \,, &\varphi_0' =  \varphi_0 \\ 
         \rho_m^{\mathbb{Z}_N}(w) e^{2\pi i \frac{w}{N}\frac{\Delta \varphi_0}{\pi}}\,, &\varphi_0' \neq  \varphi_0 .
    \end{cases}
\end{align}
where we used $v=w/N$ for compatibility with \eqref{subgrouplike_defects} and $\varrho_m^{\Z_N}$ are $\Z_N$ characters for representation $m\in\Z$. These signal a regular and a projective representation. 

 \section{Categorical Phrasing}\label{app:CatPhrasing}
In this appendix, we translate our \quotes{physicists' language} into the categorical framework in which most of the tools employed in the main text were originally derived; see \cite{Fuchs:2002cm, Frohlich_2007} for more details. 

It is useful to recall how the various labels in the pictures present in this article can be expressed in terms of chiral data. This allows us to simplify the RDMs through defect manipulations like fusion, crossing etc. 

Given a chiral algebra ${\cA}$, the associated Modular Tensor Category (MTC) of representations is denoted ${\cC}={\rm Rep}({\cA})$. Conformal families are labeled by a set $\confFam$ and correspond to irreducible ${\cA}$-modules, which are the simple objects of $\cC$. 
A phase, or choice of spectrum for the CFT, is labeled by a symmetric special Frobenius algebra object $P$ in ${\cC}$. The boundary conditions $\a,\a',\b,\b'$ are labelled by simple objects in the category of left $P$-modules ${}_P{\cC}$, whose label set we call $\cB_{P}$. Accordingly, the boundary orientation is conventionally chosen so that the worldsheet lies on its left. 

Topological interfaces $a,b \in \TopInt{PQ}$ are labeled by simple objects in the category of left-right $P$-$Q$-bimodules $_{P}{\cC}_{Q}$. Accordingly, the orientation of the interface should be such that the phase $P$ lies on its left and the phase $Q$ on its right. If $\mu$ is a defect within a single CFT labeled by phase $P$, then it corresponds to a simple object in the category of $P$-$P$-bimodules $_{P}{\cC}_{P}$. Notice that the \quotes{identity line defect} corresponds to the object $P$ in $_{P}{\cC}_{P}$. 

Topological trivalent junctions have labels corresponding to the different ways these objects can fuse into each other: for example, the junction $a$ is a basis element in the space ${\Hom}_{P}(\alpha,\mu\otimes_{Q}\alpha')$ within $_{P}{\cC}$ and $a^+$ is its dual element in ${\Hom}_{P}(\mu\otimes_{Q}\alpha',\alpha)$.

Lastly, an interface field $\phi_{ij}^{(a,b)}$, see \eqref{InterfaceField}, corresponding to a state in $\Hjunc{a}{b}{PQ}$ with $(i,j)\in\confFam\times\confFam$ and $a,b\in\TopInt{PQ}$ corresponds to an element of ${\Hom}_{P|Q}(U_{i}\otimes^{+}\I_a\otimes^{-}U_{j},\I_b)$. We remark that when the phases $P,Q\cong{\id}$ are labeled by the trivial Frobenius algebra, the associated CFT has diagonal spectrum, e.g. A-series minimal models, and all the Hom spaces above reduce to the ones in the category ${\cC}$. 

\end{appendix}

\bibliographystyle{ytphys}
\bibliography{entanglement_entropy_biblio.bib}
\end{document}